\documentclass[12pt]{article}
\usepackage{a4wide}
\usepackage[centertags]{amsmath}
\usepackage{amssymb}
\usepackage[sort&compress,numbers]{natbib}
\usepackage{graphicx}
\newcommand{\cdt}{\!\cdot\!}

\newcommand{\prn}[1]{\left ( #1 \right )}

\newcommand{\brk}[1]{\left [ #1 \right ]}
\newcommand{\abs}[1]{\left\lvert #1 \right\rvert}
\newcommand{\nrm}[1]{\left\lVert #1 \right\rVert}

\newcommand{\diff}[3][\rule{0mm}{0mm}]{\frac{\mathrm{d}^{#1} #2}{\mathrm{d}{#3}^{#1}}}

\newcommand{\pdiffc}[3][\rule{0mm}{0mm}]{\left (\frac{\partial #2}{\partial {#3}}\right )_{\!\!#1}}

\newcommand{\dr}{\mathrm{d}}
\newcommand{\e}{\mathrm{e}}

\DeclareMathOperator{\Tr}{Tr}

\newlength{\lingap}
\newcommand{\sgap}{\\[-\lingap]}

\newcommand{\CO}{\mathcal{O}}

\newcommand{\CN}{\mathcal{N}}

\newcommand{\R}{\mathbb{R}}

\newcommand{\p}{\partial}

\newcommand{\half}{\frac{1}{2}}

\newcommand{\tc}{\tau}
\newcommand{\tloc}{\mathcal{T}}

\newcommand{\eloc}{\mathcal{E}}
\newcommand{\sloc}{\mathcal{S}}
\newcommand{\ploc}{\mathcal{P}}
\newcommand{\rloc}{\mathcal{R}}
\newcommand{\gpf}{\mathcal{Z}_\mathrm{gc}}
\title{Large rotating AdS black holes from fluid mechanics}
\author{ Sayantani Bhattacharyya$^{(a)}$, Subhaneil Lahiri$^{(b)}$,\\
 R. Loganayagam$^{(a)}$ and Shiraz Minwalla$^{(a)}$
\\
\small{\emph{$^{(a)}$Department of Theoretical Physics, Tata Institute of Fundamental Research,}}\\
\small{\emph{Homi Bhabha Rd, Mumbai 400 005, India}}\sgap
\small{\emph{$^{(b)}$Jefferson Physical Laboratory, Harvard University, Cambridge MA 02138, USA}}\sgap
}

\begin{document}

\maketitle

%% Preprint numbers, etc.
%\preprintno{8cm}{6cm}{
%    \texttt{arXiv:yymm.nnnn [hep-th]}
%}

%%%%%%%%%%%%%%%%%%%%%%%%%%%%%%%%%%%%%%%%%%%%%%%%%%%%%%%%%%%%%%%%%%%%%%%%%%

\begin{abstract}
We use the AdS/CFT correspondence to argue that large rotating black holes in global AdS$_D$ spaces are dual to stationary solutions of the relativistic Navier-Stokes equations on $S^{D-2}$. Reading off the equation of state of this fluid from the thermodynamics of non-rotating black holes, we proceed to construct the nonlinear spinning solutions of fluid mechanics that are dual to rotating black holes. In all known examples, the thermodynamics and the local stress tensor of our solutions are in precise agreement with the thermodynamics and boundary stress tensor of the spinning black holes. Our fluid dynamical description applies to large non-extremal black holes as well as a class of large non-supersymmetric extremal black holes, but is never valid for supersymmetric black holes. Our results yield predictions for the thermodynamics of all large black holes in all theories of gravity on AdS spaces, for example, string theory on $AdS_5 \times S^5$ and M theory on $AdS_4 \times S^7$ and $AdS_7 \times S^4$.
\end{abstract}

\tableofcontents

%%%%%%%%%%%%%%%%%%%%%%%%%%%%%%%%%%%%%%%%%%%%%%%%%%%%%%%%%%%%%%%%%%%%%%%%%%
% Beginning of Article:
%%%%%%%%%%%%%%%%%%%%%%%%%%%%%%%%%%%%%%%%%%%%%%%%%%%%%%%%%%%%%%%%%%%%%%%%%%

\section{Introduction}\label{sec:intro}

In this paper, we predict certain universal features in the thermodynamics and other classical properties of large rotating black holes in global AdS$_D$ spaces for arbitrary $D$. Our analysis applies to black holes in any consistent theory of gravity that admits an AdS$_D$ background; for example, IIB theory on $AdS_5 \times S^5$ or M theory on $AdS_7 \times S^4$ or $AdS_4 \times S^7$.

All theories of gravity on an AdS$_D$ background are expected to admit a dual description as a quantum field theory on $S^{D-2} \times $ time \cite{Maldacena:1997re,Witten:1998qj}. Moreover, it is expected to be generally true that quantum field theories at sufficiently high energy density admit an effective description in terms of fluid dynamics. Putting together these facts, we propose that large, rotating black holes in arbitrary global AdS$_{D}$ spaces admit an accurate dual description as rotating, stationary configurations of a conformal fluid on $S^{D-2}$.

Assuming our proposal is indeed true, we are able to derive several properties of large rotating AdS black holes as follows: We first read off the thermodynamic equation of state of the dual `fluid' from the properties of large, static, non-rotating AdS black holes. Inputting these equations of state into the Navier-Stokes equations, we are then able to deduce the thermodynamics of rotating black holes. In the rest of this introduction, we will describe our proposal and its consequences, including the tests it successfully passes, in more detail.

Consider a theory of gravity coupled to a gauge field (based on a
gauge group of rank $c$) on AdS$_D$. In an appropriate limit, the
boundary theory is effectively described by conformal fluid dynamics
with $c$ simultaneously conserved, mutually commuting $U(1)$ charges
$R_i$ ($i=1 \ldots c$). Conformal invariance and extensivity force the
grand canonical partition function of this fluid to take the form
\begin{equation} \label{pffom:eq}
 \frac{1}{V}\ln \gpf=  T^{d-1}\ h(\zeta/T)\,,
\end{equation}
where $\zeta$ represents the set of the $c$ chemical potentials conjugate to the $U(1)$ charges of the fluid, $V$ and $T$ represent the volume and the overall temperature of the fluid respectively and $d=D-1$ is the spacetime dimensions of the boundary. As we have explained above, the as yet unknown function $h(\zeta/T)$ may be read off from the thermodynamics of large, charged, static black holes in AdS.

The thermodynamic equation of state described above forms an input into the relativistic Navier-Stokes equations that govern the effective dynamics of the boundary conformal fluid. The full equations of fluid dynamics require more data than just the equation of state; for example we need to input dissipative parameters like viscosity. However, fluid dynamics on $S^{d-1}$ admits a distinguished $c+n+1$ parameter set of stationary solutions (the parameters are their energy $E$, $c$ commuting charges $R_i$ and $n=\operatorname{rank} (SO(d))=\left[\frac{d}{2}\right]$ commuting angular momenta\footnote{Here, we use the notation $[x]$ to denote the integer part of the real number $x$. For a list of notation, see appendix \ref{app:notation}.} on $S^{d-1}$). These solutions are simply the configurations into which any fluid initial state eventually settles down in equilibrium. They have the feature that their form and properties are independent of the values of dissipative parameters.

Although these solutions are nonlinear (i.e.\ they cannot be thought of as a small fluctuation about a uniform fluid configuration), it turns out that they are simple enough to be determined explicitly. These solutions turn out to be universal (i.e.\ they are independent of the detailed form of the function $h(\zeta/T)$). Their thermodynamics is incredibly simple; it is summarised by the partition function
\begin{equation}\label{final:eq}
 \ln \gpf =
 \ln \Tr \exp \left[ -\frac{(H - \zeta_i R_i - \Omega_a  L_a)}{T} \right]
 = \frac{V_{d} T^{d-1} h(\zeta/T)}
       {\prod_{a=1}^{n}(1-\Omega_a^2)}\,,
\end{equation}
where $H$,$L_a$ and $\Omega_a$ represent the energy, angular momenta and the angular velocities of the fluid respectively and $V_d=\operatorname{Vol}(S^{d-1})$ is the volume of the sphere
$S^{d-1}$.

We now turn to the gravitational dual interpretation of the fluid
dynamical solutions we have described above. A theory of a rank $c$ gauge field,
interacting with gravity on AdS$_D$, possesses a $c+n+1$ parameter set
of black hole solutions, labelled by the conserved charges
described above. We propose that these black holes (when large) are
dual to the solutions of fluid dynamics described in the previous
paragraph. Our proposal yields an immediate prediction about the
thermodynamics of large rotating black holes: the grand canonical
partition function of these black holes must take the form of
\eqref{final:eq}.

Notice that while the dependence of the partition function \eqref{final:eq} on $\zeta/T$ is arbitrary, its dependence on $\Omega_a$ is completely fixed. Thus, while our approach cannot predict thermodynamic properties of the static black holes, it does allow us to predict the thermodynamics of large rotating black holes in terms of the thermodynamics of their static counterparts.\footnote{ The analogue of our procedure in an asymptotically flat space (which we unfortunately do not have) would be a method to deduce the thermodynamics and other properties of the charged Kerr black hole, given the solution of static charged black holes.} Further, our solution of fluid dynamics yields a detailed prediction for the boundary stress tensor and the local charge distribution of the corresponding black hole solution, which may be compared with the boundary stress tensor and currents calculated from black hole solutions (after subtracting the appropriate counterterms \cite{Kraus:1999di, Henningson:1998gx, de-Haro:2000xn, Skenderis:2000in, Papadimitriou:2005ii, Cheng:2005wk, Olea:2005gb, Olea:2006vd}).

Our proposal is highly reminiscent of the membrane paradigm in black hole physics (see, for instance, \cite{Thorne:1986iy, Cardoso:2007ka}). However, we emphasise that our fluid dynamical description of black holes is not a guess; our proposal follows directly from the AdS/CFT correspondence in a precisely understood regime (see \cite{Aharony:1999ti} for a review of AdS/CFT correspondence).\footnote{Alternatively, one could regard the agreement between fluid dynamics and gravity described below as a test of the AdS/CFT correspondence (provided we are ready to assume in addition the applicability of fluid mechanics to quantum field theories at high density).}

We have tested the thermodynamical predictions described above on every class of black
hole solutions in AdS$_D$ spaces that we are aware of. These solutions
include the most general uncharged rotating black holes in arbitrary
AdS$_D$ space \cite{Hawking:1998kw, Gibbons:2004uw, Gibbons:2004ai}, various classes
of charged rotating black holes in $AdS_5 \times S^5$
\cite{Cvetic:2004ny, Chong:2005hr, Chong:2005da,Chong:2006zx}, in
$AdS_7\times S^4$ and in $AdS_4 \times S^7$ \cite{Chong:2004dy,
Chong:2004na, Cvetic:2005zi}. In the strict fluid dynamical limit, the
thermodynamics of each of these black holes exactly reproduces\footnote{See, however, \S\S\ref{sec:nlo} for a puzzle regarding the first subleading corrections for a class of black holes in AdS$_5$.} \eqref{final:eq}. In all the cases we have checked, the boundary
stress tensor and the charge densities of these black holes (read off
from the black hole solutions using the AdS/CFT dictionary) are also
in perfect agreement with our fluid dynamical solutions.
The agreement described in this paragraph occurs only when one
would expect it to, as we now explain in detail.

Recall that the equations of fluid mechanics describe the evolution of local energy densities, charge densities and fluid velocities as functions of spatial position. These equations are applicable only under certain conditions. First, the fluctuations about mean values (of variables like the local energy density) must be negligible. In the situations under study in our paper, the neglect of fluctuations is well justified by the `large $N$' limit of the field theory, dual to the classical limit of the gravitational bulk.

Second, the equations of fluid mechanics assume that the fluid is in local thermodynamic equilibrium at each point in space, even though the energy and the charge densities of the fluid may vary in space. Fluid mechanics applies only when the length scales of variation of thermodynamic variables - and the length scale of curvatures of the manifold on which the fluid propagates - are large compared to the equilibration length scale of the fluid, a distance we shall refer to as the mean free path $l_\mathrm{mfp}$.

The mean free path for any fluid may be estimated as \cite{Son:2007vk} $l_\mathrm{mfp}\sim\frac{\eta}{\rho}$ where $\eta$ is the shear viscosity and $\rho$ is the energy density of the fluid. For fluids described by a gravitational dual, $\eta =\frac{s}{4 \pi}$ where $s$ is the entropy density \cite{Son:2007vk}. Consequently, for the fluids under study in this paper, $l_\mathrm{mfp} \sim \frac{s}{4\pi\rho}$. As we will see in \S\S\ref{sec:validity}, the most stringent bound on $l_\mathrm{mfp}$, for
the solutions presented in this paper, comes from requiring that $l_\mathrm{mfp}$ be small compared to the radius of the $S^{d-1}$, which we set to unity. Consequently, fluid dynamics should be an accurate description for our solutions whenever $\frac{s}{4\pi\rho} \ll 1$. In every case we have studied, it turns out that this condition is met whenever the horizon radius, $R_H$, of the dual black hole is large compared to the AdS radius , $R_\mathrm{AdS}$. Black holes that obey this condition include all black holes whose temperature is large compared to unity, but also includes large radius extremal black holes in $AdS_5 \times S^5$, $AdS_7 \times S^4$ and $AdS_4 \times S^7$. It, however, never includes supersymmetric black holes in the same backgrounds, whose horizon radii always turn out to be of the same order as the AdS radius.

It follows that we should expect the Navier-Stokes equations to reproduce the thermodynamics of only large black holes. In all the cases we have studied, this is indeed the case. It is possible to expand the formulae of black hole thermodynamics (and the stress tensor and charge distribution) in a power series in $R_\mathrm{AdS}/R_H$. While the leading order term in this expansion matches the results of fluid dynamics, we find deviations from the predictions of Navier-Stokes equations at subleading orders.

Starting with the work of Policastro, Son and Starinets \cite{Policastro:2001yc}, there have been several fascinating studies over the last few years, that have computed fluid dynamical dissipative and transport coefficients from gravity (see the review \cite{Son:2007vk} and the references therein). The work reported in this paper differs from these analyses in several ways. Firstly, the solutions of fluid mechanics we study are nonlinear; in general they cannot be thought of as small fluctuations about the uniform fluid configuration dual to static black holes. Second, all our solutions are stationary: dissipative parameters play no role in our work.

Indeed our work rather follows the same line of investigation as the one applied to
plasmaballs and plasmarings in \cite{Aharony:2005bm, Lahiri:2007ae}.
These investigations used the boundary fluid dynamics to make detailed
predictions about the nature and phase structure of the black holes
and black rings in Scherk-Schwarz compactified AdS spaces. The
predictions of these papers have not yet been quantitatively verified
as the corresponding black hole and black ring solutions have not yet
been constructed. The perfect agreement between fluid dynamics and
gravity in the simpler and better studied context of this paper lends
significant additional support to those predictions of
\cite{Aharony:2005bm, Lahiri:2007ae} that were made using fluid
mechanics.

While in this paper we have used fluid dynamics to make predictions for black hole physics, the reverse view point may also prove useful. Existing black hole solutions in AdS spaces provide exact equilibrium solutions to the equations of fluid dynamics to all orders in $l_\mathrm{mfp}$. A study of the higher order corrections of these solutions (away from the $l_\mathrm{mfp} \to 0$ limit) might yield useful information about the nature of the fluid dynamical approximation of quantum field theories.

The plan of our paper is as follows - In \S\ref{sec:stress}, we set up
the basic fluid mechanical framework necessary for our work. It is
followed by \S\ref{sec:rotate} in which we consider in detail a
specific example of rigidly rotating fluid - a conformal fluid in $S^3
\times \R$. A straightforward generalisation gives us a succinct way of formulating fluid mechanics in
spheres of arbitrary dimensions in \S\ref{sec:otherspheres}.

We proceed then to compare the fluid mechanical predictions with
various types of black holes in arbitrary dimensions. First, we
consider uncharged rotating black holes in arbitrary dimensions in
\S\ref{sec:arbitrary}. Their thermodynamics, stress tensors and charge
distributions are computed and are shown to exactly match the fluid
mechanical predictions. In \S\ref{sec:bhcomp}, we turn to the large
class of rotating black hole solutions in $AdS_5 \times S^5$. Many
different black holes with different sets of charges and angular
momenta are considered in the large horizon radius limit and all of
them are shown to fit exactly into our proposal in the strict fluid dynamical limit. However we also find deviations from the predictions of the Navier-Stokes equations at first subleading order in $l_\mathrm{mfp}$ for black holes with all SO(6) Cartan charges nonzero (these deviations vanish when the angular velocities, or one of the SO(6) charges is set to zero). This finding is at odds with naive expectations from fluid dynamics, which predict the first deviations from the Navier-Stokes equations to occur at ${\cal O}(l_\mathrm{mfp}^2)$ and is an as yet unresolved puzzle.

This is followed by \S\ref{sec:m2m5}, dealing with large rotating
black holes in $AdS_4 \times S^7$ and $AdS_7 \times S^4$ backgrounds
which are dual to field theories on M2 and M5 branes respectively. The
thermodynamics of the rotating black hole solutions in these spaces
are derived from their static counterparts using the duality to fluid
mechanics and it is shown how the known rotating black hole solutions
agree with the fluid mechanical predictions in the large horizon
radius limit. In each of these cases, the formulae of black hole
thermodynamics deviate from the predictions of the Navier-Stokes equations
only at $\CO(l_\mathrm{mfp}^2)$ in accord with general expectations.
In the final section, we conclude our work and discuss further directions.

In appendix \ref{sec:conffluidmech}, we discuss the constraints imposed by conformal invariance on the equations of fluid mechanics. In appendix \ref{sec:free}, we discuss the thermodynamics of free theories on spheres. In appendix \ref{sec:bhstrapps}, we present our computations of the boundary stress tensor for two classes of black holes in AdS spaces. In appendix \ref{app:notation}, we summarise the notation used throughout this paper.

\section{The equations of fluid mechanics}\label{sec:stress}

\subsection{The equations}\label{sec:basiceq}

The fundamental variables of fluid dynamics are the local proper energy density $\rho$, local charge densities $r_i$ and fluid velocities $u^\mu=\gamma(1,\vec{v})$. Assuming local thermodynamic equilibrium, the rest frame entropy density $s$, the pressure $\ploc$, the local temperature $\tloc$ and the chemical potentials $\mu_i$ of the fluid can be expressed as functions of $\rho$ and $r_i$ using the equation of state and the first law of thermodynamics, as in \S\S\ref{sec:cnftherm}, \S\S\ref{sec:thermid} below. In what follows, we will often find it convenient to express the above thermodynamic quantities in terms of $\tloc$ and $\mu_i$ rather than as functions of $\rho$ and $r_i$.

The equations of fluid dynamics are simply a statement of the conservation of the stress tensor $T^{\mu \nu}$ and the charge currents $J^\mu_i$.
\begin{equation}\label{Epconsv:eq} \begin{split}
  \nabla_\mu T^{\mu\nu} &= \p_\mu T^{\mu\nu}
                        + \Gamma^\mu_{\mu\lambda} T^{\lambda\nu}
                        + \Gamma^\nu_{\mu\lambda} T^{\mu\lambda}
                        = 0\,, \\
\nabla_\mu J^\mu_i&=\p_\mu J_i^\mu + \Gamma^\mu_{\mu \alpha} J^\alpha_i=0\,.
\end{split}
\end{equation}

\subsection{Perfect fluid stress tensor}\label{sec:blkstr}

The dynamics of a fluid is completely specified once the stress tensor and charge currents are given as functions of $\rho, r_i$ and $u^\mu$. As we have explained in the introduction, fluid mechanics is an effective description at long distances (i.e, it is valid only when the fluid variables vary on distance scales that are large compared to the mean free path $l_\mathrm{mfp}$). As a consequence it is natural to expand the stress tensor and charge current in powers of derivatives. In this subsection we briefly review the leading (i.e.\ zeroth) order terms in this expansion.

It is convenient to define a projection tensor
\begin{equation}\label{proj:eq}
  P^{\mu\nu} = g^{\mu\nu} + u^\mu u^\nu.
\end{equation}
$P^{\mu\nu}$ projects vectors onto the 3 dimensional submanifold orthogonal to $u^\mu$. In other words, $P^{\mu\nu}$ may be thought of as a projector onto spatial coordinates in the rest frame of the fluid. In a similar fashion, $- u^\mu u^\nu$ projects vectors onto the time direction in the fluid rest frame.

To zeroth order in the derivative expansion, Lorentz invariance and the correct static limit uniquely determine the stress tensor, charge and the entropy currents in terms of the thermodynamic variables. We have
%%
%\begin{equation*}
%\begin{split}
%  T^{\mu\nu}_\mathrm{perfect}& = A u^\mu u^\nu + B P^{\mu\nu}\, \\
%(J^\mu_i)_\mathrm{perfect}&=Cu^\mu
%\end{split}
%\end{equation*}
%%
%Transforming to the rest frame, we may interpret the Lorentz
%scalars $A, B, C$ respectively as $\rho$, $\ploc$ and $r_i$
%where $\rho$ is the proper energy density, $\ploc$ is the pressure
%and $r_i$ the proper charge density of the fluid. We conclude
%%
\begin{equation}\label{currents:eq}
\begin{split}
  T^{\mu\nu}_\mathrm{perfect}& = \rho u^\mu u^\nu + \ploc P^{\mu\nu}, \\
  (J^\mu_i)_\mathrm{perfect}&=r_iu^\mu, \\
  (J^\mu_S)_\mathrm{perfect}&=s u^\mu,
\end{split}
\end{equation}
where $\rho=\rho(\tloc,\mu_i)$ is the rest frame energy density, $s=s(\tloc, \mu_i)$ is the rest frame entropy density of the fluid and $\mu_i$ are the chemical potentials. It is not difficult to verify that in this zero-derivative (or perfect fluid) approximation, the entropy current is conserved. Entropy production (associated with dissipation) occurs only at the first subleading order in the derivative expansion, as we will see in the next subsection.

\subsection{Dissipation and diffusion}\label{sec:visc}

Now, we proceed to examine the first subleading order in the derivative expansion. In the first subleading order, Lorentz invariance and the physical requirement that entropy be non-decreasing determine the form of the stress tensor and the current to be (see, for example, \S\S14.1 of \cite{Andersson:2006nr})
\begin{equation}\label{extraTvisc:eq}
\begin{split}
  T^{\mu\nu}_\mathrm{dissipative} &= -\zeta \vartheta P^{\mu\nu} -
  2\eta\sigma^{\mu\nu} + q^\mu u^\nu + u^\mu q^\nu,\\
 (J^\mu_i)_\mathrm{dissipative} &= q^\mu_{i},\\
(J^\mu_S)_\mathrm{dissipative} &= \frac{q^\mu-\mu_i q^\mu_{i}}{\tloc}\,.
\end{split}
\end{equation}
where
\begin{equation}\label{fluidtensors:eq}
\begin{split}
  a^\mu &= u^\nu \nabla_\nu u^\mu, \\
  \vartheta &= \nabla_\mu u^\mu, \\
%  P^{\mu\nu} &= g^{\mu\nu} + u^\nu u^\mu, \\
  \sigma^{\mu\nu} &= \half \prn{P^{\mu\lambda} \nabla_\lambda u^\nu
                   + P^{\nu\lambda} \nabla_\lambda u^\mu}
                   - \frac{1}{d-1} \vartheta P^{\mu\nu}, \\
%  \omega^{\mu\nu} &= \half \prn{P^{\mu\lambda} \nabla_\lambda u^\nu
%                   - P^{\nu\lambda} \nabla_\lambda u^\mu} \\
 q^\mu &= -\kappa P^{\mu\nu} (\p_\nu\tloc + a_\nu\tloc)\,, \\
 q^\mu_{i}
   & = - D_{ij} P^{\mu\nu}\p_\nu\! \prn{\frac{\mu_j}{\tloc}},
\end{split}
\end{equation}
are the acceleration, expansion, shear tensor, heat flux
and diffusion current respectively.

These equations define a set of new fluid dynamical parameters in addition to those of the previous subsection: $\zeta$ is the bulk viscosity, $\eta$ is the shear viscosity, $\kappa$ is the thermal conductivity and $D_{ij}$ are the diffusion coefficients. Fourier's law of heat conduction $\vec{q} = -\kappa \vec{\nabla} \tloc$ has been relativistically modified to
\begin{equation}\label{heatcond:eq}
  q^\mu = -\kappa P^{\mu\nu} (\p_\nu\tloc + a_\nu\tloc)\,,
\end{equation}
with an extra term that is related to the inertia of flowing heat. The diffusive contribution to the charge current is the relativistic generalisation of Fick's law.

At this order in the derivative expansion, the entropy current is no longer conserved; however, it may be checked \cite{Andersson:2006nr} that
\begin{equation}\label{increase:eq}
\tloc\nabla_\mu J^\mu_S = \frac{q^\mu q_\mu}{\kappa \tloc}
+ \tloc (D^{-1})^{ij} q_i^\mu q_{j\mu}  + \zeta \theta ^2
+ 2 \eta \sigma_{\mu \nu} \sigma^{\mu \nu}.
\end{equation}
As $q^\mu$, $q^\mu_i$ and $\sigma^{\mu \nu}$ are all spacelike
vectors and tensors, the RHS of \eqref{increase:eq} is positive provided
$\eta, \zeta, \kappa$ and $D$ are positive parameters, a condition we further assume. This establishes that (even locally) entropy can only be produced but never destroyed. In equilibrium, $\nabla_\mu J^\mu_S$ must vanish. It follows that, $q^\mu$, $q^\mu_i$, $\theta$ and $\sigma^{\mu \nu}$
each individually vanish in equilibrium.

From the formulae above, we see that the ratio of $T_\mathrm{dissipative}^{\mu \nu}$ to $T_\mathrm{perfect}^{\mu\nu}$ is of the order $\eta/(R \rho)$ where $\eta$ is the shear viscosity, $\rho$ is the rest frame energy density and $R$ is the typical length scale of the flow under consideration. Consequently, the Navier-Stokes equations may be thought of as the first term in a series expansion of the microscopic equations in $l_\mathrm{mfp}/R$ , where $l_\mathrm{mfp}\sim\frac{\eta}{\rho}$. In this sense, $l_\mathrm{mfp}$ plays a role analogous to the string scale in the derivative expansion of the effective action in string theory. This length scale may plausibly be identified with the thermalisation length scale of the fluid.\footnote{This may be demonstrated within the kinetic theory, where $l_\mathrm{mfp}$ is simply the mean free path of colliding molecules, but is expected to apply to more generally to any fluid with short range interactions.}

When studying fluids on curved manifolds (as we will proceed to do in this paper), one could add generally covariant terms, built out of curvatures, to the stress tensor. For instance, we could add a term proportional to $R^{\mu \nu}$ to the expression for $T^{\mu\nu}$. We will ignore all such terms in this paper for a reason we now explain. In all the solutions of fluid mechanics that we will study, the length scale over which fluid quantities vary is the same as the length scale of curvatures of the manifold. Any expression built out of a curvature contains at least two spacetime derivatives of the metric; it follows that any contribution to the stress tensor proportional to a curvature is effectively at least two orders subleading in the derivative expansion, and so is negligible compared to all the other terms we have retained in this paper.

%where we have used the thermodynamic identity
%$\dr \sloc = \dr Q/\tloc$ where $Q$
%is the heat flow into the system.

\subsection{Conformal fluids}\label{sec:fluidtherm}

We will now specialise our discussion to a conformal fluid -- the fluid of the `stuff' of a conformal field theory in $d$ dimensions. Conformal invariance imposes restrictions on both the thermodynamics of the fluid and the derivative expansion of the stress tensor discussed in the previous subsection.

To start with, conformal invariance requires that the stress tensor be traceless.\footnote{More accurately, conformal invariance relates the nonzero trace of the stress tensor to certain curvature forms; for example, in two dimension $T^\mu_\mu =\frac{c}{12} R$ where $R$ is the scalar curvature. However, as we have described above, curvatures are effectively zero in the one derivative expansion studied in this paper. All formulae through the rest of this paper and in the appendices apply only
upon neglecting curvatures. We thank R.~Gopakumar for a discussion of this point.} This requirement relates the pressure of a conformal fluid to its density as $\ploc=\frac{\rho}{d-1}$ (this requirement may also be deduced from conformal thermodynamics, as we will see in the next subsection) where $d$ is the dimension of the spacetime in which the fluid lives. Further, the tracelessness of the stress tensor also forces the bulk viscosity, $\zeta$, to be zero.

It is easy to verify that these constraints are sufficient to guarantee the conformal invariance of the fluid dynamical equations listed above. Consider a conformal transformation $g_{\mu\nu}=\e^{2\phi}\tilde{g}_{\mu\nu}$ under which fluid velocity, temperature, rest energy density, pressure, entropy density and the charge densities transform as
\begin{equation*}
\begin{split}
 u^\mu &= \e^{-\phi}\tilde{u}^\mu, \\
 \tloc &= \e^{-\phi} {\widetilde \tloc}\,, \\
 \rho &= \e^{-d\phi}{\tilde \rho}\,, \quad
 \ploc = \e^{-d\phi}\widetilde{\ploc}\,, \\
 s &= \e^{-(d-1)\phi}{\tilde s}\,, \\
 r_i &= \e^{-(d-1)\phi}{\tilde r_i}\,.
\end{split}
\end{equation*}

It is easy to verify that these transformations induce the following transformations on the stress tensors and currents listed in the previous subsection\footnote{Note that under such a scaling, the viscosity, conductivity etc.\ scale as $\kappa=e^{-(d-2)\phi}\tilde{\kappa}$ , $\eta=e^{-(d-1)\phi}\tilde{\eta}$, $\mu_i=e^{-\phi}\tilde{\mu_i}$ and $ D_{ij}=e^{-(d-2)\phi} \widetilde{D}_{ij}$.}
\begin{equation} \label{tencur:eq}
\begin{split}
 T^{\mu\nu} &= \e^{-(d+2)\phi}\widetilde{T}^{\mu\nu},\\
 J^\mu_i & = \e^{-d \phi}\widetilde{J}^\mu_i, \\
 J^\mu_S & = \e^{-d \phi}\widetilde{J}^\mu_S.
\end{split}
\end{equation}
These are precisely the transformation properties that ensure the conformal invariance of the conservation equations \eqref{Epconsv:eq}. See appendix \ref{sec:conffluidmech} for more details.

\subsection{Conformal thermodynamics}\label{sec:cnftherm}

In this subsection, we review the thermodynamics of the conformal fluids we discuss below. The notation set up in this subsection will be used through the rest of this paper.

Define the thermodynamic potential
\begin{equation}\label{phidef:eq}
 \Phi=\eloc-\tloc\sloc -\mu_i \rloc_i\ .
\end{equation}
for which the first law of thermodynamics reads
\begin{equation}\label{firstlawphi:eq}
  \dr\Phi = - \sloc\dr\tloc
            - \ploc \dr V
            - \rloc_i\dr\mu_i\,.
\end{equation}
Let us define $\nu_i=\mu_i/\tloc$. It follows from conformal invariance and extensivity that
\begin{equation}\label{hdef:eq}
 \Phi = -V \tloc^d h(\nu)\,,
\end{equation}
where $h(\nu)$ is defined by this expression. All remaining thermodynamic expressions are easily determined in terms of the function $h(\nu_i)$
\begin{equation}\label{gcterm:eq}
\begin{split}
 \rho &= (d-1) \ploc= (d-1) \tloc^d h(\nu)\,, \\
 r_i &= \tloc^{d-1}h_i(\nu)\,, \\
 s &= \tloc^{d-1}(dh-\nu_i h_i)\,,
\end{split}
\end{equation}
where
\begin{equation*}
h_i =\frac{\partial h}{\partial \nu_i}
\end{equation*}
denotes the derivative of $h$ with respect to its $i^{th}$ argument.
%Of course we can pass from knowledge of the function $g$ of the
%previous subsubsection to knowledge of $h$ of this subsubsection.

%\subsubsection{Constraints on Dissipative Terms}

\subsection{A thermodynamic identity}\label{sec:thermid}

We will now derive a thermodynamic identity that will be useful
in our analysis below. Define
\begin{equation}\label{Gammadef:eq}
  \Gamma = \eloc - \tloc\sloc + \ploc V - \mu_i\rloc_i\ = \Phi+ \ploc V\,,
\end{equation}
the first law of thermodynamics implies that
\begin{equation}\label{firstlaw:eq}
  \dr\Gamma = -\sloc\dr\tloc + V\dr\ploc - \rloc_i\dr\mu_i\,.
\end{equation}
Consider scaling the system by a factor $(1+\epsilon)$. Under such a scaling, extensivity implies that
\begin{equation*}
  \dr\Gamma=\epsilon\Gamma\,,
  \qquad
  \dr\tloc=\dr\ploc=\dr\mu_i=0\,,
\end{equation*}
which when substituted into \eqref{firstlaw:eq} tells us that $\Gamma=0$. Then we can divide \eqref{Gammadef:eq} and \eqref{firstlaw:eq} by $V$ to get
\begin{equation}\label{gentherm:eq}
\begin{split}
  \rho + \ploc &= s\tloc + \mu_i r_i\,, \\
  \dr\ploc  &= s\dr\tloc + r_i\dr\mu_i\,.
\end{split}
\end{equation}
%
%Using the above equation, we can determine the pressure $\ploc$ of the fluid in terms of the other thermodynamic variables.

\section{Equilibrium configurations of rotating conformal fluids on $S^3$}\label{sec:rotate}

In this section and in the next, we will determine the equilibrium solutions of fluid dynamics equations for conformal fluids on spheres of arbitrary dimension. In this section, we work out the fluid dynamics on $S^3$ plus a time dimension in detail.\footnote{In this case, the dimensions of the spacetime in which the fluid lives is $d=3+1=4$. The number of mutually commuting angular momenta is $n=2$. The black hole dual lives in AdS space of dimensions $D=d+1=5$.} In the next section, we generalise the results of this section to spheres of arbitrary dimension.

\subsection{Coordinates and conserved charges}\label{sec:coord}

Consider a unit $S^3$ embedded in $\R^4$ as
\begin{equation}\label{S3inR4:eq}
\begin{split}
  x^1 &= \sin\theta \cos\phi_1 \\
  x^2 &= \sin\theta \sin\phi_1 \\
  x^3 &= \cos\theta \cos\phi_2 \\
  x^4 &= \cos\theta \sin\phi_2 \\
\end{split}
\end{equation}
with $\theta\in[0,\frac{\pi}{2}]$, $\phi_a\in[0,2\pi)$. The metric of the spacetime $S^3\times \R$ is
\begin{equation}\label{metric:eq}
  \dr s^2 = -\dr t^2 + \dr \theta^2 + \sin^2\theta\, \dr \phi_1^2 +
            \cos^2\theta\, \dr \phi_2^2\,.
\end{equation}
This gives the following non-zero Christoffel symbols:
\begin{equation}\label{chrst:eq}
  \Gamma^\theta_{\phi_1\phi_1} = -\Gamma^\theta_{\phi_2\phi_2}
         = -\cos\theta \sin\theta \,,
 \quad
  \Gamma^{\phi_1}_{\theta\phi_1} = \Gamma^{\phi_1}_{\phi_1\theta}
         = \cot\theta \,,
 \quad
  \Gamma^{\phi_2}_{\theta\phi_2} = \Gamma^{\phi_2}_{\phi_2\theta}
         = -\tan\theta \,.
\end{equation}

For the stationary, axially symmetric configurations under
consideration, $\p_t T^{\mu \nu}= \p_{\phi_a} T^{\mu \nu} = 0$. Using
\eqref{chrst:eq}, \eqref{Epconsv:eq} becomes
\begin{alignat}{2}\label{tEpconsv:eq}
    0 &= \nabla_\mu T^{\mu t} &
      &= \p_\theta T^{\theta t}
       + (\cot\theta-\tan\theta)T^{\theta t},
\\ \label{qEpconsv:eq}
    0 &= \nabla_\mu T^{\mu\theta} &
      &= \p_\theta T^{\theta\theta}
       + (\cot\theta-\tan\theta)T^{\theta\theta}
       +\cos\theta \sin\theta \prn{ T^{\phi_1\phi_1}
                                  - T^{\phi_2\phi_2} },
\\ \label{f1Epconsv:eq}
    0 &= \nabla_\mu T^{\mu \phi_1} &
      &= \p_\theta T^{\theta\phi_1}
       + (\cot\theta-\tan\theta)T^{\theta\phi_1}
       +2\cot\theta\, T^{\theta\phi_1},
\\ \label{f2Epconsv:eq}
    0 &= \nabla_\mu T^{\mu \phi_2} &
      &= \p_\theta T^{\theta\phi_2}
       + (\cot\theta-\tan\theta)T^{\theta\phi_2}
       -2\tan\theta\, T^{\theta\phi_2}.
\end{alignat}

The Killing vectors of interest are $\p_t$ (Energy) and $\p_{\phi_a}$ (SO(4) Cartan angular momenta). Using the formula for the related conserved charge, $\int\dr^3x\sqrt{-g}\,T^{0\mu}g_{\mu\nu}k^{\nu}$, we get:
\begin{equation}\label{charge:eq}
\begin{split}
  E &= \int\!\dr\theta\dr\phi_1\dr\phi_2\,\cos\theta\sin\theta
       \,T^{tt}, \\
  L_1 &= \int\!\dr\theta\dr\phi_1\dr\phi_2\,\cos\theta\sin^3\theta
       \,T^{t\phi_1}, \\
  L_2 &= \int\!\dr\theta\dr\phi_1\dr\phi_2\,\cos^3\theta\sin\theta
       \,T^{t\phi_2}.
\end{split}
\end{equation}
Assuming $q^\mu = q^\mu_{i} = 0$ (as will be true for stationary solutions we study in this paper), the entropy and the R-charges corresponding to the currents in \eqref{currents:eq} are given as
\begin{equation}\label{charge2:eq}
\begin{split}
  S &= \int\!\dr\theta\dr\phi_1\dr\phi_2\,\cos\theta\sin\theta
       \,\gamma s\,, \\
  R_i &= \int\!\dr\theta\dr\phi_1\dr\phi_2\,\cos\theta\sin^3\theta
       \,\gamma r_i\,.
\end{split}
\end{equation}

\subsection{Equilibrium solutions}\label{sec:eqmot}

As we have explained in the \S\S\ref{sec:visc}, each of the three quantities
$\sigma^{\mu \nu}, q^\mu, q^\mu_i$ must vanish on any stationary
solution of fluid dynamics. The requirement that $\sigma^{\mu \nu}=0$ has a unique solution - the fluid motion should be just a rigid rotation. By an $SO(4)$ rotation we can choose the two orthogonal two planes of this rotation as the (1-2) and (3-4) planes (see \eqref{S3inR4:eq}). This implies that $u^\mu = \gamma(1,0,\omega_1,\omega_2)$ (where we have listed the $(t,\theta,\phi_1,\phi_2)$ components of the velocity) with $\gamma=\prn{1-v^2}^{-1/2}$ and $v^2=\omega_1^2\sin^2\theta+\omega_2^2\cos^2\theta$, for some constants $\omega_1$ and $\omega_2$.

Our equilibrium fluid flow enjoys a symmetry under translations of $t$, $\phi_1$ and $\phi_2$; consequently all thermodynamic quantities are functions only of the coordinate $\theta$.

Evaluating the tensors in \eqref{fluidtensors:eq}, we find
\begin{equation}\label{ourfluidtensors:eq}
\begin{split}
  a^\mu &= (0,-\p_\theta\ln\gamma,0,0), \\
  \vartheta &= 0, \\
  \sigma^{\mu\nu} &= 0, \\
%  \omega^{\mu\nu} &= P^{\mu\lambda}\nabla_\lambda u^\nu,\\
  q^\mu &= -\kappa\gamma \prn{0,\diff{}{\theta}\brk{\frac{\tloc}{\gamma}},0,0},\\
  q_{i}^\mu &= -D_{ij} \prn{0,\diff{}{\theta}\brk{\frac{\mu_j}{\tloc}},0,0}.\\
\end{split}
\end{equation}

The requirement that $q^\mu$ and $q^\mu_i$ vanish forces us to set
\begin{equation} \label{solution:eq}
\tloc = \tc \gamma\,, ~~~\mu_i = \tloc \nu_i\,,
\end{equation}
for constant $\tc$ and $\nu_i$. These conditions completely determine all the thermodynamic quantities as a function of the coordinates on the sphere. We will now demonstrate that this configuration solves the Navier-Stokes equations.

First note that for an arbitrary rigid rotation, the dissipative part of the stress tensor evaluates to
\begin{equation}\label{dissT:eq}
  T^{\mu\nu}_\mathrm{dissipative} =
   -\kappa\gamma^2
   \begin{pmatrix}
     0 & 1        & 0        & 0        \\
     1 & 0        & \omega_1 & \omega_2 \\
     0 & \omega_1 & 0        & 0        \\
     0 & \omega_2 & 0        & 0        \\
   \end{pmatrix}
   \diff{}{\theta}\brk{\frac{\tloc}{\gamma}},
\end{equation}
an expression which simply vanishes once we impose \eqref{solution:eq}. Consequently, all nonzero contributions to the stress tensor come from the `perfect fluid piece' and are given by
\begin{equation}\label{blkstrour:eq}
  T^{\mu\nu}_\mathrm{perfect} = \gamma^2
   \begin{pmatrix}
     (\rho+v^2\ploc) & 0 & \omega_1(\rho+\ploc) & \omega_2(\rho+\ploc) \\
     0                           & \gamma^{-2}\ploc & 0 & 0 \\
     \omega_1(\rho+\ploc)      & 0 & \omega_1^2\rho+(\csc^2\theta-\omega_2^2\cot^2\theta)\ploc & \omega_1\omega_2(\rho+\ploc) \\
     \omega_2(\rho+\ploc)      & 0 & \omega_1\omega_2(\rho+\ploc) & \omega_2^2\rho+(\sec^2\theta-\omega_1^2\tan^2\theta)\ploc \\
   \end{pmatrix}.
\end{equation}
The only non-trivial equation of motion, \eqref{qEpconsv:eq}, can be
written as
\begin{equation}\label{eom:eq}
  \diff{\ploc}{\theta} - \frac{\rho+\ploc}{\gamma}\diff{\gamma}{\theta} = 0\,.
\end{equation}
Now using the thermodynamic identity \eqref{gentherm:eq} we may recast
\eqref{eom:eq} as
\begin{equation}\label{eomtherm:eq}
  \gamma s \diff{}{\theta}\brk{\frac{\tloc}{\gamma}} +
  \gamma r_i \diff{}{\theta}\brk{\frac{\mu_i}{\gamma}} = 0\,,
\end{equation}
an equation which is automatically true from \eqref{solution:eq}. Consequently,
rigidly rotating configurations that obey \eqref{solution:eq} automatically obey the Navier-Stokes equations.

In a similar fashion, it is easy to verify that all nonzero contributions to the charge currents come from the perfect fluid piece of that current, and that the conservation of these currents holds for our solutions.

In summary the $3+c$ parameter set of stationary solutions to fluid mechanics listed in this subsection (the parameters are $\tc,\omega_a$ and $\nu_i$ where $i=1\ldots c$ ) constitute the most general stationary solutions of fluid mechanics.

\subsection{Stress tensor and currents}\label{sec:chspin}

Using the equations of state \eqref{gcterm:eq}, we find that
\begin{equation}\label{chdensityh:eq}
\begin{split}
  \rho = 3\ploc &= 3 \tc^4 \gamma^4 h(\nu), \\
  s &= \tc^3  \gamma^3 [4h(\nu) - \nu_i h_i(\nu)], \\
  r_i &= \tc^3 \gamma^3 h_i(\nu).
\end{split}
\end{equation}

The stress tensor is
\begin{multline}\label{chstr:eq}
  T^{\mu\nu} = \tc^4 A  \gamma^6% \times \\
   \begin{pmatrix}
     3+v^2 & 0 & 4\omega_1 & 4\omega_2 \\
     0                           & 1-v^2 & 0 & 0 \\
     4\omega_1      & 0 & 3\omega_1^2+\csc^2\theta-\omega_2^2\cot^2\theta & 4\omega_1\omega_2 \\
     4\omega_2      & 0 & 4\omega_1\omega_2 & 3\omega_2^2+\sec^2\theta-\omega_1^2\tan^2\theta \\
   \end{pmatrix}.
\end{multline}
Charge and entropy currents are given by
\begin{equation}\label{chcurrent:eq}
\begin{split}
 J^\mu_i &= \tc^3 \gamma^4 C_i(1, 0, \omega_1, \omega_2)\,,  \\
 J^\mu_S &= \tc^3 \gamma^4 B (1, 0, \omega_1, \omega_2)\,,
\end{split}
\end{equation}
where we have defined
\begin{equation}\label{chcoeffs:eq}
\begin{split}
 A   &= h(\nu) \,,                 \\
 B   &= 4h(\nu) -\nu_i h_i(\nu) \,, \\
 C_i &= h_i(\nu) = \frac{\partial h}{\partial \nu_i} \,.
\end{split}
\end{equation}

\subsection{Charges}\label{sec:plch}

The energy, angular momentum, entropy and R-charges may now easily be
evaluated by integration: we find
\begin{equation}\label{chch:eq}
\begin{split}
  E &= \frac{V_4 \tc^4 A}
               {(1-\omega_1^2)(1-\omega_2^2)}
               \brk{\frac{2\omega_1^2}{1-\omega_1^2}+\frac{2\omega_2^2}{1-\omega_2^2}+3 },\\
  L_1 &= \frac{V_4 \tc^4 A}
              {(1-\omega_1^2)(1-\omega_2^2)}
               \brk{\frac{2\omega_1}{1-\omega_1^2}},\\
  L_2 &= \frac{V_4 \tc^4 A }
              {(1-\omega_1^2)(1-\omega_2^2)}
               \brk{\frac{2\omega_2}{1-\omega_2^2}},\\
  S &= \frac{V_4 \tc^3 B}
            {(1-\omega_1^2)(1-\omega_2^2)}\,,\\
  R_i &= \frac{V_4 \tc^3 C_i}
            {(1-\omega_1^2)(1-\omega_2^2)}\,,
\end{split}
\end{equation}
where $V_4= \operatorname{Vol}(S^3) = 2\pi^2$ is the volume of $S^3$. These formulae constitute a complete specification of the thermodynamics of stationary rotating conformal fluids on $S^3$.

\subsection{Potentials}\label{sec:plchempots}

In the previous subsection we have evaluated all the thermodynamic
charges of our rotating fluid solutions. It is also useful to
evaluate the chemical potentials corresponding to these solutions.
To be specific we define these chemical potentials via the grand canonical partition function defined in the introduction
\begin{equation} \label{gcp:eq}
 \gpf =\Tr \exp \prn{ \frac{1}{T}( -H + \Omega_a L_a + \zeta_i R_i ) }
 = \exp\prn{- \frac{ E-TS-\Omega_a L_a - \zeta_i R_i }{ T } },
\end{equation}
where the last expression applies in the thermodynamic limit. In other words
\begin{equation}\label{chpotdef:eq}
  T = \pdiffc[L_b,R_j]{E}{S}, \quad
  \Omega_a = \pdiffc[S,L_b,R_j]{E}{L_a}, \quad
  \zeta_i = \pdiffc[S,L_b,R_j]{E}{R_i} .
\end{equation}
It is easy to verify that\footnote{We can express
$\dr E -\tc\dr S - \omega_a\dr L_a - \tc\nu_i\dr R_i$
in terms of $\dr\tc,\dr\omega_a,\dr\nu_i$ and check that it vanishes,
or we can check the Legendre transformed statement $\dr (E -\tc S - \omega_a L_a - \tc\nu_i R_i) =
-S\dr\tc - L_a\dr\omega_a - R_i \dr(\tc\nu_i)$.}
\begin{equation} \label{pot:eq}
 T = \tc \,,\qquad
 \Omega_a = \omega_a \,,\qquad
 \zeta_i = \tc \nu_i \,.
\end{equation}

Note that $T$, $\Omega_a$ and $\zeta_i$ are distinct from $\tloc$, $\omega_a$ and $\mu_i$. While the former quantities are thermodynamic properties of the whole fluid configuration, the latter quantities are local thermodynamic properties of the fluid that vary on the $S^3$. In a similar fashion, the energy $E$ of the solution is, of course, a distinct concept from the local rest frame energy density $\rho$ which is a function on the sphere. In particular, $E$ receives contributions from the kinetic energy of the fluid as well as its internal energy, $\eloc$.

\subsection{Grand canonical partition function}\label{sec:gcpf}

The grand canonical partition function \eqref{gcp:eq} is easily
computed; we find
\begin{equation}
 \ln \gpf
 =  \frac{V_4 T^3 h(\zeta/T)}{(1-\Omega_1^2)(1-\Omega_2^2) }\,,
\end{equation}
where $V_4=V(S^3)=2\pi^2$ is the volume of $S^3$.

In other words, the grand canonical partition function of the
rotating fluid is obtained merely by multiplying
the same object for the non-rotating fluid by a universal angular
velocity dependent factor.

\subsection{Validity of fluid mechanics}\label{sec:validity}

A systematic way to estimate the domain of validity of the Navier-Stokes equations would be to list all possible higher order corrections to these equations, and to check under what circumstances the contributions of these correction terms to the stress tensor and currents are small compared to the terms we have retained. Rather than carrying out such a detailed (and worthwhile) exercise, we present in this section a heuristic physical estimate of the domain of validity of fluid dynamics.

Consider a fluid composed of a collection of interacting `quasiparticles', that move at an average speed $v_p$ and whose collisions are separated (on the average) by the distance $l_\mathrm{mfp}$ in the fluid rest frame. Consider a particular quasiparticle that undergoes two successive collisions: the first at the coordinate location $x_1$ and subsequently at $x_2$. In order for the fluid approximation to hold, it must be that
\begin{enumerate}
\item The fractional changes in thermodynamic quantities between
the two collision points (e.g.\ $[T(x_1)-T(x_2)]/T(x_1)$) are small. This condition is
necessary in order for us to assume local thermal equilibrium.

\item The distance between the two successive collisions is small compared to the curvature/compactification scales of the manifold on which the fluid propagates. This approximation is necessary, for example, in order to justify the neglect of curvature corrections to the Navier-Stokes equations.
\end{enumerate}

Let us now see when these two conditions are obeyed on our solutions. Recall that the local temperatures in our solutions take the form $\tloc= T \gamma$ where $T$ is the overall temperature of the solution. If we treat the free path $l_\mathrm{mfp}$ as a function of temperature and chemical potentials, conformal invariance implies that
\begin{equation*}
l_\mathrm{mfp}( \tloc, \nu_i )=
\frac{1}{\gamma}l_\mathrm{mfp}( T, \nu_i)\,.
\end{equation*}

Hence, the first condition listed above is satisfied when the
fractional variation in (say) the temperature is small over the rest
frame mean free path $l_\mathrm{mfp}( \tloc, \nu_i )$, i.e.\ provided
\begin{equation}\label{valid:eq}
 \frac{l_\mathrm{mfp}(T, \nu_i)}{\gamma} \ll
\gamma\prn{\frac{\partial\gamma}{\partial\theta}}^{-1},
\end{equation}
which must hold for all points of the sphere.\footnote{Recall that all variations in the temperature are perpendicular to fluid velocities, so that the typical scale of variation in both the rest frame and the lab frame coincide.} The strictest condition one obtains from this is
\begin{equation} \label{firstcond:eq}
 l_\mathrm{mfp}(T, \nu_i) \ll
   \frac{1}{\abs{\sqrt{1-\omega_1^2}-\sqrt{1-\omega_2^2}}}\,.
\end{equation}

It turns out that the second condition listed above is always more stringent, especially when applied to fluid quasiparticles whose rest frame motion between two collisions is in the same direction as the local fluid velocity. It follows from the formulae of Lorentz transformations that the distance on the sphere between two such collisions is $l_\mathrm{mfp}(\tloc, \nu_i) \gamma( 1+v/v_p)=l_{\mathrm{mfp}} (T, \nu_i)( 1+v/v_p)$, where $v_p$ is the quasiparticle's velocity in the rest frame of the fluid and $v$ the fluid velocity. As the factor $(1+v/v_p)$ is bounded between 1 and 2, we conclude that the successive collisions happen at distances small compared to the radius of the sphere provided
\begin{equation}\label{realcond:eq}
l_{\mathrm{mfp}} (T, \nu_i) \ll 1\,.
\end{equation}

Hence, we conclude that the condition \eqref{realcond:eq} (which is always
more stringent than \eqref{firstcond:eq}) is the condition for the
applicability of the equations of fluid mechanics.

Of course the model (of interacting quasiparticles) that we have used
to obtain \eqref{realcond:eq} need not apply to the situations of our
interest. However the arguments that led to \eqref{realcond:eq} were
essentially kinematical which leads us to believe that the result will be universal.
Nonetheless, it would be useful to verify this result by performing the
detailed analysis alluded to at the beginning of this subsection.

\section{Rotating fluids on spheres of arbitrary dimension}\label{sec:otherspheres}

We now generalise the discussion of the previous section to the study
of conformal fluids on spheres of arbitrary dimension.

%\subsection{Stress tensor, currents and charges}\label{sec:arbitdim}

Let us embed $S^{2n}$ in $\R^{2n+1}$ as
\begin{equation}\label{otcoord:eq}
 \begin{split}
   x^{2a-1} &= \prn{\prod_{b=1}^{a-1} \cos\theta_b}
               \sin\theta_a \cos\phi_a\,, \\
   x^{2a} &= \prn{\prod_{b=1}^{a-1} \cos\theta_b}
               \sin\theta_a \sin\phi_a\,, \\
   x^{2n+1} &= \prn{\prod_{b=1}^{n} \cos\theta_b},
 \end{split}
\end{equation}
Where $\theta_n\in[0,\pi]$, all other $\theta_a\in[0,\frac{\pi}{2}]$ and $\phi_a\in[0,2\pi)$. Any products with the upper limit smaller than the lower limit should be set to one. Although we appear to have specialised to even dimensional spheres above, we can obtain all odd dimensional sphere, $S^{2n-1}$, simply by setting $\theta_n=\pi/2$ in all the formulae of this section.

The metric on $S^{2n}\times$ time is given by
\begin{equation}\label{otsphmet:eq}
  \dr s^2 = -\dr t^2
            + \sum_{a=1}^n\prn{\prod_{b=1}^{a-1} \cos^2\theta_b} \dr\theta_a^2
            + \sum_{a=1}^n\prn{\prod_{b=1}^{a-1} \cos^2\theta_b}
                               \sin^2\theta_a \dr\phi_a^2\,.
\end{equation}

We choose a rigidly rotating velocity
\begin{equation}\label{otrigid:eq}
\begin{gathered}
  u^t = \gamma \qquad
  u^{\theta_a} = 0 \qquad
  u^{\phi_a} = \gamma\omega_a \\
  \gamma = (1-v^2)^{-1/2} \qquad
  v^2 = \sum_{a=1}^n\prn{\prod_{b=1}^{a-1} \cos^2\theta_b}
                               \sin^2\theta_a \omega_a^2
\end{gathered}
\end{equation}
As in \S\S\ref{sec:eqmot}, the equations of motion are solved, without
dissipation, by setting
\begin{equation}\label{otsol:eq}
  \frac{\tloc}{\gamma} = \tc = \text{constant}, \qquad
  \frac{\mu_i}{\tloc} = \nu_i = \text{constant},
\end{equation}
which gives the densities
\begin{equation}\label{otdens:eq}
\begin{split}
  \rho = (d-1)\ploc &= (d-1) \tc^d \gamma^d h(\nu_i), \\
  s &= \tc^{d-1}  \gamma^{d-1} [dh(\nu) - \nu_i h_i(\nu)], \\
  r_i &= \tc^{d-1} \gamma^{d-1} h_i(\nu),
\end{split}
\end{equation}

This gives a stress tensor
\begin{equation}\label{otstr:eq}
\begin{gathered}
  T^{tt} = \tc^d A (d\gamma^{d+2} - \gamma^d) \qquad
  T^{t\phi_a} = T^{\phi_a t} = \tc^d A d\gamma^{d+2} \omega_a \\
  T^{\theta_a\theta_a} = \tc^d A\gamma^d
                   \prn{\prod_{b=1}^{a-1} \sec^2\theta_b} \\
  T^{\phi_a\phi_a} = \tc^d A\brk{ d\gamma^{d+2}\omega_a^2
             +\gamma^d \prn{\prod_{b=1}^{a-1} \sec^2\theta_b} \csc^2\theta_a} \qquad
  T^{\phi_a\phi_b} = \tc^d A d\gamma^{d+2} \omega_a \omega_b
\end{gathered}
\end{equation}
and currents
\begin{equation}\label{otcurrents:eq}
\begin{aligned}
  J_S^t &= \tc^{d-1} B \gamma^d &
  J_S^{\theta_a} &= 0 &
  J_S^{\phi_a} &= \tc^{d-1} B \gamma^d \omega_a \,,
  \\
  J_i^t &= \tc^{d-1} C_i \gamma^d &
  J_i^{\theta_a} &= 0 &
  J_i^{\phi_a} &= \tc^{d-1} C_i \gamma^d \omega_a \,,
\end{aligned}
\end{equation}
where
\begin{equation}\label{otchcoeffs:eq}
\begin{split}
 A   &= h(\nu)\,,                  \\%= g(\chi)- \chi_i g_i(\chi)   \\
 B   &= dh(\nu) -\nu_i h_i(\nu)\,, \\%= 4g(\chi) -3 \chi_i g_i(\chi)\\
 C_i &= h_i(\nu)\,.                %= \chi_i
\end{split}
\end{equation}

Integrating these gives\footnote{In deriving these formulae we have `conjectured' that $\int_{S^{d-1}} \gamma^d= \frac{V_d}{\prod_{b=1}^{[d/2]}(1-\omega_b^2)}$. It is easy to derive this formula for odd spheres. We have also analytically checked this formula for $S^2$ and $S^4$. We are ashamed, however, to admit that we have not yet found an analytic derivation of this integral for general even spheres.}
\begin{equation}\label{otints:eq}
\begin{split}
  E &= \frac{ V_d\, \tc^d\, h(\nu) }{ \prod_b(1-\omega_b^2) }
       \brk{2\sum_a\frac{\omega_a^2}{1-\omega_a^2} + d - 1} \,,\\
  S &= \frac{ V_d\, \tc^{d-1} [dh(\nu) - \nu_i h_i(\nu)] }
            { \prod_b(1-\omega_b^2) } \,,\\
  L_a &= \frac{ V_d\, \tc^d\, h(\nu) }{ \prod_b(1-\omega_b^2) }
       \brk{\frac{2\omega_a}{1-\omega_a^2}} \,,\\
  R_i &= \frac{ V_d\, \tc^{d-1}\,  h_i(\nu) }
            { \prod_b(1-\omega_b^2) } \,,
\end{split}
\end{equation}
where
\begin{equation*}
  V_d = \operatorname{Vol}(S^{d-1}) = \frac{2\cdt\pi^{d/2}}{\Gamma(d/2)}\,.
\end{equation*}

Differentiating these gives
\begin{equation}\label{otpot:eq}
  T = \tc \qquad
  \Omega_a = \omega_a \qquad
  \zeta_i = \tc \nu_i\,.
\end{equation}
and the grand partition function
\begin{equation}\label{otgpf:eq}
  \ln\gpf = \frac{V_d\,T^{d-1}\,h(\zeta/T)}{\prod_b(1-\Omega_b^2)}\,.
\end{equation}
As in the previous subsection, the fluid dynamical approximation is expected to be valid provided $l_\mathrm{mfp}(T, \nu_i) \ll 1$.

In appendix \ref{sec:free}, we have computed the thermodynamics of a free charged scalar field on a sphere, and compared with the general results of this section.

\section{Comparison with uncharged black holes in arbitrary dimensions}\label{sec:arbitrary}

In the rest of this paper, we will compare the predictions from fluid dynamics derived above with the thermodynamics, stress tensors and charge distributions of various classes of large rotating black hole solutions in AdS spaces. We start with uncharged rotating black holes on $D$ dimensional AdS spaces (where $D$ is arbitrary), which are dual to rotating configurations of uncharged fluids on spheres of dimension $(D-2)$.

\subsection{Thermodynamics and stress tensor from fluid mechanics}

In case of uncharged fluids the function $h(\nu)$ in the above section is a constant $h(\nu)=h$. Therefore $h_i(\nu) = \frac{\partial h(\nu)}{\partial \nu_i}$ are all equal to zero. It follows from equations \eqref{otints:eq} and \eqref{otpot:eq} that
\begin{equation}\label{nochartherm:eq}
\begin{split}
E &= \frac{ V_d\, T^d\, h }{ \prod_b(1-\Omega_b^2) }
       \brk{\sum_a\frac{2\Omega_a^2}{1-\Omega_a^2} + d - 1} \,,\\
  S &= \frac{ V_d\, T^{d-1}h d  }
            { \prod_b(1-\Omega_b^2) } \,,\\
  L_a &= \frac{V_d\, T^d\, h }{ \prod_b(1-\Omega_b^2) }
       \brk{\frac{2\Omega_a}{1-\Omega_a^2}} \,,\\
  R_i &= 0\,.
\end{split}
\end{equation}
The partition function is given by
\begin{equation}\label{uncharpar:eq}
  \ln\gpf = \frac{V_d\,T^{d-1}\,h}{\prod_b(1-\Omega_b^2)}\,.
\end{equation}
The stress tensor becomes
\begin{equation}\label{unchrstr:eq}
\begin{gathered}
  T^{tt} = h T^d (d\gamma^{d+2} - \gamma^d) \qquad
  T^{t\phi_a} = T^{\phi_a t} = h T^d d\gamma^{d+2} \Omega_a \\
  T^{\theta_a\theta_a} = h T^d \gamma^d
                   \prn{\prod_{b=1}^{a-1} \sec^2\theta_b} \\
  T^{\phi_a\phi_a} = h T^d \brk{ d\gamma^{d+2}\Omega_a^2
             +\gamma^d \prn{\prod_{b=1}^{a-1} \sec^2\theta_b} \csc^2\theta_a} \qquad
  T^{\phi_a\phi_b} = h T^d  d\gamma^{d+2} \Omega_a \Omega_b\,.
\end{gathered}
\end{equation}

The mean free path in fluid dynamics can be estimated by taking the ratio of shear viscosity to energy density. As mentioned in the introduction, for fluids with gravity duals we can equivalently estimate $l_\mathrm{mfp}$ by taking the ratio of entropy to $4 \pi$ times the energy (because of the universal relation $s=4\pi\eta$).
\begin{equation}\label{lmfparbitads:eq}
 \left.l_\mathrm{mfp}(T,\nu)\right|_{\Omega=0} \sim  \brk{\frac{S}{4\pi E}}_{\Omega=0}
 = \frac{d}{4\pi T(d-1)}.
\end{equation}
Consequently the expansion in $l_\mathrm{mfp}$ translates simply to an expansion in inverse powers of the temperature of our solutions.

\subsection{Thermodynamics from black holes}

The most general solution for uncharged rotating black holes in AdS$_D$  was obtained in \cite{Gibbons:2004uw, Gibbons:2004ai}. These solutions are labelled by the $n+1$ parameters\footnote{Recall that $n$ denotes the number of commuting angular momenta and is given by the expression $n= \operatorname{rank}\brk{SO(D-1)}$ on AdS$_D$.} $a_i$ and $r_+$ (these are related to the $n$ angular velocities and the horizon radius (or equivalently the mass parameter) of the black holes). The surface gravity $\kappa$ and the horizon area $A$ of these black holes are given by\footnote{In the expression of $\kappa$ for even dimension, the sign inside the second term in equation (4.7) of \cite{Gibbons:2004ai} is different form the sign given in equation (4.18) of \cite{Gibbons:2004uw}; we believe the latter sign is the correct one.}
\begin{equation}
\begin{split}
 \kappa &=
 \left\{\begin{aligned}
  &r_+(1+r_+^2)\sum_{i=1}^n\frac{1}{r_+^2 +a_i^2}-\frac{1}{r_+}
   &\quad \text{when}\ D &= 2n+1\,,\\
  & r_+(1+r_+^2)\sum_{i=1}^n\frac{1}{r_+^2 +a_i^2}-\frac{1-r_+^2}{2r_+}
   & \text{when}\ D &= 2n+2\,,
 \end{aligned}\right.\\
 A &=
 \left\{\begin{aligned}
  &\frac{V_d}{r_+}\prod_{i=1}^n\frac{r_+^2 + a_i^2}{1-a_i^2}
   &\quad \text{when}\ D &= 2n+1\,,\\
  &V_d\prod_{i=1}^n\frac{r_+^2 + a_i^2}{1-a_i^2}
   & \text{when}\ D &= 2n+2\,.
 \end{aligned}\right.
\end{split}
\end{equation}
We will be interested in these formulae in the limit of large $r_+$. In this limit the parameter $m$ (which appears in the formulae of \cite{Gibbons:2004uw, Gibbons:2004ai}) and the temperature $T = \kappa/2\pi$ are given as functions of $r_+$ by
\begin{equation}\label{thori:eq}
\begin{split}
 T &= \brk{\frac{(D-1)r_+}{4\pi}}\left( 1+ \mathcal{O}(1/r_+^2) \right), \\
 2m &= r_+^{D-1} \left( 1 +\mathcal{O}(1/r_+^2) \right).
\end{split}
\end{equation}
From these equations, it follows that the parameter $m$ is related to the temperature $T$ as
\begin{equation}\label{mTarbitbh:eq}
2m=T^{D-1}\brk{\frac{4 \pi}{D-1}}^{D-1}\left( 1 +\mathcal{O}(1/T^2) \right).
\end{equation}
To leading order in $r_+$, the thermodynamic formulae take the form
\begin{equation}\label{arbtherm:eq}
\begin{split}
\Omega_i &= a_i\,,\\
 E &=\frac{V_{D-1} T^{D-1}}
    {16\pi G_D \prod_{j=1}^{n}(1-a_j^2)}
    \brk{\frac{4\pi}{D-1}}^{D-1}
    \brk{\sum_{i=1}^{n}\frac{2a_i^2}{1-a_i^2}+D-2} ,\\
L_i &= \frac{V_{D-1} T^{D-1}}
    {16\pi G_D \prod_{j=1}^{n}(1-a_j^2)}
    \brk{\frac{4\pi}{D-1}}^{D-1}
     \brk{\frac{2 a_i}{1-a_i^2}} ,\\
 S &= \frac{V_{D-1} T^{D-2}(D-1)}
    {16\pi G_D \prod_{j=1}^{n}(1-a_j^2)}
     \brk{\frac{4\pi}{D-1}}^{D-1},\\
R_i &= 0\,,
\end{split}
\end{equation}
where $V_{D-1}$ is the volume of $S^{D-2}$ and $G_D$ is Newton's constant in $D$ dimensions. The corrections to each of these expressions are suppressed by factors of $\mathcal{O}( 1/r_+^2) =\mathcal{O}(1/T^2)$ relative to the leading order results presented above (i.e.\ there are no next to leading order corrections).

These thermodynamic formulae listed in \eqref{arbtherm:eq} are in perfect agreement with the fluid mechanics expressions in \eqref{nochartherm:eq} upon making the following identifications: the spacetime dimensions of the boundary theory $d=D-1$, the black hole angular velocities $a_i$ are identified with $\Omega_a$ and the constant $h$ is identified as
\begin{equation}\label{hval:eq}
h = \frac{1}{16\pi G_D}\brk{\frac{4\pi}{D-1}}^{D-1}.
\end{equation}
In the next subsection, we will see that this agreement goes beyond the global thermodynamic quantities. Local conserved currents are also in perfect agreement with the black hole physics.

\subsection{Stress tensor from rotating black holes in AdS$_D$}

The uncharged rotating black holes both in odd dimensions ($D = 2n + 1$) and even dimensions ($D =  2n + 2$) are presented in detail in \cite{Gibbons:2004uw}, equation (E-3) and \cite{Gibbons:2004ai}, equation (4.2) . After performing some coordinate transformations (see appendix \ref{sec:bhstruc}) that take the metric of that paper to the standard form of AdS$_{D}$ at the boundary, we have computed the stress tensor of this solution.

Our calculation, presented in appendix \ref{sec:bhstruc} uses the standard AdS/CFT dictionary. In more detail, we foliate the solution in boundary spheres, compute the extrinsic curvature $\Theta^\mu_\nu$ of these foliations near the boundary, subtract off the appropriate counter terms contributions \cite{Kraus:1999di, Henningson:1998gx, de-Haro:2000xn, Skenderis:2000in, Papadimitriou:2005ii, Cheng:2005wk, Olea:2005gb, Olea:2006vd}, and finally multiply the answer by the $r^{D-1}$ to obtain the stress tensor on a unit sphere.

We find that the stress tensor so calculated takes the form (see appendix \ref{sec:bhstruc})
\begin{equation}\label{uncharblstodd:eq}
\begin{split}
\Pi^{tt} &= \frac{2m}{16 \pi G_D}[(D-1)\gamma^{D+1}-\gamma^{D-1}]\\
\Pi^{\phi_a\phi_a} &=
  \frac{2m}{16\pi G_D}[(D-1)\gamma^{D+1}\omega_a^2+\gamma^{D-1}\mu_a^{-2}]\\
\Pi^{t\phi_a} &= \Pi^{\phi_a t} =  \frac{2m}{16\pi G_D}(D-1)\gamma^{D+1}\omega_a\\
\Pi^{\phi_a\phi_b} &= \Pi^{\phi_b\phi_a}
  =  \frac{2m}{16 \pi G_D}\gamma^{D+1}\omega_a\omega_b\\
\Pi^{\theta_a\theta_a
} &=
  \frac{2m}{16 \pi G_D}\gamma^{D-1}\left(\prod_{b=1}^{a-1}\sec^2\theta_b\right).\\
\end{split}
\end{equation}
Here $\gamma^{-2} = 1 -\sum_{a=1}^n\omega_a^2\mu_a^2 $ where $\mu_a = \left(\prod_{b=1}^{a-1}\cos\theta_b\right)\sin\theta_a$.

Note that the functional form of these expressions (i.e.\ dependence of various components of the stress tensor on the coordinates of the sphere) agrees exactly with the predictions of fluid dynamics even at finite values of $r_+$. In the large $r_+$ limit (using \eqref{mTarbitbh:eq} and \eqref{hval:eq}) , we further have
\begin{equation*}
\begin{split}
\Omega_a &=\omega_a\,,\\
 D-1 &= d \,,\\
\frac{2m}{16 \pi G_D}& = T^d h\,. \\
\end{split}
\end{equation*}
With these identifications, \eqref{uncharblstodd:eq} coming from gravity agrees precisely with \eqref{unchrstr:eq} from fluid mechanics.

% \subsubsection{AdS Spaces of even dimension ($D = 2n +2$)}
%
%  Following the same procedure described
% above, we find the following expressions for the boundary stress tensor
% corresponding to these solutions (see Appendix \ref{sec:bhstruc} for details)
% %
% \begin{equation}\label{uncharblsteven:eq}
% \begin{split}
% T^{tt} &= -\frac{2m}{16\pi G_D}\gamma^{(2n+3)}[2n + 1-\gamma^{-2}]\\
% T^{\phi_i\phi_i} &= \frac{2m}{16\pi G_D}\gamma^{(2n+3)}[(2n+1)a_i^2\mu_i^2-\gamma^{-2}]\\
% T^{t\phi_i} &= T^{\phi_i t} =  \frac{2m}{16\pi G_D}\gamma^{(2n+3)}(2n +1)a_i\\
% T^{\phi_i\phi_j} &= T^{\phi_j\phi_i} =  \frac{2m}{16\pi G_D}\gamma^{(2n+3)}a_ia_j\\
% T^{\theta_i\theta_i} &=  \frac{2m}{16\pi G_D}\gamma^{(2n+1)}\left(\prod_{j=1}^{i-1}\sec^2\theta_j\right)\\
% \end{split}
% \end{equation}
% %
% Here  $\mu_a = \left(\prod_{b=1}^{a-1}\cos\theta_b\right)\sin\theta_a$,
% $\gamma^{-2} = 1 -\sum_{a=1}^n\omega_a^2\mu_a^2 $,
% These expressions match with the expressions derived from fluid mechanics
% in \eqref{unchrstr:eq} upon setting $m=\half (\frac{4 \pi T}{D-1})^{D-1}$ and
% using \eqref{hval:eq}.

We proceed now to estimate the limits of validity of our analysis above. From the black hole side, since we have expanded the formulae of black hole thermodynamics in $1/r_+$ to match them with fluid mechanics, this analysis is valid if $r_+$ is large. From the fluid mechanics side, we expect corrections of the order of $l_\mathrm{mfp}$. To estimate $l_\mathrm{mfp}$ in this case, we substitute \eqref{thori:eq} into \eqref{lmfparbitads:eq} to get
\begin{equation*}
l_\mathrm{mfp} \sim \frac{1}{r_+(d-1)} \ll 1\,.
\end{equation*}
Hence, we see that the condition from fluid mechanics is exactly the same as taking large horizon radius limit: the expansion of black hole thermodynamics in a power series in $ \frac{1}{r_+}$ appears to be exactly dual to the fluid mechanical expansion as a power series in $l_\mathrm{mfp}$.

\section{Comparison with black holes in $AdS_5 \times S^5$}\label{sec:bhcomp}

Large $N$, $\CN=4$ Yang-Mills, at strong 't Hooft coupling on $S^{3}\times \R$, is dual to classical gravity on $AdS_5 \times S^5$. Hence, we can specialise the general fluid dynamical analysis presented above to the study of equilibrium configurations of the rotating $\CN=4$ plasma on $S^3$ and then compare the results with the physics of classical black holes in $AdS_5\times S^5$.

Large black holes in $AdS_5\times S^5$ are expected to appear in a six parameter family, labelled by three SO(6) Cartan charges ($c=3$), two $SO(4)$ rotations ($n=2$) and the mass. While the most general black hole in $AdS_5\times S^5$ has not yet been constructed, several sub-families of these black holes have been determined.

In this section, we will compare the thermodynamic predictions of fluid mechanics with all black hole solutions that we are aware of and demonstrate that the two descriptions agree in the large horizon radius limit. For one class of black holes we will also compare black hole stress tensor and charge distributions with that of the fluid mechanics and once again find perfect agreement (in the appropriate limit).

We begin this section with a review of the predictions of fluid mechanics for strongly coupled $\mathcal{N}=4$ Yang-Mills on $S^3$. Note that this is a special case of the conformal fluid dynamics of previous sections with $d=D-1=4$.

% \footnote{Further, for comparison with the previous section, we note that Newton's constant in $AdS_5$ is given by $\frac{1}{G_5}=\frac{2 N^2}{\pi}$. Substituting this into \eqref{hval:eq} , we get the value of $h$ at $\nu=0$ case for $AdS_5$ as
% \begin{equation*}
% h(\nu=0,D=5) =\frac{\pi^4}{16\pi G_5} =\frac{\pi^2 N^2}{8}
% \end{equation*}}

\subsection{The strongly coupled $\mathcal{N}=4$ Yang-Mills Plasma }

The gravity solution for SO(6) charged black branes (or, equivalently, large SO(6) charged but non-rotating black holes in $AdS_5 \times S^5$) has been used to extract the equation of state of $\CN=4$ Yang-Mills (see \cite[\S2]{Son:2006em} for the thermodynamic expressions in the infinite radius limit).

Rather than listing all the thermodynamic variables, we use the earlier parametrisation of \eqref{gcterm:eq} to state our results. The thermodynamics of the $\CN=4$ Yang-Mills is described by the following equations\footnote{Note that our convention for the gauge field differs from \cite[\S2]{Son:2006em} by a factor of $\sqrt{2}$.}
\begin{equation}\label{bbeosparam:eq}
\begin{split}
  h(\nu) = \frac{\ploc}{\tloc^4} &=
     2 \pi^2 N^2 \frac{ \prod_j(1+\kappa_j)^3}
                      {(2+\sum_j \kappa_j - \prod_j\kappa_j)^4} \,,\\
  \nu_i = \frac{\mu_i}{\tloc} &=
    \frac{2\pi \prod_j (1+\kappa_j)}
         {\prn{2+ \sum_j \kappa_j - \prod_j \kappa_j}}
    \frac{\sqrt{\kappa_i}}{1+\kappa_i} \,,\\
  h_i(\nu) = \frac{r_i}{\tloc^3} &=
    \frac{2\pi N^2 \prod_j (1+\kappa_j)^2}
         {\prn{2+ \sum_j \kappa_j - \prod_j \kappa_j}^3}
    \sqrt{\kappa_i} \,.%,\\
%  \nu_j h_j(\nu) &=
%    \frac{4\pi^2 N^2 \prod_j (1+\kappa_j)^2}
%         {\prn{2+ \sum_j \kappa_j - \prod_j \kappa_j}^4}
%    \brk{2\prod_j (1+\kappa_j) -
%         \prn{2+ \sum_j \kappa_j - \prod_j \kappa_j}}\,.
\end{split}
\end{equation}
where the auxiliary parameters $\kappa_i$ have a direct physical interpretation in terms of entropy and charge densities (see \S2 of \cite{Son:2006em}) -
\begin{equation}\label{kappaqs:eq}
\kappa_i = \frac{4\pi^2 R_i^2}{S^2}\,.
\end{equation}

$\kappa_i$ are constrained by $\kappa_i \geq 0$ and by the condition\footnote{Which is obtained by requiring that the temperature $\tloc \geq 0$ in the expression for $\tloc$ in \cite[\S2]{Son:2006em}.}
$$\frac{2+ \sum_j \kappa_j - \prod_j \kappa_j}{\prod_j (1+\kappa_j)} =\brk{\sum_j\frac{1}{1+\kappa_j} -1} \geq 0.$$
It follows from \eqref{kappaqs:eq} that $\kappa_i$ is finite for configurations with finite charge and non-zero entropy. The configurations with $\kappa_i\to\infty$ (for any $i$) are thermodynamically singular, since in this limit, the $i^{th}$ charge density is much larger than the entropy density. Hence, in the following, we shall demand that $\kappa_i$ be finite.

The general analysis presented before now allows us to construct the most general stationary solution of the $\mathcal{N}=4$ fluid rotating on a 3-sphere. The thermodynamic formulae and currents of these solutions follow from \eqref{chcurrent:eq}, \eqref{chstr:eq} and \eqref{chch:eq} upon setting
\begin{equation}\label{bbchcoeffs:eq}
\begin{split}
 A &= h(\nu) = 2\pi^2 N^2 \frac{ \prod_j(1+\kappa_j)^3}
                       {(2+\sum_j \kappa_j - \prod_j\kappa_j)^4}
 \,, \\
 B &= 4h(\nu)- \nu_i h_i(\nu) =4\pi^2 N^2  \frac{\prod_j (1+\kappa_j)^2}
         {({2+ \sum_j \kappa_j - \prod_j \kappa_j})^3}
 \,, \\
 C_i &= h_i(\nu) = 2\pi N^2 \sqrt{\kappa_i} \frac{\prod_j (1+\kappa_j)^2}
         {( {2+ \sum_j \kappa_j - \prod_j \kappa_j})^3 }
 \,,
\end{split}
\end{equation}
which leads to
%%
%\begin{equation}\label{bbchch:eq}
%\begin{split}
%  E &= \frac{4\pi^4 N^2 \tc^4 [4-(1+\omega_1^2)(1+\omega_2^2)]}
%               {(1-\omega_1^2)^2(1-\omega_2^2)^3}
%       \frac{\prod_j (1+\kappa_j)^3}
%         {\prn{2+ \sum_j \kappa_j - \prod_j \kappa_j}^4}
%\,,\\
%  L_1 &= \frac{8\pi^4 N^2 \tc^4 \omega_1}
%              {(1-\omega_1^2)^2(1-\omega_2^2)}
%       \frac{\prod_j (1+\kappa_j)^3}
%         {\prn{2+ \sum_j \kappa_j - \prod_j \kappa_j}^4}
%\,,\\
%  L_2 &= \frac{8\pi^4 N^2 \tc^4 \omega_2}
%              {(1-\omega_1^2)(1-\omega_2^2)^2}
%       \frac{\prod_j (1+\kappa_j)^3}
%         {\prn{2+ \sum_j \kappa_j - \prod_j \kappa_j}^4}
%\,,\\
%  S &= \frac{8\pi^4 N^2 \tc^3}
%            {(1-\omega_1^2)(1-\omega_2^2)}
%       \frac{\prod_j (1+\kappa_j)^2}
%         {\prn{2+ \sum_j \kappa_j - \prod_j \kappa_j}^3}
%\,,\\
%  R_i &= \frac{4\pi^3 N^2 \tc^3 \sqrt{\kappa_i}}
%            {(1-\omega_1^2)(1-\omega_2^2)}
%       \frac{\prod_j (1+\kappa_j)^2}
%         {\prn{2+ \sum_j \kappa_j - \prod_j \kappa_j}^3}
%\,.\\
%\end{split}
%\end{equation}
%%
%
\begin{equation}\label{genchpot:eq}
  \zeta_i = \frac{2\pi T \prod_j (1+\kappa_j)}
         {\prn{2+ \sum_j \kappa_j - \prod_j \kappa_j}}
    \frac{\sqrt{\kappa_i}}{1+\kappa_i} \,,
\end{equation}
and
\begin{equation}\label{gpfch:eq}
\begin{split}
  \ln\gpf &%\equiv \ln\Tr\brk{ \e^{- \beta H + \beta \Omega_a L_a
%                                + \beta\zeta_i R_i}}
%         = S  - \beta E + \beta\Omega_a L_a + \beta\zeta_i R_i\\
%         &= \frac{2\pi^2 T^3 [g(\chi) - \chi_jg_j(\chi)]}
%                {(1-\omega_1^2)(1-\omega_2^2)}
         = \frac{2\pi^2 N^2 V_4 T^3  \prod_j(1+\kappa_j)^3}
                {(1-\Omega_1^2)(1-\Omega_2^2)
                 \prn{2+ \sum_j \kappa_j - \prod_j \kappa_j}^4}\,,
\end{split}
\end{equation}
where we have used the notation $V_4=\operatorname{Vol}(S^3)=2\pi^2$ as before.

As before, the mean free path in fluid mechanics can be estimated as
\begin{equation}\label{ads5lmfp:eq}
\begin{split}
l_\mathrm{mfp} \sim \brk{\frac{S}{4\pi E}}_{\Omega=0} &= \frac{B}{(d-1)4\pi TA} = \frac{\prn{2+ \sum_j \kappa_j - \prod_j \kappa_j}}{6\pi T\prod_j(1+\kappa_j)}\\
&= \frac{1}{6\pi T} \brk{\sum_j\frac{1}{1+\kappa_j} -1}.
\end{split}
\end{equation}

\subsection{The extremal limit}\label{sec:exttherm}

The strongly coupled ${\cal N}=4$ Yang-Mills plasma has an interesting feature; it has interesting and nontrivial thermodynamics even at zero temperature. In this subsection, we investigate this feature and point out that it implies the existence of interesting zero temperature solutions of fluid dynamics which will turn out to be dual to large, extremal black holes.

\subsubsection{Thermodynamics}\label{sec:extthrm}

In the above section, we presented the thermodynamics of strongly coupled $\CN=4$ Yang-Mills plasma in terms of the parameters $\kappa_i$. These parameters are constrained by the conditions $\kappa_i \geq 0$ and $\sum_i \frac{1}{1+\kappa_i} \geq 1$ with $\kappa_i$ finite. In order to visualise the allowed range over which the variables $\kappa_i$'s can vary, it is convenient to define a new set of variables
\begin{equation}\label{xyzkappa:eq}
\begin{split}
X_i &=\frac{1}{1+\kappa_i}\,,\quad X_i = X,\ Y,\ Z ,\\
\chi &=\frac{T}{X+Y+Z-1}\,.
\end{split}
\end{equation}

\begin{figure}[tbh]
 \centering
 \includegraphics[width=0.4\textwidth]{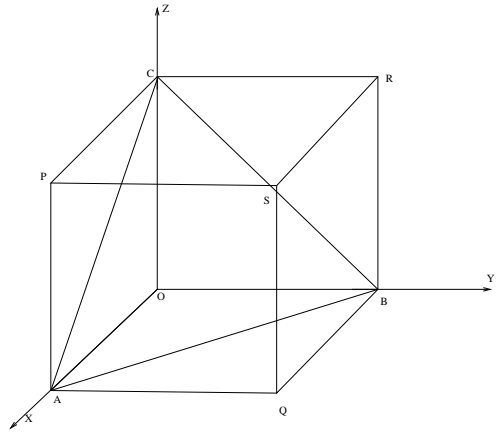}
\caption{The space of allowed $\kappa_i$'s. The axes correspond to $X=\frac{1}{1+\kappa_1}$, $Y=\frac{1}{1+\kappa_2}$ and $Z=\frac{1}{1+\kappa_3}$. The $X_i$'s can lie anywhere in the cube outside the ``extremal" plane $X+Y+Z = 1$.}
 \label{fig:cube}
\end{figure}

The constraints $\kappa_i \geq 0$ and $\sum_i \frac{1}{1+\kappa_i} \geq 1$ with $\kappa_i$ finite translate into the constraints $0 < X_i \leq 1$ and $X+Y+Z \geq 1$ . Geometrically, this is just the statement that $X_i$'s can lie anywhere inside the cube shown in fig.\ref{fig:cube}, away from the planes $X_i=0$ and on or above the plane $X+Y+Z = 1$.

The energy density, the entropy density and the charge densities of the Yang-Mills plasma may be rewritten as a function of $X, Y, Z$ and $\chi$ as
\begin{equation}\label{thermxyz:eq}
\begin{split}
  \rho &= 6 \pi^2  N^2 XYZ \chi^4,\qquad
         s = 4 \pi^2  N^2 XYZ \chi^3,\\
  r_i &=  2 \pi N^2 XYZ  \chi^3 \sqrt{\frac{1-X_i}{X_i}}.
\end{split}
\end{equation}
The condition for the validity of fluid mechanics becomes
\begin{equation}\label{ads5lmfpxyz:eq}
l_\mathrm{mfp} \sim \ \frac{1}{6\pi \chi} \ll 1 \qquad \text{or} \qquad \chi \gg 1\,.
\end{equation}

Consider now the case in which $\chi$ is large, but finite and $X,Y,Z$ take values close to the interior of the triangle $ABC$ in fig.\ref{fig:cube}. From \eqref{xyzkappa:eq} and \eqref{thermxyz:eq}, it is evident that this is equivalent to taking an extremal limit $T \to 0$  with appropriate chemical potentials. All thermodynamic quantities listed above are smooth in this limit and the fluid mechanics continues to be valid.

The $\CN=4$ Yang-Mills plasma with three nonzero R-charges always has a nonsingular extremal limit. In the case that one of the charges say $r_3$ is zero, then we are constrained to move on the $X_3 =1$ plane in the space of $X_i$'s. Hence, we can never approach the `extremal triangle' $X+Y+Z=1$.\footnote{Remember that we have already excluded, on physical grounds, the point $X_1=X_2=0,\ X_3 =1$ which lies in the intersection of $X_3 =1$ plane and the extremal plane $X+Y+Z=1$.} Thus, we have no nonsingular extremal limit if any one of the three R-charges is zero. By a similar argument, no nonsingular extremal limit exists if two of the R-charges were zero.

We note that Gubser and Mitra have previously observed that
charged black branes near extremality are sometimes thermodynamically unstable \cite{Gubser:2000ec}. Although we have not performed a careful analysis of the thermodynamic stability of the charged fluids we study in this paper (see however \cite{Son:2006em}), we suspect that these fluids all have Gubser-Mitra type thermodynamic instabilities near extremality. If this is the case, the near extremal fluid solutions we study in this section  and the next -- and the black holes that these are dual to -- are presumably unstable to small fluctuations. Whether stable or not, these configurations are valid solutions of fluid dynamics. We postpone a serious discussion of stability to future work.\footnote{We thank Sangmin Lee for discussion of these issues.}

\subsubsection{Fluid mechanics}\label{sec:extfm}

The thermodynamic expressions for the charges of a rotating Yang-Mills plasma
take the form
\begin{equation}\label{ads5thermxyz:eq}
\begin{split}
  E &= \frac{2\pi^2 N^2 XYZ  V_4 }
               {(1-\omega_1^2)(1-\omega_2^2)}
               \brk{\frac{2\omega_1^2}{1-\omega_1^2}+\frac{2\omega_2^2}{1-\omega_2^2}+3 }
               \brk{\frac{T}{X+Y+Z-1}}^4,\\
  L_1 &= \frac{2\pi^2 N^2 XYZ V_4}
              {(1-\omega_1^2)(1-\omega_2^2)}
               \brk{\frac{2\omega_1}{1-\omega_1^2}}
               \brk{\frac{T}{X+Y+Z-1}}^4,\\
  L_2 &= \frac{V_4 \tc^4 A }
              {(1-\omega_1^2)(1-\omega_2^2)}
               \brk{\frac{2\omega_2}{1-\omega_2^2}}
               \brk{\frac{T}{X+Y+Z-1}}^4,\\
  S &= \frac{4\pi^2 N^2 XYZ V_4}
            {(1-\omega_1^2)(1-\omega_2^2)}
            \brk{\frac{T}{X+Y+Z-1}}^3,\\
  R_i &= \frac{2\pi N^2 XYZ  V_4}
            {(1-\omega_1^2)(1-\omega_2^2)}
             \brk{\frac{T}{X+Y+Z-1}}^3
             \sqrt{\frac{1-X_i}{X_i}}\,,
\end{split}
\end{equation}
and the mean free path
\begin{equation}\label{ads5lmfpxyz1:eq}
l_\mathrm{mfp} \sim \ \frac{X+Y+Z-1}{6\pi T} \ll 1\,.
\end{equation}
We see that all thermodynamical charges of our rotating fluid configurations are nonsingular, and that fluid mechanics is a valid approximation for these solutions, in the extremal limit described in the previous subsection, provided only that $\chi \gg 1$.\footnote{In greater generality, in order for fluid mechanics to be a valid approximation for our solutions it is necessary that either $T \gg 1$ (which is by itself sufficient) or that $X+Y+Z -1 \rightarrow 0$ (under which condition the ratio $\chi$ of the previous section must be large and (conservatively) none of $X$, $Y$ or $Z$ be very small).}

The solution so obtained describes a rotating fluid whose local temperature vanishes everywhere, but whose rest frame charge density is a function of location on the $S^3$ (it scales like $\gamma^3$). As we will see below these extremal configurations of rotating fluid on $S^3$ are exactly dual to large, rotating, extremal black holes in AdS$_5$.

\subsection{Predictions from fluid mechanics in special cases}\label{sec:jalbidhan}

As mentioned in the beginning of this section, the most general black hole in $AdS_5\times S^5$ has not yet been constructed, but several subfamilies of these black holes are known. To facilitate the comparison between fluid mechanics on $S^3$ on one hand and these subfamilies of black holes on the other, in this subsection, we specialise the general predictions of the previous subsection to various specific cases.

\subsubsection{All SO(6) charges equal: arbitrary angular velocities}\label{sec:eqch}

Consider first the case of a fluid with equal SO(6) charges (with the rotational parameters arbitrary). That is we set $\kappa_1=\kappa_2=\kappa_3=\kappa$ in the general formulae above. Noting that $(2+3  \kappa-\kappa^3)=(\kappa+1)^2(2 -\kappa)$ we find that the stress tensor and currents are given by \eqref{chstr:eq} and \eqref{chcurrent:eq} with
\begin{equation}\label{eqchcoeff:eq}
\begin{split}
 A   &= \frac{ 2 \pi^2 N^2 (1+\kappa) }{ (2 - \kappa)^4 }\,, \\
 B   &= \frac{ 4 \pi^2 N^2 }{ (2-\kappa)^3 }\,, \\
 C_i &= \frac{ 2 \pi N^2 \sqrt{\kappa} }{ (2-\kappa)^3 }\,.
\end{split}
\end{equation}
The thermodynamics can be summarised by
\begin{equation}\label{eqchtherm:eq}
  \zeta_i = \frac{ 2\pi T\sqrt{\kappa} }{ (2-\kappa) }\,, \qquad
  \ln\gpf(T,\Omega,\zeta)  = \frac{  2\pi^2 N^2 V_4 T^3 (1+\kappa)}
                       { (1-\Omega_1^2)(1-\Omega_2^2)(2-\kappa)^4 }\,.
\end{equation}
The formula for mean free path \eqref{ads5lmfp:eq} reduces to
\begin{equation}\label{lmfpthreech:eq}
l_\mathrm{mfp} \sim \frac{1}{6\pi T} \brk{\frac{2-\kappa}{1+\kappa}}.
\end{equation}

Let us specialise the extremal thermodynamics of $\CN=4$ Yang-Mills fluid presented before to this case. In terms of the variables introduced in \S\S\ref{sec:exttherm}, we have $X=Y=Z$ which is a straight line in the $X_i$ space. The extremal limit is obtained when this line cuts the extremal plane $X+Y+Z=1$ , i.e, at the point $X=Y=Z=1/3 $. This corresponds to the extremal limit $\kappa \to 2 $.

More explicitly, in the extremal limit
\begin{equation}
 T \to 0\,, \qquad
 (2 - \kappa) = \frac{T}{\chi}\,,
\end{equation}
with $\chi$ large but finite. The thermodynamic quantities obtained by differentiating the grand partition function \eqref{largechargetherm:eq},
\begin{equation}\label{eqchext:eq}
\begin{split}
 S &=\frac{2 N^2 \pi^2 V_4 T^{3}} {(2-\kappa)^3}
     \frac{1}{(1-a^2)(1-b^2)}\\
 L_1 &=\frac{4 N^2 \pi^2 V_4 T^4 (1+\kappa)}{(2-\kappa)^4}
       \frac{a}{(1-a^2)^2(1-b^2)}\\
 L_2 &=\frac{4 N^2 \pi^2 V_4 T^4 (1+\kappa)}{(2-\kappa)^4}
       \frac{b}{(1-a^2)(1-b^2)^2} \\
 R &=\frac{2 \pi N^2 V_4 T^3 \sqrt{\kappa}}{(2-\kappa)^3}
     \frac{1}{(1-a^2)(1-b^2)}\\
 E &=\frac{ 2 \pi^2  N^2 V_4 T^4 (1+\kappa) }{ (2-\kappa)^4 }
     \frac{ [4-(1+a^2)(1+b^2)] }{ (1-a^2)^2(1-b^2)^2 }\,,
\end{split}
\end{equation}
are all smooth; and they describe a fluid configuration whose energy, angular momentum, charge and entropy scale as $N^2 \chi^4, ~ N^2 \chi^4, ~ N^2 \chi^3$ and $~N^2 \chi^3$ respectively.

\subsubsection{Independent SO(6) charges: equal rotations}\label{sec:eqrot}

Consider the special case $\omega_1=\omega_2=\Omega$
(the three SO(6)) chemical potentials are left arbitrary).
The stress tensor and currents are given by
\eqref{chstr:eq} and \eqref{chcurrent:eq} with
\begin{equation}\label{eqrotcoeff:eq}
\begin{split}
 A &= 2\pi^2 N^2 \frac{ \prod_j(1+\kappa_j)^3}
                       {(2+\sum_j \kappa_j - \prod_j\kappa_j)^4}
 \,, \\
 B &= 4\pi^2 N^2  \frac{\prod_j (1+\kappa_j)^2}
         {({2+ \sum_j \kappa_j - \prod_j \kappa_j})^3}
 \,, \\
 C_i &= 2\pi N^2 \sqrt{\kappa_i} \frac{\prod_j (1+\kappa_j)^2}
         {( {2+ \sum_j \kappa_j - \prod_j \kappa_j})^3 }
 \,, \\
 \gamma &= \frac{1}{\sqrt{1-\Omega^2}}\,, \qquad v = \Omega\,.
\end{split}
\end{equation}
The thermodynamics can be summarised by
\begin{equation}\label{eqrottherm:eq}
  \zeta_i = \frac{2\pi T \prod_j (1+\kappa_j)}
         {\prn{2+ \sum_j \kappa_j - \prod_j \kappa_j}}
    \frac{\sqrt{\kappa_i}}{1+\kappa_i} \,,
 \qquad
  \ln\gpf
         = \frac{ 2\pi^2 N^2 V_4 T^3 \prod_j(1+\kappa_j)^3}
                {(1-\Omega^2)^2
                 \prn{2+ \sum_j \kappa_j - \prod_j \kappa_j}^4}\,.
\end{equation}
The expression for mean free path \eqref{ads5lmfp:eq} reduces to
\begin{equation}\label{lmfpindepch:eq}
l_\mathrm{mfp} \sim \frac{1}{6\pi T} \brk{\sum_j\frac{1}{1+\kappa_j} -1}.
\end{equation}

The extremal limit $\sum_j(1+\kappa_j)^{-1} \to 1 $ with all $\kappa_i$ kept finite, is nonsingular, and yields solutions that are well described by fluid dynamics when $l_\mathrm{mfp}$ is small.

\subsubsection{Two equal nonzero SO(6) charges:
arbitrary angular velocities}\label{sec:twoch}

Consider now the case when $\kappa_1=\kappa_2=\kappa$, $\kappa_3=0$. We find that the stress tensor and currents are given by \eqref{chstr:eq} and \eqref{chcurrent:eq} with
\begin{equation}\label{twochcoeff:eq}
\begin{split}
 A   &= \frac{ \pi^2 N^2 (1+\kappa)^2 }{ 8 }\,, \\
 B   &= \frac{ \pi^2 N^2 (1+\kappa) }{ 2 }\,, \\
 C_1 = C_2 &= \frac{ \pi N^2\sqrt{\kappa}(1+\kappa) }{ 4 }\,,\\
 C_3 &= 0\,.
\end{split}
\end{equation}
The thermodynamics can be summarised by
\begin{equation}\label{twochtherm:eq}
  \zeta_1=\zeta_2 = \pi T \sqrt{\kappa} \,, \qquad
  \ln\gpf(T,\Omega,\zeta)  = \frac{  \pi^2 N^2 V_4 T^3  (1+\kappa)^2}
                       { 8 (1-\Omega_1^2)(1-\Omega_2^2) }\,.
\end{equation}
The expression for mean free path, from \eqref{ads5lmfp:eq}, is
\begin{equation}\label{lmfptwoch:eq}
l_\mathrm{mfp} \sim \frac{1}{3\pi T(1+\kappa)}\,.
\end{equation}
It follows from \eqref{lmfptwoch:eq} that fluid mechanics is a good approximation when $T$ is large. Though this equation would appear to suggest that the fluid dynamical approximation is also valid (for instance) at fixed $T$ and large $\kappa$, we have emphasised before, the limit of large $\kappa$ is thermodynamically suspect. Conservatively, thus, fluid mechanics applies only at large temperatures.

\subsubsection{A single nonzero charge: arbitrary angular velocities}
\label{sec:onech}

We now set $\kappa_1=\kappa$, $\kappa_2=\kappa_3=0$ leaving angular velocities arbitrary. The stress tensor and currents are given by \eqref{chstr:eq} and \eqref{chcurrent:eq} with
\begin{equation}\label{onechcoeff:eq}
\begin{split}
 A   &= \frac{ 2 \pi^2 N^2 (1+\kappa)^3 }{ (2 + \kappa)^4 }\,, \\
 B   &= \frac{ 4 \pi^2 N^2 (1+\kappa)^2 }{ (2+\kappa)^3 }\,, \\
 C_1 &= \frac{ 2 \pi N^2 \sqrt{\kappa}(1+\kappa)^2 }{ (2+\kappa)^3 }\,,\\
 C_2 = C_3 &= 0\,.
\end{split}
\end{equation}
The thermodynamics can be summarised by
\begin{equation}\label{onechtherm:eq}
  \zeta = \frac{ 2\pi T\sqrt{\kappa} }{ (2+\kappa) }\,, \qquad
  \ln\gpf(T,\Omega,\zeta)  = \frac{ 2\pi^2 N^2 V_4 T^3 (1+\kappa)^3}
                       { (1-\Omega_1^2)(1-\Omega_2^2)(2+\kappa)^4 }\,.
\end{equation}
The mean free path, from \eqref{ads5lmfp:eq} is given by
\begin{equation}\label{lmfponech:eq}
l_\mathrm{mfp} \sim \frac{1}{6\pi T}\brk{\frac{2+\kappa}{1+\kappa}}.
\end{equation}

As in the previous subsection, this particular case does not admit thermodynamically nonsingular zero temperature (or extremal) configurations.

\subsection{Black holes with all R-charges equal}\label{sec:bheqch}

Having derived the fluid mechanics predictions for various different black holes, we now proceed to examine the black hole solutions. First, we will focus on the case of black holes with arbitrary angular momenta in AdS$_5$ but equal SO(6) charges. The relevant solution has been presented in \cite{Chong:2005hr}.

\subsubsection{Thermodynamics}

The black holes presented in \cite{Chong:2005hr} are labelled by two angular velocities $a, b$, and three more parameters $q, m$ and $r_+$. These five parameters are not all independent; they are constrained by one equation  relating horizon radius to the parameter $m$ ($\Delta_r=0$ in that paper). We thus have a four parameter set of black holes.\footnote{We work in conventions in which the AdS radius and hence the parameter $g$ of \cite{Chong:2005hr} is set to unity.}

The relatively complicated black hole thermodynamic formulae of \cite{Chong:2005hr} simplify if the parameter $r_+$ (which may be interpreted as the horizon radius) is taken to be large. In particular, consider the limit
\begin{equation}\label{conditions:eq}
  r_+ \gg 1 \qquad \text{and} \quad y=q/r_+^3 \quad \text{fixed.}
\end{equation}
In this limit, to leading order,we have
\begin{equation} \label{rpmtc:eq}
\begin{split}
  T &= \frac{r_+}{2\pi}(2-y^2)\,,\\
 2m &= r_+^4 (1 + y^2)\,.
\end{split}
\end{equation}
From the positivity of $T$ and $r_+$ it follows immediately that
$0\leq y^2 \leq 2$.

%  $r_+$ may be reexpressed in terms of the black hole temperature $T$ and the variable $y$ using
% %
% \begin{equation} \label{rptc:eq}
%   r_+= \frac{2 \pi T}{2-y^2}\,.
% \end{equation}
% %
% In this limit the parameter $m$ of \cite{Chong:2005hr} is determined (via the equation $\Delta_r=0$ of the same paper) to be
% %
% \begin{equation} \label{mtc:eq}
%   2m= r_+^4 (1 + y^2)
% \end{equation}

%
% Over the next few paragraphs we will explore black hole thermodynamics
% in the limit \eqref{conditions:eq}; at the end of this subsection we
% will present the fluid dynamic interpretation of this limit.

Multiplying all thermodynamic integrals in \cite{Chong:2005hr}
by $\frac{R_\mathrm{AdS}^3}{G_5} = \frac{2N^2}{\pi}$ and noting that
our charge $R$ is equal to their $Q/ \sqrt{3}$, the black hole
thermodynamic formulae reduce to (to leading order in $r_+$)
\begin{equation}\label{largechargetherm:eq}
\begin{gathered}
\begin{aligned}
 \Omega_1 &= a\,,\\
 \Omega_2 &= b\,,\\
%  T &= \frac{r_+}{2 \pi} (2 - y^2)\,,  &
 \zeta_i
%&=r_+y
    &=\frac{2 \pi y T}{(2-y^2)}\,,\\
%  S
% %   &={N^2 \pi r_+^{3}} \frac{1}{ (1-a^2)(1-b^2)}&
%    &=\frac{4 \pi^2  N^2 } {(2-y^2)^3}
%      \brk{\frac{V_4 T^{3}}{(1-a^2)(1-b^2)}}\\
%  L_1
% %    &=\frac{N^2 r_+^4 (1+y^2)}{2}  \frac{a}{(1-a^2)^2(1-b^2)}&
%    &=\frac{2 \pi^2 N^2  (1+y^2)}{(2-y^2)^4}
%      \brk{\frac{2 a V_4 T^4}{(1-a^2)^2(1-b^2)}}\\
%  L_2
% %   &=\frac{N^2 r_+^4 (1+y^2)}{2} \frac{b}{(1-a^2)(1-b^2)^2}&
%    &=\frac{2 \pi^2 N^2   (1+y^2)}{(2-y^2)^4}
%      \brk{\frac{2 b V_4 T^4}{(1-a^2)(1-b^2)^2}} \\
%  R
% %   &=\frac{ N^2 r_+^3 y }{ 2 } \frac{1}{(1-a^2)(1-b^2)}&
%    &=\frac{2 \pi N^2  y}{(2-y^2)^3}
%      \brk{\frac{V_4 T^3}{(1-a^2)(1-b^2)}}\\
%  E
% %   &= \frac{ N^2 r_+^4 (1+y^2) }{ 4 }
% %      \frac{ [4-(1+a^2)(1+b^2)] }{ (1-a^2)^2 (1-b^2)^2 }&
%     &=\frac{ 2 \pi^2  N^2  (1+y^2) }{ (2-y^2)^4 }
%        \brk{\frac{ V_4 T^4 }{ (1-a^2)(1-b^2) }}
%        \brk{\frac{2a^2}{1-a^2}+\frac{2b^2}{1-b^2}+3}\\
 \end{aligned}\\
 \ln\gpf
% = \frac{\pi N^2 r_+^3 (1+y^2)}{2}
%       \frac{1}{(1-a^2) (1-b^2)}
    =\frac{ 2 \pi^2  N^2  (1+y^2) }{ (2-y^2)^4 }
     \brk{\frac{ V_4 T^3 }{ (1-\Omega_1^2)(1-\Omega_2^2) }}.
\end{gathered}
\end{equation}
Once we identify the black hole parameter $y^2$ with the fluid parameter $\kappa$, these formula take precisely the form of fluid mechanics formulae \eqref{eqchtherm:eq} with the equation of state coming from \eqref{eqchcoeff:eq}.\footnote{ The functions $h(\nu)$ and its derivatives have simple expressions as functions of bulk parameters. Comparing with \eqref{hval:eq} we find
\begin{equation}\label{h3ch:eq}
\begin{split}
h(\nu)&= \frac{2N^2}{\pi}\brk{\frac{2m}{16\pi T^4}}= \frac{m}{8\pi G_5 T^4} \\
h_i(\nu) &=\frac{2N^2}{\pi} \frac{q}{8\pi T^3} = \frac{q}{8\pi G_5 T^3}
\end{split}
\end{equation}
}

We can now compute the fluid mechanical mean free path $l_\mathrm{mfp}$ as a function of bulk black hole parameters. From equations \eqref{rpmtc:eq} and \eqref{lmfpthreech:eq}, we find (assuming that $r_+$ is large)
\begin{equation*}
l_\mathrm{mfp} \sim \frac{1}{3r_+(1+\kappa)}\,.
\end{equation*}
As $1+\kappa=1+y^2$ is bounded between 1 and 2, it appears from this equation that the expansion in powers of $1/r_+$ is simply identical to the fluid dynamical expansion in powers of $l_\mathrm{mfp}$. This explains why black hole thermodynamics agrees with the predictions of the Navier-Stokes equations when (and only when) $r_+$ is large.

% In the case that
% $\kappa_1=\kappa_2=\kappa_3=\kappa$ which we focus on for simplicity, it
% is easily verified using   \eqref{bbeos:eq} that the fluid dynamical
% approximation is valid provided
% %
% \begin{equation}\label{validity:eq}
%  \frac{\tloc}{(2 - \kappa)}  \gg 1
% \end{equation}
%
% Recall that the agreement between black hole and fluid dynamics holds
% if and only if \eqref{conditions:eq} applies; it is now easy to see why
% this had to be the case. Once we identify $x^2$ with the fluid
% parameter $\kappa$, \eqref{conditions:eq} is precisely
% \eqref{validity:eq}, a condition for the validity
% of fluid mechanics.\footnote{As discussed in
% \S\S\ref{sec:validity}, the factor of $\gamma$ relating $\tloc$
% and $T$ is unimportant.} As a consequence, an AdS black hole may validly
% be described in fluid mechanics if and only if horizon radius of the black
% hole (defined as the cube root of the area of the horizon) is large
% in units of the AdS radius.

\subsubsection{Stress tensor and charge currents}

In appendix \ref{sec:bhstr}, we have computed the boundary
stress tensor corresponding to this black hole solution (by foliating the
space into $S^3$ 's at infinity, computing the extrinsic curvature of these
sections, and subtracting the appropriate counterterms).
At leading order in $\frac{1}{r_+}$
\begin{equation}\label{stenmain:eq}
\begin{split}
 \Pi^{tt} &= \frac{m}{8\pi G_5}\gamma^4(4\gamma^2-1) \\
 \Pi^{\phi\phi} &= \frac{m}{8\pi G_5}\gamma^4\left(4\gamma^2a^2 +
                              \frac{1}{\sin^2\theta}\right)\\
\Pi^{\psi\psi} &= \frac{m}{8\pi G_5}\gamma^4\left(4\gamma^2a^2 +
                               \frac{1}{\cos^2\theta}\right)\\
 \Pi^{t\phi} &= \Pi^{\phi t} = \frac{4m}{8\pi G_5}a\gamma^6 \\
\Pi^{t\psi} &= \Pi^{\psi t} = \frac{4m}{8\pi G_5}b\gamma^6 \\
\Pi^{\phi\psi} &= \Pi^{\psi \phi} = \frac{4m}{8\pi G_5}ab\gamma^6 \\
 \Pi^{\theta\theta} &= \frac{m}{8\pi G_5}\gamma^4.
\end{split}
\end{equation}
In a similar fashion, the charge currents on $S^3$ may be computed from
$J_i^{\mu} = - r^4 g^{\mu\nu}A_{\nu}|_{r\rightarrow\infty}$ where
the indices $\mu, \nu$ are tangent to the $S^3\times $ time foliations and
and the bulk gauge field $A_{\nu}$ is given in the equation (2) of
\cite{Chong:2005hr}. We find
\begin{equation} \label{chargecurtc:eq}
\begin{split}
J_1^t &= J_2^t = J_3^t = \frac{q}{8\pi G_5}\gamma^4\\
J_1^{\theta} &= J_2^{\theta} = J_3^{\theta} = 0\\
J_1^{\phi} &= J_2^{\phi} = J_3^{\phi} = \frac{q}{8\pi G_5}\gamma^4 a\\
J_1^{\psi} &= J_2^{\psi} = J_3^{\psi} = \frac{q}{8\pi G_5}\gamma^4 b\,.\\
\end{split}
\end{equation}
Using \eqref{h3ch:eq}, it is evident that the expressions in
\eqref{chargecurtc:eq} are in precise agreement with the predictions
\eqref{otcurrents:eq} of fluid dynamics.

\subsection{Black holes with independent SO(6) charges and two equal rotations}\label{sec:bheqrot}

The most general (five parameter) black hole solutions with the two angular velocities set equal can be found in \cite{Cvetic:2004ny}. The thermodynamics of these black holes was computed in \cite{Cvetic:2005zi}.

The black hole solutions depend on the parameters $\delta_1, \delta_2,
\delta_3, a, m, r_+$ that are related by the equation $Y(r)=0$.
The thermodynamics of these black holes simplify in the limit
\begin{equation*}
 r_+ \gg 1\,,\qquad \frac{2ms_i^2}{r_+^2} = H_i -1\quad \text{fixed.}
\end{equation*}
Then solving the equation $Y = 0$ in this limit, one can express $m$ as
\begin{equation*}
 2m = \frac{(H_1H_2H_3)r_+^4}{(1-a^2)}\,.
\end{equation*}
%
%where $r_+$ is related to the temperature as
%%
%\begin{equation*}
% r_+ = \frac{2\pi T}{\sqrt{1-a^2}}\left(\frac{\sqrt{H_1H_2H_3}}{H_1H_2 + H_2H_3 +H_3H_1-H_1H_2H_3}\right)
%\end{equation*}
%
The various thermodynamic quantities in this limit\footnote{We believe that \cite{Cvetic:2005zi} has a typo: (3.10) should read $\Phi_i=\frac{2m}{r^2H_i}(s_ic_i+\half a\Omega (c_is_js_k - s_ic_jc_k))$. Note that they also use coordinates $\psi=\phi_1+\phi_2$ and $\varphi=\phi_1-\phi_2$ so that $\Omega\p_\psi = \frac{\Omega}{2}\p_{\phi_1} + \frac{\Omega}{2}\p_{\phi_2}$ so that $\Omega_a=\frac{\Omega}{2}$.} (after multiplying integrals by $\frac{R_\mathrm{AdS}^3}{G_5} = \frac{2N^2}{\pi}$) can be summarised by
%%
%\begin{equation}\label{threedif}
%\begin{split}
%S &= 8N^2\pi^4T^3\left(\frac{(H_1H_2H_3)^2}{(H_1H_2 + H_2H_3 +H_3H_1-H_1H_2H_3)^3}\right)\frac{1}{(1-a^2)^2}\\
%E &= 4N^2\pi^4T^4\left(\frac{(H_1H_2H_3)^3}{(H_1H_2 + H_2H_3 +H_3H_1-H_1H_2H_3)^4}\right)\frac{3+a^2}{(1-a^2)^3}\\
%J &= 8N^2\pi^4t^4\left(\frac{(H_1H_2H_3)^3}{(H_1H_2 + H_2H_3 +H_3H_1-H_1H_2H_3)^4}\right)\frac{a}{(1-a^2)^3}\\
%Q_i &= 4N^2\pi^3T^3\left(\frac{(H_1H_2H_3)^2\sqrt{H_i-1}}{(H_1H_2 + H_2H_3 +H_3H_1-H_1H_2H_3)^3}\right)\frac{1}{(1-a^2)^2}\\
%\nu_i &= 2\pi \left(\frac{H_1H_2H_3}{(H_1H_2 + H_2H_3 +H_3H_1-H_1H_2H_3)}\right)\left(\frac{\sqrt{H_i -1}}{H_i}\right)\frac{1+a^2/2}{1-a^2}(???)\\
%\end{split}
%\end{equation}
%%
%
\begin{equation}\label{bheqrottherm:eq}
\begin{gathered}
  \Omega_1=\Omega_2 =a\,, \qquad
  T = \frac{r_+\sqrt{1-a^2}}{2\pi}
     \left(\sum_jH_j^{-1}-1\right)\prod_j\sqrt{H_j} \,,\\
  \zeta_i = r_+\sqrt{1-a^2} \prn{\frac{\sqrt{H_i-1}}{H_i}} \prod_j\sqrt{H_j}
  = \frac{2\pi T}{\sum_jH_j^{-1}-1} \prn{\frac{\sqrt{H_i-1}}{H_i}}  \,,\\
  \ln\gpf = \frac{\pi N^2 r_+^3}{2\sqrt{1-a^2}}
                           \prn{\frac{\prod_j\sqrt{H_j}}{\sum_jH_j^{-1}-1}}
    = \frac{4\pi^4 N^2 T^3}
           {(1-\Omega^2)^2\prn{\prod_jH_j}\prn{\sum_jH_j^{-1}-1}^4}
\end{gathered}
\end{equation}
These expressions match with \eqref{eqrottherm:eq} if $\kappa_i$ is identified with $H_i -1$, demonstrating perfect agreement between black hole and fluid dynamical thermodynamics.

Translating the estimate for the mean free path into the black hole variables, we find
\begin{equation*}
l_\mathrm{mfp} \sim \frac{1}{3r_+\prod_j\sqrt{H_j}} \ll 1\,,
\end{equation*}
(an equation that is valid only in the large $r_+$ limit). Notice that $l_\mathrm{mfp}$ is automatically small in the large $r_+$ limit, explaining
why black hole thermodynamics agrees with the predictions of the Navier-Stokes equations in this limit.

Notice that the fluid mechanical expansion parameter $l_\mathrm{mfp}$ appears to differ from the expansion parameter of black hole thermodynamics used above, $1/r_+$, by a factor of $1/\sqrt{ \prod_i H_i}$. When the three charges of the black hole are in any fixed ratio $a:b:c$, with none of $a, b$ or $c$ either zero or infinity, it may easily be verified that this additional factor is bounded between a nonzero number (which depends on $a, b, c$) and unity. In this case the two expansion parameters - $l_\mathrm{mfp}$ and $1/r_+$ - are essentially the same.

However when one of the black hole charges (say $R_1$) vanishes $H_2$ and/or $H_3$ can formally take arbitrarily large values. In this extreme limit
$l_\mathrm{mfp}$ appears to differ significantly from the bulk expansion parameter $1/r_+$. However large $H_i$ implies large $\kappa_i$, a limit that we have argued above to be thermodynamically singular. Keeping away from the suspicious large $\kappa_i$ limit, it is always true that $l_\mathrm{mfp}$ is essentially identical $1/r_+$, the parameter in which we have expanded the formulas of black hole thermodynamics.

Finally we emphasise that the black hole studied in this subsection include a large class of perfectly nonsingular zero temperature or extremal black holes
with finite $\kappa_i$ and large $r_+$ which perfectly reproduce the predictions of extremal fluid mechanics of \S\S\ref{sec:exttherm}.

In more detail, the thermodynamical quantities of a general solution in this subsection is given in terms of $X,Y,Z$ (defined as in \eqref{xyzkappa:eq}) as
\begin{equation}\label{3chtherm:eq}
\begin{split}
S &=  \frac{N^2\pi r_+^3}{\sqrt{XYZ(1-a^2)}}\,,\\
E &=  \frac{N^2 r_+^4(3+a^2)}{4XYZ(1-a^2)}\,,\\
L &= \frac{N^2 r_+^4 a}{2XYZ(1-a^2)}\,,\\
R_i &= \frac{N^2 r_+^3}{2 \sqrt{XYZ(1-a^2)}} \sqrt{\frac{1-X_i}{X_i}}\,,\\
\zeta_i &=  r_+ \sqrt{\frac{X_i(1-X_i)(1-a^2)}{XYZ}}\,.
\end{split}
\end{equation}
From these expressions, together with the formula for temperature in
\eqref{bheqrottherm:eq} it follows that the limit $X+Y+Z \to 1$ (with none of $X, Y, Z$ zero) is extremal (the temperature goes to zero) and non-singular
(all thermodynamic quantities are finite and well defined). Note that $r_+$ is an arbitrary parameter for these extremal black holes. When $r_+$ is large
the fluid dynamical description is valid. The black holes so obtained are exactly dual to the extremal fluid configurations described in \S\S\ref{sec:exttherm}.

\subsection{Black holes with two equal large R-charges and third R-charge small}\label{sec:twoeqonedep}

Chong et al.\ \cite{Chong:2005da} have determined a class of black hole solutions
with two SO(6) charges held equal, while the third charge is
varied as a function of these two equal charges. In the large radius limit,
it turns out that this third charge is negligible compared to the first two,
so for our purposes these solutions can be thought of as black holes with
two equal SO(6) charges, with arbitrary rotations and the third SO(6)
charge set to zero. The parameters of this black hole solution are
$a, b, m, r_+, s$, which are related by the equation $X(r_+)=0$.

Black hole formulae simplify in the limit
\begin{equation*}
 r_+ \gg 1 \quad \text{and}\quad k= \frac{2ms^2}{r_+^2} \quad \text{fixed,}
\end{equation*}
in units where the inverse AdS radius $g=1$, which leads to
\begin{equation*}
  2m = r_+^4 (1+k)^2\,.
\end{equation*}

Multiplying all thermodynamic integrals in \cite{Chong:2005da}
by $\frac{R_\mathrm{AdS}^3}{G_5} = \frac{2N^2}{\pi}$,
in this limit, the thermodynamics can be summarised by
\begin{equation}\label{bhtwochtherm:eq}
\begin{gathered}
\begin{aligned}
  \Omega_1 &= a
  \,,\quad
  \Omega_2 = b
  \,,&
  T &= \frac{r_+}{\pi}
  \,,\\
  \zeta_1=\zeta_2 &= \pi T \sqrt{k}
  \,,&
  \zeta_3 &\sim {\cal{O}}\left(\frac{1}{r_+^2}\right)
  \,,
\end{aligned}\\
%\begin{split}
%  S &= (2\pi^2\ T^3) \frac{\pi^2 N^2 (1-\sqrt{1-4y^2}) }
%            { 4y^2(1-a^2)(1-b^2)}
% \,,\\
%  L_1 &= 2(2\pi^2\  T^4)\frac{\pi^2 N^2 (1-\sqrt{1-4y^2})^2 }
%              { 16 y^4(1-a^2)(1-b^2) }
%         \frac{ a }{ 1-a^2 }
%  \,,\\
%  L_2 &= 2(2\pi^2\ T^4) \frac{\pi^2 N^2 (1-\sqrt{1-4y^2})^2 }
%              { 16 y^4(1-a^2)(1-b^2) }
%         \frac{ b }{ 1-b^2 }
%  \,,\\
%  R_1 = R_2 &=(2\pi^2\ T^3) \frac{ \pi N^2 (1-\sqrt{1-4y^2})^2 }
%                    { 16 y^3(1-a^2)(1-b^2) }
%  \,,\\
%  R_3 &= -\frac{ r_+^2 N^2 ab (1-\sqrt{1-4y^2})^2 }
%               { 4 y^2  (1-a^2)(1-b^2) }
%  \,,\\
%  E &= (2\pi^2\ T^4)\frac{ \pi^2 N^2 (1-\sqrt{1-4y^2})^2 }
%            { 16 y^4(1-a^2)(1-b^2) }
%         \brk{\frac{ 2 }{ 1-a^2 }+\frac{ 2 }{ 1-b^2 }-1}
%  \,,
%\end{split}\\
  \ln\gpf = \frac{  \pi^2 N^2 V_4 T^3  (1+k)^2}
                       { 8 (1-a^2)(1-b^2) }\,.
\end{gathered}
\end{equation}
Note that $\zeta_3$ and $R_3$ are subleading in $r_+$.
These formulae are in perfect agreement with \eqref{twochtherm:eq}
if we identify
\begin{equation*}
\kappa= k\,.
%\leq 1
\end{equation*}
From the expression for the temperature, it follows that all extremal or zero temperature black holes have $r_+=0$. Consequently all extremal black holes (of the class of black holes described in this subsection) are singular, dual to the fact that the fluid mechanics has no thermodynamically nonsingular zero temperature solutions.

Translating the estimate for the fluid dynamical mean free path into the black hole variables we find (assuming $r_+ \gg 0$)
\begin{equation*}
l_\mathrm{mfp} \sim \frac{1}{3r_+(1+\kappa)}\,.
\end{equation*}
It follows that the fluid dynamical expansion parameter is essentially the same as $1/r_+$, provided we stay away from the thermodynamically suspect parameter regime of large $\kappa$ .

%Since $\kappa$ is bounded, it is evident that the large $r_+$ expansion in black hole physics is exactly equivalent %to a small $l_\mathrm{mfp}$ expansion in fluid mechanics.

\subsection{Black holes with two R-charges zero}\label{sec:onechads5}

The solution for the most general black hole with two R-charges set to zero relevant solution has was presented in \cite{Chong:2006zx}. The parameters of this black hole are $x_0, m, \delta, a, b$ related by $X(x_0)=0$.

The thermodynamics of these black holes simplifies in the limit
\begin{equation*}
  x_0 \gg 1 \,,\qquad y= \sqrt{x_0} \delta \quad \text{fixed,}
\end{equation*}
in units where $g=1$, which leads to
\begin{equation*}
  2m = \frac{x_0^2}{(1-y^2)}\,.
\end{equation*}
This gives an upper bound on $y$: $y\leq 1$.

Multiplying all thermodynamic integrals in \cite{Chong:2006zx} by $\frac{R_\mathrm{AdS}^3}{G_5} = \frac{2N^2}{\pi}$, in this limit, the thermodynamic formulae can be summarised by
\begin{equation}\label{bhonechtherm:eq}
\begin{gathered}
\begin{aligned}
  \Omega_1 &= a
  \,,&
  \Omega_2 &= b
  \,,\\
  T &= \frac{\sqrt{x_0}(2-y^2)}{2\pi\sqrt{1-y^2}}
  \,,&
  \zeta &= \sqrt{x_0}y
     =\frac{2\pi T y\sqrt{1-y^2}}{2-y^2}
  \,,
\end{aligned}\\
%\begin{split}
%  S &= \frac{ x_0^{3/2} \pi N^2 }
%            { \sqrt{1-y^2}(1-a^2)(1-b^2) }
%     =\frac{8\pi^4 N^2 T^3 (1-y^2)}{(2-y^2)^3(1-a^2)(1-b^2)}
%  \,,\\
%  L_1 &= \frac{ x_0^2 N^2 }
%              { 2 (1-y^2) (1-a^2)(1-b^2) }
%         \brk{\frac{ a }{ 1-a^2 }}
%     =\frac{8\pi^4 N^2 T^4 (1-y^2)}{(2-y^2)^4(1-a^2)(1-b^2)}
%      \brk{\frac{ a }{ 1-a^2 }}
%  \,,\\
%  L_2 &= \frac{ x_0^2 N^2 }
%              { 2 (1-y^2) (1-a^2)(1-b^2) }
%         \brk{\frac{ b }{ 1-b^2 }}
%     =\frac{8\pi^4 N^2 T^4 (1-y^2)}{(2-y^2)^4(1-a^2)(1-b^2)}
%      \brk{\frac{ b }{ 1-b^2 }}
%  \,,\\
%  R &= \frac{ x_0^{3/2} N^2 y }
%            { 2 (1-y^2) (1-a^2)(1-b^2) }
%     =\frac{4\pi^3 N^2 T^3 \sqrt{1-y^2}}{(2-y^2)^3(1-a^2)(1-b^2)}
%  \,,\\
%  E &= \frac{ x_0^2 N^2 }
%            { 4 (1-y^2) (1-a^2)(1-b^2) }
%         \brk{\frac{ 2 }{ 1-a^2 }+\frac{ 2 }{ 1-b^2 }-1}
%     =\frac{4\pi^4 N^2 T^4 (1-y^2)}{(2-y^2)^4(1-a^2)(1-b^2)}
%  \,,
%\end{split}\\
  \ln\gpf = \frac{ x_0^{3/2} \pi N^2 }
            { 2\sqrt{1-y^2}(2-y^2)(1-a^2)(1-b^2) }
   = \frac{4\pi^4 N^2 T^3 (1-y^2)}{(1-\Omega_1^2)(1-\Omega_2^2)(2-y^2)^4}
\end{gathered}
\end{equation}
Upon identifying $\kappa=\frac{y^2}{1-y^2}$, we find perfect agreement with \eqref{onechtherm:eq}. Under this identification, the expression for temperature becomes
\begin{equation*}
 T = \frac{\sqrt{x_0}(2+\kappa)}{2\pi\sqrt{1+\kappa}}\,.
\end{equation*}
As in the subsection above, it follows immediately from this equation that
the black hole temperature vanishes only for the singular black holes with $x_0=0$. This matches with the fact that there are no nonsingular extremal fluid dynamical solutions in this case.

The fluid dynamical mean free path may be evaluated as a function of bulk parameters as
\begin{equation*}
 l_\mathrm{mfp} \sim \frac{1}{3\sqrt{x_0(1+\kappa)}}\,.
\end{equation*}

Note that $l_\mathrm{mfp}$ is small whenever $\sqrt{x_0}=r_+$ is large, an observation that explains the agreement of black hole thermodynamics in the large $r_+$ limit with the Navier-Stokes equations. In more generality we see that $l_\mathrm{mfp}$ is essentially the same as $1/r_+$, provided we keep away from the thermodynamically suspicious parameter regime of $\kappa$ large.

\subsection{Extremality and the attractor mechanism}

As discussed in the previous subsections, there exists a duality between extremal large rotating AdS black holes on one hand and the extremal configurations of the fluid dynamics on the other. This implies that the thermodynamic properties of these large rotating extremal black holes are completely determined by the corresponding properties of large static extremal black holes. As an application of this observation, let us recall the suggestion \cite{Sen:2005wa, Goldstein:2005hq, Astefanesei:2006dd} that the attractor mechanism for black holes implies the non-renormalisation of the entropy of all extremal configurations, as a function of the 't~Hooft coupling $\lambda$. It follows immediately from the fluid mechanical description at large charges, that were any such non-renormalisation theorem be proved for static extremal configurations, it would immediately imply a similar result for rotating extremal configurations.

\subsection{BPS bound and supersymmetric black holes}

All solutions of IIB supergravity on $AdS_5 \times S^5$, and all
configurations of $\mathcal{N}=4$ Yang-Mills on $S^3$ obey the BPS bound
\begin{equation}\label{bpsbound:eq}
E \geq L_1+L_2 + \sum_iR_i= L_1+L_2+3 R\,.
\end{equation}
Within the validity of the fluid dynamical approximation, described in
this paper,
\begin{equation}\label{wc:eq}
E-L_1-L_2=  2 \pi^2 T^4  A \frac{3 + \omega_1 + \omega_2 - \omega_1 \omega_2}
{(1+\omega_1) (1+\omega_2)} \,;
\end{equation}
notice that the RHS of this equation is positive definite. The BPS
bound is obeyed provided
\begin{equation}\label{bpsmass:eq}
 \tau  A \frac{3 + \omega_1 + \omega_2 - \omega_1 \omega_2}
{(1+\omega_1) (1+\omega_2)}
\geq  C_i\,.
\end{equation}
Plugging in the explicit expressions for $A$ and $C_i$ from
\eqref{eqchcoeff:eq}, we find
this condition is satisfied provided
\begin{equation}\label{bpsmassn:eq}
r_+= \frac{ 2 \pi  T}{2-\kappa}
\geq \frac{\sqrt{\kappa} (1+\omega_1) (1+\omega_2)} {(1+\kappa)
(3 + \omega_1 + \omega_2 - \omega_1 \omega_2)}\,.
\end{equation}
The RHS of \eqref{bpsmassn:eq} is of order unity. It follows that \eqref{bpsmassn:eq} is saturated only when $r_+$ of unit order. It follows that when $r_+ \gg 1$ (so that fluid dynamics is a valid approximation) the BPS bound is always obeyed as a strict inequality. Supersymmetric black holes are never reliably described within fluid mechanics.\footnote{Although it is possible to make the energy of supersymmetric black holes parametrically larger than their entropy, this is achieved by scaling either $\omega_1$ or $\omega_2$ to unity with $r_+$ kept at unit order. It is easy to verify that in this limit the local, rest frame mean free path of the fluid is of unit order in regions of the $S^3$ and so fluid mechanics may not be used to describe these configurations.} The extremal black holes with large horizon radius, that are well described by fluid mechanics\footnote{Note that the `physical' radius $\text{(Area)}^{1/3}$ of the black hole is distinct from the parameter $r_+$ which determines the validity of fluid dynamics. The physical radius can be made arbitrarily large, nevertheless fluid mechanics is only valid if $r_+$ is large.} (see the previous subsection) are always far from supersymmetry.

We have noted above that a large class of extremal configurations in strongly interacting Yang-Mills -- all those that admit a fluid dynamic description -- are not BPS. This is in sharp contrast with the results of computations in free Yang-Mills theory, in which all extremal configurations are supersymmetric \cite{Kinney:2005ej}. This difference is related to the fact, noted previously, the divergent mean free path prevents a fluid mechanical description from applying to free theories. A practical manifestation of this fact is that the function $h(\nu)$, which appears in the analysis of free Yang-Mills in equation (5.2) of \cite{Kinney:2005ej}, and plays the role of $r_+$ in our discussion here, is always of order unity for all allowed values of the chemical potential, and so can never become large.

%that Bose condensation prevents fluid dynamical description from applying charged free theories

\subsection{Fluid dynamics versus black hole physics at next to leading order}\label{sec:nlo}

As we have explained above, the formulae for all thermodynamic charges
and potentials of black holes of temperature $T$ and chemical potentials
$\nu_i$, in $AdS_5 \times S^5$, may be expanded as a Taylor series
in $1/r_+ \sim l_\mathrm{mfp}(T, \nu_i)$. As we have verified
above, for every known family of large AdS black holes, the leading
order results in this expansion perfectly match the predictions of the
Navier-Stokes equations. Higher order terms in this expansion represent
corrections to Navier-Stokes equations. In this subsection we
investigate the structure of these corrections.

Let us first investigate the case of black holes with at least one SO(6) charge set equal to zero (the black holes studied in \S\S\ref{sec:twoeqonedep} and \S\S\ref{sec:onechads5}). It is not difficult to verify that the first deviations from the large radius thermodynamics of these black holes occur at $\mathcal{O}(1/r_+^2) \sim l_\mathrm{mfp}^2$. This result is in perfect accord with naive expectations from fluid mechanics. As we have explained above, the fluid dynamical configurations presented in this paper are exact solutions to the equations of fluid mechanics with all one derivative terms, i.e.\ to the first order in $l_\mathrm{mfp}$. In general we would expect our solutions (and their thermodynamics) to be modified at $\mathcal{O}( l_\mathrm{mfp}^2)$, exactly as we find from the black hole formulae.

However when we turn our attention to black holes with all three SO(6)
charges nonzero we run into a bit of a surprise. It appears that the
thermodynamics (and stress tensor and charge currents) of these black holes
receives corrections at order $\mathcal{O}(1/r_+) \sim l_\mathrm{mfp}$.
This result is a surprise because, for the reason we have explained in the
previous paragraph, we would have expected the first corrections to
our fluid mechanical configuration to occur at $\mathcal{O}({l_\mathrm{mfp}^2})$.

We do not have a satisfactory resolution to this puzzle. In this subsection
we will simply present the expressions for the first order corrections
to black hole thermodynamics in a particular case (the case of black holes
with all SO(6) charges equal), and leave the explanation of these
formulae to future work.

As we have mentioned above, the thermodynamics of a charged rotating
black hole in $AdS_5\times S^5$ with three equal charges and two
different angular momenta can be found in \cite{Chong:2005hr}. To calculate
next to leading order (NLO) corrections to the thermodynamics of large
black holes, we systematically expand the thermodynamic quantities.

We find it convenient to shift to a new parametrisation
in which there are no NLO corrections to the intensive quantities.
This allows us to cast the NLO corrections entirely in terms of the
intensive quantities. The parameters we choose are related to the
parameters in \cite{Chong:2005hr} in the following way
\begin{equation}\label{nloparam:eq}
\begin{split}
 a &=\omega_a -\frac{\sqrt{\kappa}(1-\omega_a^2)\omega_b}{\ell}\,, \\
 b &=\omega_b -\frac{\sqrt{\kappa}(1-\omega_b^2)\omega_a}{\ell}\,, \\
 r_+ &= \ell\ + \sqrt{\kappa}\omega_a\omega_b\,, \\
 q &= \sqrt{\kappa}\ell^3 + 3 \kappa \ell^2\omega_a\omega_b\,.\\
\end{split}
\end{equation}

In terms of these parameters, the intensive quantities can be written as
\begin{equation}
\begin{split}
 \Omega_a &= \omega_a + \CO\!\brk{\frac{1}{\ell^2}} ,\\
 \Omega_b &= \omega_b + \CO\!\brk{\frac{1}{\ell^2}} ,\\
 T &= \brk{\frac{2-\kappa}{2\pi}} \ell + O\brk{\frac{1}{\ell}} ,\\
 \nu &= \frac{2\pi \sqrt{\kappa}}{2-\kappa}+ \CO\!\brk{\frac{1}{\ell^2}} ,\\
\end{split}
\end{equation}
where we have calculated up to NLO and confirmed that the intensive
quantities do not get corrected in this order.

This in turn means that the new parameters can be directly interpreted in
terms of the intensive quantities.
\begin{equation*}
 \omega_a = \Omega_a + \CO\!\brk{l_\mathrm{mfp}^2}, \qquad
 \omega_b = \Omega_b+ \CO\!\brk{l_\mathrm{mfp}^2},
\end{equation*}

where $l_\mathrm{mfp} \sim \frac{ 2- \kappa}{T}$.
\begin{equation*}
 \ell= T\brk{\frac{\sqrt{\pi^2+2\nu^2}+\pi}{2}}+\CO\!\brk{\frac{1}{T^2}}, \qquad
 \sqrt{\kappa}= \frac{\sqrt{\pi^2+2\nu^2}-\pi}{\nu}+ \CO\!\brk{\frac{1}{T^2}}.
\end{equation*}

Now, we calculate NLO corrections to the extensive quantities in terms
of the new parameters.
\begin{equation}
\begin{split}
2m &= (1+\kappa) \ell^4 +4\sqrt{\kappa}(1+\kappa)\omega_a\omega_b \ell^3 + \CO[\ell^2]\,, \\
S &= \frac{T^3}{G_5(1-\omega_a^2)(1-\omega_b^2)}\brk{\frac{4\pi^5}{(2-\kappa)^3}+\CO\!\brk{\frac{1}{T^2}}} ,\\
L_a &= \frac{T^4}{G_5(1-\omega_a^2)(1-\omega_b^2)}\brk{\frac{2\omega_a}{1-\omega_a^2}\brk{\frac{2\pi^5(1+\kappa)}{(2-\kappa)^4}}-\frac{\pi\nu^3\omega_b}{4T}\brk{\frac{1+\omega_a^2}{1-\omega_a^2}}+\CO\!\brk{\frac{1}{T^2}}} ,\\
L_b &= \frac{T^4}{G_5(1-\omega_a^2)(1-\omega_b^2)}\brk{\frac{2\omega_b}{1-\omega_b^2}\brk{\frac{2\pi^5(1+\kappa)}{(2-\kappa)^4}}-\frac{\pi\nu^3\omega_a}{4T}\brk{\frac{1+\omega_b^2}{1-\omega_b^2}}+\CO\!\brk{\frac{1}{T^2}}} ,\\
R &= \frac{T^3}{G_5(1-\omega_a^2)(1-\omega_b^2)}\brk{\frac{2\pi^4 \sqrt{\kappa}}{(2-\kappa)^3}-\frac{\pi\nu^2}{4T}\omega_a\omega_b+\CO\!\brk{\frac{1}{T^2}}},\\
E &= \frac{T^4} {G_5(1-\omega_a^2)(1-\omega_b^2)} \left[\frac{2\pi^5(1+\kappa)}{(2-\kappa)^4} \brk{\frac{2}{1-\omega_a^2}+\frac{2}{1-\omega_b^2}-1}\right.\\
&\left.\qquad - \frac{\pi\nu^3\omega_a\omega_b}{4T} \brk{\frac{2}{1-\omega_a^2}+\frac{2}{1-\omega_b^2}} + \CO\!\brk{\frac{1}{T^2}}\right],
\end{split}
\end{equation}
where $G_5 = \pi R^3_\mathrm{AdS}/(2N^2)$ is the Newton's constant in AdS$_5$.

In particular, the subleading terms can be isolated and written as
\begin{equation}
\begin{split}
 \Delta S &= 0\,, \\
 \Delta E &= -\frac{\pi\zeta^3\omega_a\omega_b}{4G_5(1-\omega_a^2)
(1-\omega_b^2)} \brk{\frac{2}{1-\omega_a^2}+\frac{2}{1-\omega_b^2}} ,\\
 \Delta L_a &= -\frac{\pi\zeta^3\omega_b(1+\omega_a^2)}{4G_5(1-\omega_a^2)^2(1-\omega_b^2)}\,,\\
 \Delta L_b &= -\frac{\pi\zeta^3\omega_a(1+\omega_b^2)}{4G_5(1-\omega_a^2)(1-\omega_b^2)^2}\,,\\
 \Delta R &=  -\frac{\pi\zeta^2\omega_a\omega_b}{4G_5(1-\omega_a^2)(1-\omega_b^2)}\,,\\
 \Delta\ln\gpf &= \frac{\pi\zeta^3\omega_a\omega_b}{4G_5T(1-\omega_a^2)(1-\omega_b^2)}\,.
 \end{split}
\end{equation}

\section{Comparison with black holes in $AdS_4 \times S^7$ and $AdS_7 \times S^4$}\label{sec:m2m5}

In this section we compare solutions of rotating fluids of the M5 or M2 brane
conformal field theory on $S^2$ or $S^5$ to the classical physics of black holes in M theory on $AdS_4 \times S^7$ and $AdS_7 \times S^5$ respectively.
Our results turn out to be qualitatively similar to those of the previous section with
one difference: the puzzle regarding the next to leading order agreement between fluid dynamics and black hole physics seems to be absent in this case.

\subsection{Predictions from fluid mechanics}

The equations of state of the strongly coupled M2 and M5 brane fluids
were computed from spinning brane solutions in \cite{Harmark:1999xt}.
Our parameters are related to theirs by $\kappa_i = l_i^2/r_H^2$.

\subsubsection{M2 branes}\label{sec:m2fluid}

We define our R-charges to be half of the angular momenta of \cite{Harmark:1999xt} to agree with gauged supergravity conventions. The equation of state is
\begin{equation}\label{m2eos:eq}
\begin{split}
  h(\nu) &= \frac{4\pi^2(2N)^{3/2} \prod_j(1+\kappa_j)^{5/2}}
     {3(3+2\sum_j\kappa_j +\sum_{j<k}\kappa_j\kappa_k - \prod_j\kappa_j)^3}
     \,, \\
  \nu_i  &= \frac{4\pi \prod_j(1+\kappa_j)}
     {(3+2\sum_j\kappa_j +\sum_{j<k}\kappa_j\kappa_k - \prod_j\kappa_j)}
     \prn{\frac{\sqrt{\kappa_i}}{1+\kappa_i}}
     \,,\\
  h_i(\nu) &= \frac{\pi(2N)^{3/2} \prod_j(1+\kappa_j)^{3/2}}
     {3(3+2\sum_j\kappa_j +\sum_{j<k}\kappa_j\kappa_k - \prod_j\kappa_j)^2}
     \sqrt{\kappa_i}
     \,,
\end{split}
\end{equation}
where $i,j,k=1\ldots4$.

The stress tensor and currents are given by \eqref{otstr:eq} and \eqref{otcurrents:eq} with
\begin{equation}\label{m2cur:eq}
\begin{split}
  A &= \frac{4\pi^2(2N)^{3/2} \prod_j(1+\kappa_j)^{5/2}}
     {3(3+2\sum_j\kappa_j +\sum_{j<k}\kappa_j\kappa_k - \prod_j\kappa_j)^3}
     \,,\\
  B  &= \frac{4\pi^2(2N)^{3/2} \prod_j(1+\kappa_j)^{3/2}}
     {3(3+2\sum_j\kappa_j +\sum_{j<k}\kappa_j\kappa_k - \prod_j\kappa_j)^2}
     \,,\\
  C_i &= \frac{\pi(2N)^{3/2} \prod_j(1+\kappa_j)^{3/2}}
     {3(3+2\sum_j\kappa_j +\sum_{j<k}\kappa_j\kappa_k - \prod_j\kappa_j)^2}
     \sqrt{\kappa_i}
     \,.
\end{split}
\end{equation}

The thermodynamics can be summarised by
\begin{equation}\label{m2therm:eq}
\begin{split}
  \zeta_i &= \frac{4\pi T \prod_j(1+\kappa_j)}
     {(3+2\sum_j\kappa_j +\sum_{j<k}\kappa_j\kappa_k - \prod_j\kappa_j)}
     \prn{\frac{\sqrt{\kappa_i}}{1+\kappa_i}} \,,\\
  \ln\gpf &= \frac{16\pi^3(2N)^{3/2} T^2 \prod_j(1+\kappa_j)^{5/2}}
     {3(1-\Omega^2)(3+2\sum_j\kappa_j +\sum_{j<k}\kappa_j\kappa_k - \prod_j\kappa_j)^3}\,. \\
\end{split}
\end{equation}
The mean free path in fluid dynamics is given by
\begin{equation}\label{m2lmfp:eq}
\begin{split}
l_\mathrm{mfp} \sim \brk{\frac{S}{4\pi E}}_{\Omega=0} &= \frac{B}{(d-1)4\pi TA} =  \frac{\prn{3+2\sum_j\kappa_j +\sum_{j<k}\kappa_j\kappa_k - \prod_j\kappa_j}}{8\pi T\prod_j(1+\kappa_j)}\\
&= \frac{1}{8\pi T} \brk{\sum_j\frac{1}{1+\kappa_j} -1}.
\end{split}
\end{equation}

This simplifies when the charges are pairwise equal, $\kappa_3=\kappa_1$ and $\kappa_4=\kappa_2$. In this case, with $i=1,2$:
\begin{equation}\label{m2pairtherm:eq}
\begin{split}
  \zeta_i &= \frac{4\pi T \prod_j(1+\kappa_j)}
     {(3 + \sum_j\kappa_j - \prod_j\kappa_j)}
     \prn{\frac{\sqrt{\kappa_i}}{1+\kappa_i}} \,,\\
  \ln\gpf &= \frac{16\pi^3(2N)^{3/2} T^2 \prod_j(1+\kappa_j)^2}
     {3(1-\Omega^2)(3 + \sum_j\kappa_j - \prod_j\kappa_j)^3}\,. \\
\end{split}
\end{equation}
and the mean free path becomes
\begin{equation}\label{m2pairlmfp:eq}
\begin{split}
l_\mathrm{mfp} &\sim \frac{\prn{3 + \sum_j\kappa_j - \prod_j\kappa_j}}{8\pi T\prod_j(1+\kappa_j)}
=\frac{1}{8\pi T} \brk{\sum_i\frac{2}{1+\kappa_i} -1}.
\end{split}
\end{equation}

It is evident that the thermodynamic equations of state listed above allow a set of extremal fluid configurations very similar to those discussed in \S\S\ref{sec:exttherm}. The analysis of \S\S\ref{sec:exttherm} can be easily extended to fluids on $S^2$.
%We have not performed an analysis of the stability of this fluid, along the lines of \S\S\ref{sec:5dstab}, but it would be interesting to do so.

\subsubsection{M5 branes}\label{m5fluid:eq}

We define our R-charges to be twice the angular momenta of \cite{Harmark:1999xt} to agree with gauged supergravity conventions. The equation of state is
\begin{equation}\label{m5eos:eq}
\begin{split}
  h(\nu) &= \frac{64 \pi^3 N^3 \prod_j(1+\kappa_j)^4}
     {3(3 + \sum_j\kappa_j - \prod_j\kappa_j)^6}
     \,, \\
  \nu_i  &= \frac{2\pi \prod_j(1+\kappa_j)}
     {(3 + \sum_j\kappa_j - \prod_j\kappa_j)}
     \prn{\frac{\sqrt{\kappa_i}}{1+\kappa_i}}
     \,,\\
  h_i(\nu) &= \frac{128 \pi^2 N^3 \prod_j(1+\kappa_j)^3}
     {3(3 + \sum_j\kappa_j - \prod_j\kappa_j)^5}
     \sqrt{\kappa_i}
     \,,
\end{split}
\end{equation}
where $i=1,2$.

The stress tensor and currents are given by \eqref{otstr:eq} and \eqref{otcurrents:eq} with
\begin{equation}\label{m5cur:eq}
\begin{split}
  A &= \frac{64 \pi^3 N^3 \prod_j(1+\kappa_j)^4}
     {3(3 + \sum_j\kappa_j - \prod_j\kappa_j)^6}
     \,, \\
  B  &= \frac{128 \pi^3 N^3 \prod_j(1+\kappa_j)^3}
     {3(3 + \sum_j\kappa_j - \prod_j\kappa_j)^5} \\
  C_i &= \frac{128 \pi^2 N^3 \prod_j(1+\kappa_j)^3}
     {3(3 + \sum_j\kappa_j - \prod_j\kappa_j)^5}
     \sqrt{\kappa_i}
     \,.
\end{split}
\end{equation}

The thermodynamics can be summarised by
\begin{equation}\label{m5therm:eq}
\begin{split}
  \zeta_i &= \frac{4\pi T \prod_j(1+\kappa_j)}
     {(3+2\sum_j\kappa_j +\sum_{j<k}\kappa_j\kappa_k - \prod_j\kappa_j)}
     \prn{\frac{\sqrt{\kappa_i}}{1+\kappa_i}} \,,\\
  \ln\gpf &= \frac{64\pi^6 N^3 T^5 \prod_j(1+\kappa_j)^4}
     {3\prod_a(1-\Omega_a^2)(3 + \sum_j\kappa_j - \prod_j\kappa_j)^3}\,. \\
\end{split}
\end{equation}
The mean free path in fluid dynamics is given by
\begin{equation}\label{m5lmfp:eq}
\begin{split}
l_\mathrm{mfp} \sim \brk{\frac{S}{4\pi E}}_{\Omega=0} &= \frac{B}{(d-1)4\pi TA} =  \frac{\prn{3 + \sum_j\kappa_j - \prod_j\kappa_j}}{10 \pi T\prod_j(1+\kappa_j)}\\
&= \frac{1}{10 \pi T} \brk{\sum_j\frac{2}{1+\kappa_j} -1}.
\end{split}
\end{equation}

In the case that the three rotation parameters are equal, $\Omega_1=\Omega_2=\Omega_3=\Omega$, we have $\gamma=(1-\Omega^2)^{-1/2}$ and
\begin{equation}\label{m5eqrottherm:eq}
\begin{split}
  \zeta_i &= \frac{4\pi T \prod_j(1+\kappa_j)}
     {(3+2\sum_j\kappa_j +\sum_{j<k}\kappa_j\kappa_k - \prod_j\kappa_j)}
     \prn{\frac{\sqrt{\kappa_i}}{1+\kappa_i}} ,\\
  \ln\gpf &= \frac{64\pi^6 N^3 T^5 \prod_j(1+\kappa_j)^4}
     {3(1-\Omega^2)^3(3 + \sum_j\kappa_j - \prod_j\kappa_j)^3}\,. \\
\end{split}
\end{equation}
It is evident that the thermodynamic equations of state listed above allow a set of extremal fluid configurations very similar to those discussed in \S\S\ref{sec:exttherm}. The analysis of \S\S\ref{sec:exttherm} can be easily extended to fluids on $S^5$.
%We have not performed an analysis of the stability of this fluid, along the lines of \S\S\ref{sec:5dstab}, but it would be interesting to do so.

\subsection{Black holes in AdS$_4$ with pairwise equal charges}\label{sec:sttenmtwo}

The relevant solution was found in \cite{Chong:2004na}. Its thermodynamics have been computed in \cite{Cvetic:2005zi}. We consider the limit of large $r_+$ with $\frac{2m s_i^2}{r_+} = k_i$ fixed. In this limit $m$ can be written as
\begin{equation*}
 m = \frac{r_+^3}{2}(1+k_1)^2(1+k_2)^2,
\end{equation*}
and therefore $s_i \sim \frac{1}{r_+}$.

After multiplying integrals by $\frac{R_\mathrm{AdS}^2}{G_4} = \frac{(2N)^{3/2}}{3}$, the thermodynamic quantities can be expressed as
\begin{equation}\label{mtwotherm:eq}
\begin{gathered}
\begin{aligned}
 T &= \frac{r_+(3 + \sum_j k_j - \prod_j k_j)}{4\pi}\,,
 &
 \Omega &= a\,,
 \\
 \zeta_1 &= \zeta_3 = 4\pi T\frac{(1+k_2)\sqrt{k_1}}
                                 {(3 + \sum_j k_j - \prod_j k_j)}\,,
 &
 \zeta_2 &= \zeta_4 = 4\pi T\frac{(1+k_1)\sqrt{k_2}}
                                 {(3 + \sum_j k_j - \prod_j k_j)}\,,
\end{aligned}
 \\
% S &= \frac{16 \pi^3 (2N)^{3/2} T^2}{3}
%    \left(\frac{\prod_j(1+k_j)}
%               {(3 + \sum_j k_j - \prod_j k_j)^2}\right)
%    \frac{1}{1-a^2}
% \\
% E &= \frac{32 \pi^3 (2N)^{3/2} T^3}{3}
%   \left(\frac{\prod_j(1+k_j)^2}
%              {(3 + \sum_j k_j - \prod_j k_j)^3}\right)
%   \frac{1}{(1-a^2)^2}
% \\
% L &= \frac{32 \pi^3 (2N)^{3/2} T^3}{3}
%      \left(\frac{\prod_j(1+k_j)^2}
%              {(3 + \sum_j k_j - \prod_j k_j)^3}\right)
%      \frac{a}{(1-a^2)^2}
% \\
% R_i &= \frac{4 \pi^2 (2N)^{3/2} T^2}{3}
%      \left(\frac{\prod_j(1+k_j)}
%               {(3 + \sum_j k_j - \prod_j k_j)^2}\right)
%      \frac{1}{(1-a^2)}
% \\
 \ln\gpf = \frac{16 \pi^3 (2N)^{3/2} T^2}{3}
      \left(\frac{\prod_j(1+k_j)^2}
              {(3 + \sum_j k_j - \prod_j k_j)^3}\right)
     \frac{1}{1-a^2}\,.
\end{gathered}
\end{equation}
If one identifies $\kappa_i=k_i$, then these formulae match with \eqref{m2pairtherm:eq}. It is not difficult to verify that the first corrections to the thermodynamical equations above occur at ${\cal O}(1/r_+^2)$.

It is clear from \eqref{mtwotherm:eq} that the black holes of this subsection
admit a zero temperature (extremal) limit with nonsingular thermodynamics at any every value of $r_+$. These extremal black holes are dual to extremal
solutions of fluid dynamics analogous to those described in the previous section in the context of ${\cal N}=4$ Yang-Mills.

The fluid dynamical mean free path may easily be computed as a function of black hole parameters. From \eqref{m2pairlmfp:eq} we find
\begin{equation*}
l_\mathrm{mfp} \sim \frac{1}{2r_+\prod_j(1+\kappa_j)}\,.
\end{equation*}
As in the previous section, the $l_\mathrm{mfp} \sim 1/r_+$ away from
thermodynamically suspect limits of parameters.

\subsection{Black holes in AdS$_7$ with equal rotation parameters}

The relevant solution was found in \cite{Chong:2004dy}. Its thermodynamics have been computed in \cite{Cvetic:2005zi}.\footnote{We believe that \cite{Cvetic:2005zi} has the following typos: equation (4.7) should read
\begin{equation*}
  S = \frac{\pi^3(r^2+a^2)\sqrt{f_1}}{4\Xi^3}
  \quad
  T = \frac{Y'}{4\pi r(r^2+a^2)\sqrt{f_1}}
  \quad
  \Phi_i = \frac{2ms_i}{\rho^4\Xi H_i}\brk{\Xi_-\alpha_i+\beta_i(\Omega-g)}.
\end{equation*}
}

We set the parameter $g$ in \cite{Cvetic:2005zi} to be unity and consider the limit
\begin{equation*}
 \rho_+ \gg 1 \,, \qquad
 %y_i= \rho_+ \delta_i \quad
 \text{and}\quad
  H_i = 1 + \frac{2m s^2_i}{\rho_+^6}
 \quad \text{fixed,}
\end{equation*}
where i=1,2. In this limit, the parameter $m$ is given by
\begin{equation*}
2m = \rho_+^6 H_1 H_2 \,.
\end{equation*}

In this limit, after multiplying integrals by $\frac{R_\mathrm{AdS}^5}{G_7} = \frac{16N^3}{3\pi^2}$, the thermodynamics can be summarised by
\begin{equation}\label{bhm5eqrottherm:eq}
\begin{gathered}
\begin{aligned}
  \Omega &= a
  \,, &
  T &= \frac{\rho_+}{2\pi}\prn{\frac{2\sum_jH_j-\prod_jH_j}
                                    {\prod_j\sqrt{H_j}}}
  ,\\
  \zeta_1 &= 2\pi T \frac{H_2\sqrt{H_1-1}}{2\sum_jH_j-\prod_jH_j}
  \,, &
  \zeta_2 &= 2\pi T \frac{H_1\sqrt{H_2-1}}{2\sum_jH_j-\prod_jH_j}
  \,,
\end{aligned}\\
 \ln\gpf = \frac{64\pi^6 N^3 T^5}{3(1-\Omega^2)^3}
 \prn{\frac{\prod_jH_j^4}{(2\sum_jH_j-\prod_jH_j)^6}}.
\end{gathered}
\end{equation}
These formulae agree with \eqref{m5eqrottherm:eq} upon identifying $\kappa_i=H_i-1$. The first corrections to these thermodynamical formulae
occur at ${\cal O}(1/r_+^2)$. Using this identification we can rewrite the expression for  the temperature as
\begin{equation*}
  T = \frac{\rho_+\prod_j\sqrt{1+\kappa_j}}{2\pi}\prn{\sum_j\frac{2}{1+\kappa_j}-1}.
\end{equation*}
It follows that the black holes studied in this subsection admit smooth extremal limits at any value of $\rho_+$. Extremal black holes with large
$\rho_+$ (and with no $\kappa_i$ arbitrarily large) are dual to extremal solutions of fluid mechanics.

Expressing the fluid mechanical mean free path \eqref{m5lmfp:eq} as a function of black hole parameters we find
\begin{equation*}
l_\mathrm{mfp} \sim \frac{1}{5\rho_+\prod_j\sqrt{1+\kappa_j}}\,.
\end{equation*}
Once again $l_\mathrm{mfp} \sim 1/r_+$, away from thermodynamically suspect limits.

\section{Discussion}

As we have explained in this paper, the classical properties of large black holes in AdS spaces enjoy a large degree of universality, summarised by \eqref{final:eq}. However the reasoning that led to \eqref{final:eq} applies equally to all classical theories of gravity, not just to those theories that are governed by the two derivative effective action. For instance, $\mathcal{N}=4$ Yang-Mills theory at finite $\lambda$ is dual to IIB theory on $AdS_5 \times S^5$ of finite radius in string units. Even though thermodynamics of black holes in this background will receive contributions from each of the infinite sequence of $\alpha'$ corrections to the Einstein action, we expect \eqref{final:eq} to be exact in the large horizon radius limit.\footnote{Away from the supergravity limit, the mean free path $l_\mathrm{mfp} =\nu/\rho$ is expected to be given by $f(\lambda) s/\rho$ where $f(\lambda)$ is a monotonically decreasing function that interpolates between infinity at $\lambda =0$ to unity at infinite $\lambda$. Thus the condition for the validity of fluid mechanics is modified at finite $\lambda$; in the uncharged case, for instance, it is $T \gg f(\lambda)$.}

We find it particularly interesting that (at least in several particular contexts) our fluid dynamical picture applies not just to non-extremal black holes but also to large radius extremal black holes. This fact might allow us to make connections between our approach and the interesting recent investigations of the properties of extremal black holes. In particular, Astefanesei, Goldstein, Jena, Sen and Trivedi \cite{Astefanesei:2006dd} have recently argued that the attractor mechanism applies to rotating extremal black holes, and have derived a differential equation that determines the attractor geometry (and gauge field distribution, etc.) of the near horizon region of such black holes. It would be very interesting to investigate the connection, if any, between these rotating attractor equations and our equations of rotating fluid dynamics.

It would be conceptually simple (though perhaps technically intricate) to compute the spectrum of small fluctuations about the fluid dynamical solutions presented in this paper. This spectrum should match the spectrum of the (lowest) quasinormal modes about the relevant black holes (the decay of fluid fluctuations due to viscosity maps to the decay of quasi normal modes as they fall into the black hole horizon). It would be interesting to check if this is indeed the case.

It would be interesting to better understand, purely in bulk terms, why our proposal works. Roughly speaking, it should be possible to understand the metric of a black hole in global AdS and in the large radius limit, as a superposition of patches of the metric of black branes of various different temperatures and moving at various different velocities, where the temperatures and velocities are given by the solutions to the fluid dynamical equations presented in this paper. It would be interesting and useful if these words could be converted into the first term of a systematic approximation procedure to generate black hole solutions in AdS spaces in a power series in $1/r_+$. Such a construction would constitute a bulk derivation of the boundary Navier-Stokes equations (and corrections thereof).

Relatedly, it would be interesting to ask if there are any gravitational interpretations of the local properties of fluids in our solutions. For instance, fluid mechanics yields a sharp prediction for the velocity and entropy density of the fluid as a function of position on the sphere. We have not yet been able to verify these predictions, because we do not know what gravitational construction we should compare them to. The entropy of the fluid is an integral over the boundary. The entropy of the black hole is an integral over the horizon. Perhaps there exists a natural map from the horizon to the boundary that allows one to convert horizon densities to boundary densities and vice versa. Such a map (for which we have no conjecture) would permit a gravitational prediction $s$, the local entropy density of the fluid.

The fluid velocity is another quantity for which it would be useful to have a gravitational definition. We do not really have a serious proposal for such a definition: nonetheless, in the next few paragraphs we outline a caricature proposal, in order to give the reader a sense of the types of relations that might exist (we emphasise that we do not have any physical reason to believe that this caricature has any truth to it).

In the black hole solutions, there is one special Killing vector, $K = \p_t +\Omega_a \p_{\phi_a}$, that is also the null generator of the horizon. It has the norm
\begin{equation*}
 \nrm{K}^2 \equiv -K^\mu K_\mu =
 \left\{\begin{aligned}
  & r^2 \gamma^{-2} & &\text{at the boundary,}\\
  & 0 & &\text{at the horizon.}
 \end{aligned}\right.
\end{equation*}
If we were to normalise it with respect to the metric of the conformal
boundary, the result would be $\widetilde{K} = \gamma K$. This could
be identified with the fluid velocity $u^\mu$. However, as $\gamma$ is
not constant, $\widetilde{K}$ is not a Killing vector. It also seems
unnatural to use a normalisation factor that depends on $\theta$ but
not $r$. Nonetheless, this much maligned vector field has an interesting property.

Recall that black hole temperature and chemical potentials can be computed
from the formulae
\begin{equation*}
 T = \frac{\kappa}{2\pi}
  = \left.\frac{\sqrt{(\p_\mu\nrm{K})(\p^\mu\nrm{K})}}
               {2\pi}\right\rvert_\mathrm{horizon},
 \qquad
 \zeta_i = \left. A_\mu^i K^\mu \right\rvert_\mathrm{horizon}.
\end{equation*}
If one were to replace $K$ with $\widetilde{K}$ in the formulae above,
one would obtain $\tloc$ and $\mu_i$, the local temperature and
chemical potentials of the fluid.

We end this paper by reminding the reader that, while our proposal has passed many checks, our work has left one significant puzzle unresolved. While the thermodynamics and stress tensors of uncharged rotating black holes
in every dimension, plus all known black holes in $AdS_7\times S^4$ and $AdS_4 \times S^7$, deviate from the
predictions of the Navier-Stokes equations only at second order in $l_\mathrm{mfp}$, the situation is more
complex for black holes in $AdS_5\times S^5$. In this case, black holes with at least one SO(6) charge equal
to zero also agree with the results of the Navier-Stokes equations up to $\mathcal{ O}(l_\mathrm{mfp}^2)$. However the
thermodynamics of rotating black holes with all SO(6) charges nonzero, appears to deviate from our fluid mechanical
predictions at $\mathcal{O}(l_\mathrm{mfp})$ (see \S\S\ref{sec:nlo}).

We consider this a significant puzzle as our fluid dynamical configurations solve the Navier-Stokes equations including $\mathcal{O}( l_\mathrm{mfp})$ dissipative contributions. Moreover, in appendix \ref{sec:conffluidmech} we have checked by direct enumeration that all possible vectors and traceless symmetric tensors that transform homogeneously under conformal transformation and contain a single derivative simply vanish on our solution, so it is difficult
to see how any one derivative modification to the equations of fluid dynamics could help resolve this puzzle.\footnote{It has been suggested that certain pathologies in relativistic fluid dynamics lead to the breakdown
of the derivative expansion \cite{Israel:1979wp} (See also \cite{Andersson:2006nr} and the references therein). As any such pathology should apply equally to two charge and three charge black holes, we find it difficult to see how this issue could have bearing on our puzzle. We thank S.~Gupta and H.~Liu for discussions on this issue. Another possibility is that the formulae of black hole thermodynamics receive corrections -- perhaps from Wess-Zumino type terms -- that are nonzero only in an even dimensional bulk (and so in ten but not in eleven dimensions) and only when all charges are nonzero. We thank O.~Aharony for suggesting this possibility.} Once this puzzle is resolved it would be interesting to attempt to reproduce the $\mathcal{O} (l_\mathrm{mfp}^2)$ corrections to black hole thermodynamics from appropriate additions to the equations of fluid dynamics. It is perhaps worth emphasising that black holes in AdS represent exact (to all orders in $l_\mathrm{mfp}$) solutions to a dynamical flow. A detailed study of these solutions might lead to new insights into the nature of the fluid dynamical approximations of the high energy regime of quantum field theories.

\subsection*{Acknowledgements}

We would like to thank D.~Anninos, M.~Barma, P.~Basu, A.~Dabholkar, J.~David, F.~Gelis, K.~Goldstein, R.~Gopakumar, S.~Gupta, S.~Lee, H.~Liu, H.~Reall, S.~Trivedi, S.~Wadia, T.~Wiseman and all the students in the TIFR theory room for useful discussions. We would also like to thank O.~Aharony, G.~Gibbons, R.~Gopakumar, G.~Horowitz, H.~Liu and A.~Starinets for useful comments on an advance version of this manuscript. The work of S.M.\ was supported in part by a Swarnajayanti Fellowship. We must also acknowledge our debt to the steady and generous support of the people of India for research in basic science.

%%%%%%%%%%%%%%%%%%%%%%%%%%%%%%%%%%%%%%%%%%%%%%%%%%%%%%%%%%%%%%%%%%%%%%%%%%
\section*{Appendices}
\appendix
%%%%%%%%%%%%%%%%%%%%%%%%%%%%%%%%%%%%%%%%%%%%%%%%%%%%%%%%%%%%%%%%%%%%%%%%%%

\section{Conformal fluid mechanics}\label{sec:conffluidmech}

Consider a conformal fluid in d dimensions. We seek the conformal transformations of various observables of such a fluid. To this end, consider a conformal transformation which replaces the old metric $g_{\mu\nu}$ with $\tilde{g}_{\mu\nu}$ given by
\begin{equation*}
 g_{\mu\nu} = \e^{2\phi}\tilde{g}_{\mu\nu}; \qquad
 g^{\mu\nu} = \e^{-2\phi}\tilde{g}^{\mu\nu}.
\end{equation*}
The Christoffel symbols transform as
\begin{equation*}
 \Gamma_{\lambda\mu}^{\nu}= \widetilde{\Gamma}_{\lambda\mu}^{\nu} + \delta^{\nu}_{\lambda}\partial_{\mu}\phi+ \delta^{\nu}_{\mu}\partial_{\lambda}\phi-
 \tilde{g}_{\lambda\mu}\tilde{g}^{\nu\sigma}\partial_{\sigma}\phi\,.
\end{equation*}

Let $u^{\mu}$ be the four-velocity describing the fluid motion. Using $g_{\mu\nu}u^{\mu}u^{\nu}=\tilde{g}_{\mu\nu}\tilde{u}^{\mu}\tilde{u}^{\nu}=-1$, we get $u^\mu=\e^{-\phi}\tilde{u}^\mu$. It follows that the projection tensor transforms as $P^{\mu\nu}=g^{\mu\nu}+u^{\mu}u^{\nu}=\e^{-2\phi}\widetilde{P}^{\mu\nu}$. The transformation of the covariant derivative of $u^{\mu}$ is given by
\begin{equation}
 \nabla_{\mu}u^{\nu}=\partial_{\mu}u^{\nu}+\Gamma_{\mu\lambda}^{\nu}u^{\lambda}
  =\e^{-\phi}\brk{\widetilde{\nabla}_{\mu}\tilde{u}^{\nu}+\delta^{\nu}_{\mu}\tilde{u}^{\sigma}\partial_{\sigma}\phi- \tilde{g}_{\mu\lambda}\tilde{u}^{\lambda}\tilde{g}^{\nu\sigma}\partial_{\sigma}\phi}.
\end{equation}

The above equation can be used to derive the transformation of various related quantities
\begin{equation}
\begin{split}
 \vartheta =\nabla_{\mu}u^{\mu}
  &=\e^{-\phi}\brk{\tilde{\vartheta}+(d-1)\tilde{u}^{\sigma}\partial_{\sigma}\phi},\\
 a^{\nu}= u^{\mu}\nabla_{\mu}u^{\nu}
  &=\e^{-2\phi}\brk{\tilde{a}^{\nu}+\widetilde{P}^{\nu\sigma}\partial_{\sigma}\phi},\\
 \sigma^{\mu\nu} &= \frac{1}{2} \prn{P^{\mu\lambda} \nabla_\lambda u^\nu
                   + P^{\nu\lambda} \nabla_\lambda u^\mu}
                   - \frac{1}{d-1} \vartheta P^{\mu\nu}
                   = \e^{-3\phi} \tilde{\sigma}^{\mu\nu},\\
 \omega^{\mu\nu} &= \frac{1}{2} \prn{P^{\mu\lambda} \nabla_\lambda u^\nu
                  - P^{\nu\lambda} \nabla_\lambda u^\mu}
                  = \e^{-3\phi} \widetilde{\omega}^{\mu\nu}.
\end{split}
\end{equation}
Further, the transformation of the temperature and the chemical potential can be written as $\tloc=\e^{-\phi}\widetilde{\tloc}$ and $\mu=\e^{-\phi}\tilde{\mu}$. The transformation of spatial gradient of temperature (appearing in the Fourier law of heat conduction) is
\begin{equation*}
 P^{\mu\nu}(\partial_\nu \tloc +a_\nu \tloc) =
 \e^{-3\phi} \widetilde{P}^{\mu\nu}(\partial_\nu \widetilde{\tloc} +\tilde{a}_\nu \widetilde{\tloc})\,.
\end{equation*}
The viscosity, conductivity etc.\ scale as $\kappa=\e^{-(d-2)\phi}\tilde{\kappa}$ , $\eta=\e^{-(d-1)\phi}\tilde{\eta}$, $\mu_i=\e^{-\phi}\tilde{\mu_i}$ and $ D_{ij}=\e^{-(d-2)\phi} \widetilde{D}_{ij}$.

For a fluid with $c$ charges, there are $2c+2$ vector quantities involving no more than a single derivative which transform homogeneously\footnote{In the following analysis, we will neglect pseudo-tensors which can be formed out of $\epsilon_{\mu\nu\ldots}$. Additional tensors appear if such pseudo-tensors are included in the analysis.}. They are
\begin{equation*}
\begin{split}
 &u^{\mu},\quad
 \partial_\mu\nu_i,\quad
 \partial_\mu \tloc +\left(a_\mu-\frac{\vartheta}{d-1}u_\mu\right) \tloc ,
 \quad  u^\mu u^\sigma \partial_\sigma \nu_i \quad
 \text{and} \quad
 \prn{u^{\sigma}\partial_{\sigma} \tloc + \frac{\vartheta}{d-1} \tloc} u^{\mu}.
\end{split}
\end{equation*}
In the kind of solutions we consider in this paper, all of them vanish except $u^{\mu}$.

The transformation of the stress tensor is $T^{\mu\nu}=\e^{-(d+2)\phi}\widetilde{T}^{\mu\nu}$, from which it follows that
\begin{equation*}
\nabla_\mu T^{\mu\nu} =\e^{-(d+2)\phi}(\widetilde{\nabla}_{\mu}\widetilde{T}^{\mu\nu}- \tilde{g}_{\lambda\sigma}\widetilde{T}^{\lambda\sigma}\tilde{g}^{\nu\sigma}\partial_\sigma\phi)\,.
\end{equation*}
So, for the stress tensor to be conserved in both the metrics, it is necessary that $T^{\mu\nu}$ is traceless.

To consider the possible terms that can appear in the stress tensor, we should look at the traceless symmetric second rank tensors which transform homogeneously. The tensors formed out of single derivatives which satisfy the above criterion are easily enumerated. For a fluid with $c$ charges, there are $2c+4$ such tensors and they are
\begin{equation}
\begin{split}
 &u^{\mu}u^{\nu}+\frac{1}{d}g^{\mu\nu},\quad
 \sigma^{\mu\nu},\quad
  q^{\mu}u^{\nu}+q^{\nu}u^{\mu}, \quad
 \prn{u^{\sigma}\partial_{\sigma} \tloc + \frac{\vartheta}{d-1} \tloc}
    \!\prn{u^{\mu}u^{\nu}+\frac{1}{d}g^{\mu\nu}},\\
 &\frac{1}{2}\left(u^{\mu}\partial^{\lambda}\nu_i + u^{\lambda}\partial^{\mu}\nu_i\right)
   -\frac{g^{\mu\nu}}{d}u^{\sigma}\partial_{\sigma}\nu_i
 \quad \text{and} \quad
 u^{\sigma}\partial_{\sigma}\nu_i \prn{u^{\mu}u^{\nu}+\frac{1}{d}g^{\mu\nu}}.
\end{split}
\end{equation}
Among these possibilities, the stress tensor we employ just contains the tensors in the first line. It can be shown that the other tensors which appear in the above list can be removed by a redefinition of the temperature etc. Even if they were to appear in the stress tensor, for the purposes of this paper, it suffices to notice that all such tensors except $u^{\mu}u^{\nu}+\frac{1}{d}g^{\mu\nu}$ vanish on our solutions. Hence, they would not contribute to any of the thermodynamic integrals evaluated on our solutions.

%arbitrary dimension

\section{Free thermodynamics on spheres}\label{sec:free}

In \eqref{otgpf:eq} above, we have presented a general expression for the grand canonical partition function for any conformal fluid on a sphere. In this appendix , we compare this expression with the conformal thermodynamics of a free complex scalar field on a sphere.

Strictly speaking, the fluid dynamical description never applies to free theories on a compact manifold, as the constituents of a free gas have a divergent mean free path (they never collide). Nonetheless, as we demonstrate in this subsection, free thermodynamics already displays some of the features of \eqref{otgpf:eq} - in its dependence on angular velocities, for example - together with certain pathologies unique to free theories.

Consider a free complex scalar field on $S^{d-1} \times$ time. This system has a $U(1)$ symmetry, under which $\phi$ has unit charge and $\phi^*$ has charge minus one. We define the `letter partition function' \cite{Aharony:2003sx} $Z_\mathrm{let}$ as $\Tr\exp\brk{-\beta H + \nu R +\beta \Omega_a L_a}$ evaluated over all spherical harmonic modes of the scalar field
\begin{equation} \label{letter:eq}
Z_\mathrm{let}=(\e^{\nu}+\e^{-\nu}) \e^{-\beta \frac{d-2}{2}}
\left( \frac{1- \e^{-2 \beta}} {\prod_{a=1}^n (1-\e^{-\beta -\beta \Omega_a})
(1-\e^{-\beta +\beta \Omega_a})}\right)
\end{equation}
(this formula, and some of the others in this section, are valid only for even $d$; the generalisation to odd $d$ is simple). We will now examine the high temperature limit of the grand-canonical partition function separately for $\nu = 0$ and $\nu \neq 0$.

\subsection{Zero chemical potential: ($\nu=0$) case}

The second quantised partition function, $\gpf$ for the scalar field on the
sphere is given by
\begin{equation}\label{letpf:eq}
 \gpf= \exp\left( \sum_N \frac{Z_\mathrm{let}(N\beta, N \nu,
\Omega_a)}{N} \right).
\end{equation}
For small $\beta$, we have
\begin{equation*}
 Z_\mathrm{let} \approx \frac{4}{\beta^{d-1} \prod_a(1-\Omega_a^2)}\,.
\end{equation*}
It follows that\footnote{This formula has been derived before in many contexts, for example \cite{Berman:1999mh} have derived this in $d=4$ and compared it with the thermodynamics of black holes in AdS$_5$.}
\begin{equation}\label{Zgc:eq}
 \ln\gpf = \frac{4 \zeta(d)}{\beta^{d-1} \prod_{a}(1 - \Omega_a^2)}\,.
\end{equation}
Upon identifying $ V_d h|_{\nu=0}=4 \zeta(d)$, we find that
\eqref{letpf:eq} is in perfect agreement with \eqref{otgpf:eq}.

\subsection{Nonzero chemical potential: ($\nu \neq 0$) case}

The high temperature limit of the thermodynamics of a free, charged, massless
field is complicated by the occurrence of Bose condensation. This phenomenon
occurs already when $\omega_a=0$; this is the case we first focus on.

It is useful to rewrite the letter partition function as
\begin{equation}\label{letch:eq}
Z_\mathrm{let}= (2 \cosh \nu) \ \ \e^{-\beta \frac{d-2}{2}}
\sum_N m(N)  \e^{-\beta N},
\end{equation}
where $m(N) \approx 2 N^{d-2}/ (d-2)!$ for $N \gg 1$.
The logarithm of the grand canonical partition function may then be
written as a sum over Bose factors (one per `letter')
\begin{equation} \label{freegcpf:eq}
 \ln \gpf = - \sum_N m(N) \left[ \ln(1-\e^{-\beta(N +(d-2)/2) +\nu})
+ \ln( 1-\e^{-\beta (N+(d-2)/2 ) - \nu} ) \right].
\end{equation}
The total charge in this ensemble is given by
\begin{equation} \label{freecharge:eq}
 R= \frac{\partial}{\partial\nu} \ln \gpf  =
 \sum_N m(N) \left( \frac{ 1}{\e^{\beta(N+(d-2)/2) -\nu} -1}
- \frac{ 1}{\e^{\beta(N+(d-2)/2) +\nu}-1 } \right).
\end{equation}

In order to compare with fluid dynamics, we should take $\beta$ to zero
while simultaneously scaling to large $R$ as $R=\frac{q}{\beta^{d-1}}$
with $q$ held fixed. As we will see below, in order to make the total charge
$R$ large, we will have to choose the chemical potential to be large. However it
is clear from \eqref{freegcpf:eq} that $|\nu|< \beta (d-2)/2$. Consequently, the
best we can do is to set $\nu=\beta((d-2)/2)-\epsilon$ where $\epsilon$ will be taken
to be small. We are interested in the limit when $\beta$ is also small. We may
approximate \eqref{freecharge:eq} by
\begin{equation} \label{freechargeapprox:eq}
 \frac{q}{\beta^{d-1}}= \frac{1}{\epsilon}  - \frac{ 1}{\e^{\beta
(d-2)  -\epsilon}-1 }
+ \sum_{N=1}^\infty \left( \frac{ 1}{\e^{\beta N + \epsilon }-1 }
- \frac{ 1}{\e^{\beta(N+(d-2)) - \epsilon}-1 } \right).
\end{equation}
The only solution to \eqref{freechargeapprox:eq} is
\begin{equation*}
\epsilon=\frac{\beta^{d-1}}{q}\left( 1+ \mathcal{O} (\beta) \right).
\end{equation*}

Substituting this solution into the partition function, we find
\begin{equation}\label{pfff:eq}
 \ln Z_q= \frac{4 \zeta(d)}{\beta^{d-1}} \left( 1+ \mathcal{O}(\beta) \right).
\end{equation}
Consequently, to leading order the partition function is independent of the charge $q$ ! What is going on here is that almost all of the charge of the system resides in a Bose condensate of the zero mode of the field $\phi$. This zero mode contributes very little entropy or energy to the system at leading order in $\beta$.\footnote{In particular, the contribution of the zero mode to the energy is proportional to the charge, which is suppressed by a factor of $\beta$ relative to the contribution to the energy from nonzero modes.} At high temperatures, the zero mode is simply a sink that absorbs the system charge, leaving the other thermodynamic parameters unaffected.

Upon generalising our analysis to include angular velocities, we once again find that the leading order partition function (in the limit of high temperatures and a charge $R=q/\beta^{d-1}$) is independent of $q$ and in fact is given by \eqref{Zgc:eq}. Consequently, there is a slightly trivial (or pathological) sense in which the thermodynamics of a free charged scalar field agrees with the predictions of fluid mechanics - we find agreement upon setting $h(\nu)$ to a constant.

\section{Stress tensors from black holes}\label{sec:bhstrapps}

According to the usual AdS/CFT dictionary, the boundary stress tensor on $S^{d-1}$, corresponding to any finite energy solution about an AdS$_D$ background of gravity may be read off from the metric near the boundary, using the following procedure \cite{Kraus:1999di, Henningson:1998gx, de-Haro:2000xn, Skenderis:2000in, Papadimitriou:2005ii, Cheng:2005wk, Olea:2005gb, Olea:2006vd}. First we foliate the spacetime near the boundary into a one parameter set of $d$ geometries, each of which is metrically conformal to $S^{d-1} \times \R$, to leading order in deviations from the boundary. We will find it convenient to use coordinates such that the leading order metric in the neighbourhood of the boundary takes the form
\begin{equation}\label{metform:eq}
ds^2= -r^2 d t^2 +r^2  d \Omega_{d-1}^2 +\frac{d r^2}{r^2}
\end{equation}
(here $r=\infty$ is the boundary; this metric has corrections at subleading orders
in $\frac{1}{r^2}$).

In these coordinates, our foliation surfaces are simply given by $r=\text{const}$. We next compute the extrinsic curvature $\Theta_\mu^\nu = -\nabla_\mu n^\nu$ on these surfaces, where $n^\nu$ is the unit outward normal to these surfaces. The boundary stress tensor for the dual field theory on a unit sphere is given by \cite{Kraus:1999di, Henningson:1998gx, de-Haro:2000xn, Skenderis:2000in, Papadimitriou:2005ii, Cheng:2005wk, Olea:2005gb, Olea:2006vd} as\footnote{We rescale the stress tensor of \cite{Kraus:1999di} by a factor of $\frac{1}{8 \pi G_D}$ in order that the energy of our solutions is given by $\int \sqrt{\gamma}\ \Pi^{0}_{0}$ with no extra normalisation factor. }
\begin{equation}\label{sangnya:eq}
 \Pi^{\mu}_{\nu} = \lim_{r\rightarrow\infty}\frac{r^{D-1}}{8 \pi G_D}\left(\Theta^{\mu}_{\nu} - {\delta}^{\mu}_{\nu}\Theta\right),
\end{equation}
where the coordinates $\mu,\nu$ go over time and the angles on $S^{D-1}$.

The stress tensor as defined above will contain some terms which are independent of mass and charge of the black hole. These are the terms that are nonzero even on the vacuum AdS background and they diverge in the limit $r\rightarrow\infty$. These terms are all precisely cancelled, up to a zero point Casimir energy, by counter terms presented in \S2 of \cite{Kraus:1999di}. We will simply ignore all such terms below; consequently, the stress tensors we present in this paper should be thought of as the field theory stress tensors with the contribution from the Casimir energy subtracted out.\footnote{Recall that the full stress tensor of a general $d$ dimensional conformal field theory is not traceless on an arbitrary manifold; however the trace is given by a function of the manifold curvature independent of the field theory configuration. It follows that our stress tensor with Casimir contribution subtracted must be traceless, as indeed it will turn out to be.}

In order to compute the stress tensor in \eqref{sangnya:eq}, we must retain subleading corrections to the metric in \eqref{metform:eq}. However, only those corrections that are subleading at $\CO (\frac{1}{r^{D-1}})$ (relative to the leading order metric in \eqref{metform:eq}) contribute to \eqref{sangnya:eq}.

In order to compute the stress tensor corresponding to two classes of black hole solutions below, we adopt the following procedure. First, we find a coordinate change that casts the metric at infinity in the form \eqref{metform:eq} at leading order. Next we compute all the subleading corrections to the metric at order $\CO (\frac{1}{r^{D-1}})$. Finally, we use these corrections to compute the extrinsic curvature $\Theta^{\mu}_{\nu}$ and then $\Pi^{\mu}_{\nu}$ using \eqref{sangnya:eq}.
%b

\subsection{Stress tensor from rotating uncharged black holes in AdS$_D$}
\label{sec:bhstruc}

\subsubsection{$D = 2n+1$}

The most general rotating black hole in an odd dimensional AdS space is given by equation (E.3) of \cite{Gibbons:2004uw} (we specialise to odd dimensions by setting the parameter $\epsilon$ in that equation to zero). The metric presented in \cite{Gibbons:2004uw} uses as coordinates
\begin{enumerate}
\item The $n$ Killing azimuthal angles $\phi_i$ along which the black hole rotates. These may be identified with the coordinates $\phi_i$ in \S\ref{sec:otherspheres} of our paper.

\item $n$ other unspecified variables (called `direction cosines') $\mu_i$ subject to the constraint $\sum_i \mu_i^2=1$. These may be thought of as the remaining $n-1$ coordinates on $S^{2n-1}$.

\item The radial variable $r$ and timelike variable $t$.
\end{enumerate}

In order to cast the metric of \cite{Gibbons:2004uw} into the form \eqref{metform:eq} near the boundary, we perform the following change of coordinates
\begin{equation}\label{notunp:eq}
\begin{split}
 \tilde r^2 =& \sum_{i=1}^n\frac{(r^2 +a_i^2)\mu_i^2}
             {1-a_i^2}\,,\qquad
 \tilde r^2\tilde\mu_i^2(1-a_i^2) = (r^2 +a_i^2)\mu_i^2.
 \end{split}
\end{equation}
Note that
\begin{equation*}
 \sum_{i=1}^n\mu_i^2 = \sum_{i=1}^n\tilde\mu_i^2 = 1\,.
\end{equation*}
This equation may be solved by writing ${\tilde \mu}_i$ as functions of the $n-1$ variables $\theta_j$ (which may then be identified with the coordinates used in \S\ref{sec:otherspheres})
\begin{equation*}
{\tilde \mu} _i = \left(\prod_{j=1}^{i-1}\cos^2\theta_j\right)\sin^2\theta_i\,.
\end{equation*}

In these coordinates, the metric in the neighbourhood of $r \to \infty$ becomes
%\footnote{The leading behaviour at large $r$ of the pure AdS metric given in terms of $\frac{\dr r}{r}$ and ${\dr\alpha r}$ where $\alpha$ represents the coordinates $\theta, \phi, \psi, t$. In the expression below we have retained all coefficient terms of these `forms' that are at most of order $\frac{1}{r^{D-1}$.} in the correction metric (the terms that involve the mass) the metric, in these coordinates, becomes
%
\begin{equation}\label{thikmatra:eq}
\begin{split}
 \dr s^2 = -&(1+\tilde r^2)\dr t^2 + \frac{\dr \tilde r^2}{1+\tilde r^2}
           + \tilde r^2\sum_{i=1}^n(\dr\tilde\mu_i^2 + \tilde\mu_i^2 \dr\phi_i^2)\\
           +&\frac{2m}{\tilde r^{2n-2}}\gamma^{2(n+1)}\dr t^2 + \frac{2m}{\tilde r^{2n+2}}\gamma^{2n}\dr \tilde r^2\\
           -& \sum_{i=1}^n\frac{4ma_i\tilde \mu_i^2}{\tilde r^{2n-2}}\gamma^{2(n+1)}\dr t\dr \phi_i + \sum_{i=1,j=1}^n\frac{2ma_ia_j\tilde \mu_i^2\tilde\mu_j^2}{\tilde r^{2n-2}}\gamma^{2(n+1)}\dr \phi_i\dr \phi_j\,,
             %(\dr t - a\sin^2\theta \dr\phi^2)^2
\end{split}
\end{equation}
where we have retained all terms that are subleading up to $\mathcal{O}( \frac{1}{\tilde r^{D-1}})$ compared to the metric of pure AdS. Here $\gamma^{-2} = 1-\sum_{i=1}^na_i^2\tilde\mu_i^2 $ and $\sum_{i=1}^n\tilde \mu_i^2\dr \tilde\mu_i^2 = \sum_{i=1}^{n-1}\left(\prod_{j=1}^{i-1}\cos^2\theta_j\right)\dr\theta_i^2$ as in \eqref{otsphmet:eq}.

Note that this metric separates into two parts; the first piece (on the first line of \eqref{thikmatra:eq}) is the metric of pure AdS space while the terms of the remaining lines represent correction proportional to the mass $m$.

The normal vector is given by $n_{\tilde r} =\frac{1}{\sqrt{g^{\tilde r\tilde r}}}$ (with all other components zero). As our metric contains no terms that mix $r$ with other coordinates at leading order, this is the same as
\begin{equation*}
 n^{\tilde r}= \frac{1}{\sqrt{g_{\tilde r\tilde r}}}
  = \tilde r \left(1+\frac{1}{\tilde r^2}\right)^{\frac{1}{2}}
          \brk{1-\frac{m}{\tilde r^{2n}}\gamma^{2n}}.
\end{equation*}
Since the normal vector has only the $\tilde r$ component and since the component is a function of $\tilde r$ only, to compute the extrinsic curvature tensor $\Theta^\nu_\lambda$ one needs only those components of $\Gamma$ that are of the form $\Gamma^\nu_{\lambda \tilde r}$. The Christoffel symbols (that are relevant for the calculation of the stress-tensor) as calculated from this metric up to the first subleading term in $\tilde r$ are given by
\begin{equation}\label{othanosim:eq}
\begin{aligned}
 \Gamma^t_{t\tilde r} =& \frac{\tilde r}{{\tilde r^2 +1}}
      \left(1+\frac{2nm}{\tilde r^{2n}}\gamma^{2(n+1)} \right)&
 \Gamma^{\phi_i}_{t\tilde r} =& \frac{2nma_i}{\tilde r^{2n+1}}\gamma^{2(n+1)}\\
 \Gamma^t_{\phi_i \tilde r} =& -\frac{2nma_i\tilde\mu_i^2}
                            {\tilde r^{2n+1}}\gamma^{2(n+1)}&
 \Gamma^{\phi_i}_{\phi_j \tilde r} =& \frac{1}{\tilde r}
                     \left(\delta_{ij}-\frac{2nma_ia_j\tilde\mu_j}
                                  {\tilde r^{2n}}\gamma^{2(n+1)}\right)\\
 \Gamma^{\theta}_{\theta \tilde r} =& \frac{1}{\tilde r}\,.
 %\Gamma^{\psi}_{\psi r} =& \frac{1}{r}
\end{aligned}
\end{equation}

The extrinsic curvature $\Theta^\nu_\lambda$ in this case is given by
\begin{equation*}
\Theta^{\nu}_\lambda = -\Gamma_{\lambda\tilde r}^{\nu} n^{\tilde r}
  +\mathcal{ O} \left( \frac{1}{\tilde r^{2n+1}}\right).
\end{equation*}
Ignoring all terms in $\Theta^\nu_\mu$ that are independent of mass (for the reasons explained in the introduction to this appendix) we find

\begin{equation}\label{bohirbakra:eq}
\begin{aligned}
 \Theta^t_t &= -\frac{m\gamma^{D+1}}{\tilde r^{D-1}}(D-1-\gamma^{-2}) &
 \Theta^{\phi_i}_{\phi_i} &= \frac{m\gamma^{D+1}}{\tilde r^{D-1}}
                 \left((D-1)a_i^2\tilde \mu_i^2 + \gamma^{-2}\right)\\
 \Theta^t_{\phi_i} &= \frac{(D-1)m\gamma^{D+1}}{\tilde r^{D-1}}a_i\tilde\mu_i^2 & \Theta^{\phi_i}_{\phi_j} &= \frac{(D-1)m\gamma^{D+1}}{\tilde r^{D-1}}a_ia_j\tilde\mu_j^2
 \ \ \ \ (i\neq j)
 \\
 \Theta^{\phi_i}_t &= -\frac{(D-1)m\gamma^{D+1}}{\tilde r^{D-1}}a_i&
 \Theta^{\theta_i}_{\theta_i} &= \frac{m\gamma^{(D-1)}}{\tilde r^{D-1}}\,.
\end{aligned}
\end{equation}
Here the $n$ has been replaced by $\frac{D-1}{2}$. It may easily be verified that $\Theta^\nu_\lambda$ is traceless and therefore the stress tensor is also traceless according to the definition \eqref{sangnya:eq}. Raising one index in $\Theta$ by asymptotic AdS metric, normalising it appropriately and then taking the large $\tilde r$ limit one can derive the stress tensor as given in \eqref{uncharblstodd:eq}.

\subsubsection{$D = 2n+2$}

The computation of the boundary stress tensor for the most general uncharged rotating black hole in even dimensional AdS spaces is almost identical to the he analysis presented in the previous subsection. Once again the metric is given in equation (E-3) of \cite{Gibbons:2004uw}, where we must set $\epsilon$ to 1 to specialise to even dimensions. The coordinates of the black hole solution are similar to those described in the previous subsection, except that there are $n+1$ coordinates $\mu_i$ restricted by a single equation $\sum_a \mu_i^2=1$. Repeating the computations described in the previous subsection, our final result is once again simply \eqref{bohirbakra:eq}. In summary \eqref{bohirbakra:eq} is correct no matter whether $D$ is odd or even.

\subsection{Black holes in AdS$_5$ with all R-charges equal}
\label{sec:bhstr}

In this subappendix, we compute the boundary stress tensor for a class of charged black holes, namely for black holes in AdS$_5$ with all three R-charges equal. Our computation will verify the striking prediction of \S\ref{sec:rotate} that the functional form of this stress tensor is independent of the black hole charge in the fluid dynamical limit.

The metric for rotating black holes with all R-charges equal is given by (equation (1) of \cite{Chong:2005hr})
\begin{equation}\label{metricc:eq}
\begin{split}
\dr s^2 = -&\frac{\Delta_{\tilde\theta}[(1 + y^2)\rho^2\dr t + 2q\nu]\dr
t}{\Sigma_a\Sigma_b\rho^2} + \frac{2q\nu\omega}{\rho^2} +
\frac{f}{\rho^4}\left(\frac{\Delta_{\Theta}\dr
  t}{\Sigma_a\Sigma_b}-\omega\right)^2 + \frac{\rho^2\dr y^2}{\Delta_y}\\
  +& \frac{\rho^2\dr\tilde\theta^2}{\Delta_{\tilde\theta}} + \frac{y^2 +a^2}{\Sigma_a}\sin^2\tilde\theta\dr\phi^2 + \frac{y^2 + b^2}{\Sigma_b}\cos^2\tilde\theta\dr\psi^2,
\end{split}
\end{equation}
where
\begin{equation}\label{defone:eq}
\begin{split}
 \Delta _y       =& \frac{(y^2 +a^2)(y^2 +b^2)(1+y^2) + q^2 +
  2abq}{y^2} -2m \,,\\
  \rho^2  =& y^2 + a^2\cos^2\tilde \theta + b^2\sin^2\tilde \theta \,,\\
 \Delta_{\tilde\theta} =& 1 - a^2\cos^2\tilde\theta - b^2\sin^2\tilde\theta \,,\\
\Sigma_a =& 1-a^2,\\
\Sigma_b =& 1-b^2,\\
f =& 2m\rho^2 - q^2 +2abq\rho^2,\\
\nu =& b\sin^2\tilde\theta\dr\phi + a\cos^2\tilde\theta\dr\psi\,,\\
\omega =& a\sin^2\tilde\theta\frac{\dr\phi}{\Sigma_a} + b\cos^2\tilde\theta\frac{\dr\psi}{\Sigma_b}\,.
\end{split}
\end{equation}
This metric takes the form \eqref{metform:eq} near the boundary, once we perform the change of coordinates\footnote{We thank Sangmin Lee for pointing out a typo in a previous version of this paper.}
\begin{equation}\label{changev:eq}
\begin{split}
 r^2 =& \frac{y^2(1-a^2\cos^2\tilde\theta - b^2\sin^2\tilde\theta) +
             a^2\sin^2\tilde\theta + b^2\cos^2\tilde\theta - a^2b^2}
             {\Sigma_a\Sigma_b}\,,\\
 r^2\sin^2\theta =& \frac{(y^2+a^2)\sin^2\tilde\theta}{\Sigma_a}\,,\\
 r^2\cos^2\theta =& \frac{(y^2+b^2)\cos^2\tilde\theta}{\Sigma_b}\,.
\end{split}
\end{equation}
Retaining corrections only to order $\mathcal{O}(1/r^4)$ relative to the leading order metric \eqref{metform:eq}, the metric in our new coordinates becomes\footnote{The leading behaviour at large $r$ of the pure AdS metric given in terms of $\frac{\dr r}{r}$ and ${r\dr\alpha}$ where $\alpha$ represents the coordinates $\theta, \phi, \psi, t$. In the expression below, we have retained all coefficient terms of these `forms' that are at most of order $\frac{1}{r^4}$.}
\begin{equation}\label{cmetric:eq}
\begin{split}
 \dr s^2 =& -(1+r^2)\dr t^2 + \frac{\dr r^2}{1+r^2} + r^2(\dr\theta^2
 + \cos^2\theta \dr\psi^2 + \sin^2\theta \dr\phi^2)
 + \frac{2m}{r^6\Delta_{\theta}^2}\dr r^2
 + \frac{2(m+abq)}{r^2\Delta_{\theta}^3}\dr t^2\\
 - &\frac{2(2am+bq)\sin^2\theta}{r^2\Delta_{\theta}^3}\dr t\dr\phi
 - \frac{2(2bm+aq)\cos^2\theta}{r^2\Delta_{\theta}^3}\dr t\dr\psi
 + \frac{(2ma^2+2bq)\sin^4\theta}{r^2\Delta_{\theta}^3}\dr\phi^2\\
 +& \frac{(2mb^2+2aq)\cos^4\theta}{r^2\Delta_{\theta}^3}\dr\psi^2
 + \frac{2(2abm +a^2q +b^2q)\sin^2\theta\cos^2\theta}{r^2\Delta_{\theta}^3}\dr\psi\dr\phi\,,
\end{split}
\end{equation}
where $\Delta_{\theta} = \gamma^{-2}=1 - a^2\sin^2\theta - b^2\cos^2\theta.$ with $\gamma$ as in \S\S \ref{sec:eqmot}.
%

% The energy momentum tensor is defined as
% %
% \begin{equation*}
%  \Pi^{\mu\nu} = \Theta^{\mu\nu} -h^{\mu\nu}\Theta
%    = h^{\mu\lambda}(\Theta^\nu_\lambda -\delta^\nu_\lambda\Theta^\alpha_\alpha)
% \end{equation*}
% %
The unit normal vector to constant $r$ slices is given by $n_r =\frac{1}{\sqrt{g^{rr}}}$ (with all other components zero). As our metric contains no terms that mix $r$ with other coordinates at leading order , this is the same as $n^r = \frac{1}{\sqrt{g_{rr}}}$.

The Christoffel symbols (that are relevant for the calculation of the stress-tensor) as calculated from this metric up to the first subleading term (i.e.\ up to $\mathcal{O}(\frac{1}{r^4})$ terms) in $r$ are given by
\begin{equation}\label{christrai:eq}
\begin{aligned}
 \Gamma^t_{tr} =& \frac{r}{{r^2 +1}}
      \left(1+\frac{4(m+abq)}{r^4(1-a^2\sin^2\theta-b^2\cos^2\theta)^3}\right)&
 \Gamma^{\phi}_{tr} =& \frac{2(2am+bq)}{r^5(1-a^2\sin^2\theta-b^2\cos^2\theta)^3}\\
\Gamma^{\psi}_{tr} =& \frac{2(2bm+aq)}{r^5(1-a^2\sin^2\theta-b2\cos^2\theta)^3}&
 \Gamma^t_{\phi r} =& -\frac{2(2am+bq)\sin^2\theta}
                            {r^5(1-a^2\sin^2\theta-b^2\cos^2\theta)^3}\\
 \Gamma^{\phi}_{\phi r} =& \frac{1}{r}
                     \left(1-\frac{4(a^2m + abq)\sin^2\theta}
                                  {r^4(1-a^2\sin^2\theta-b^2\cos^2\theta)^3}\right)&
\Gamma^t_{\psi r} =& -\frac{2(2bm+aq)\cos^2\theta}
                            {r^5(1-a^2\sin^2\theta-b^2\cos^2\theta)^3}\\
\Gamma^{\psi}_{\psi r} =& \frac{1}{r}
                     \left(1-\frac{4(b^2m + abq)\cos^2\theta}
                                  {r^4(1-a^2\sin^2\theta-b^2\cos^2\theta)^3}\right)&
 \Gamma^{\theta}_{\theta r} =& \frac{1}{r}\\
 \Gamma^{\phi}_{\psi r} =& -\frac{(2abm + a^2q + b^2q)\cos^2\theta}
                            {r^5(1-a^2\sin^2\theta-b^2\cos^2\theta)^3}&
\Gamma^{\psi}_{\phi r} =& -\frac{(2abm + a^2q + b^2q)\sin^2\theta}
                            {r^5(1-a^2\sin^2\theta-b^2\cos^2\theta)^3}\,.
\end{aligned}
\end{equation}

The extrinsic curvature, $\Theta_{\mu}^\nu$, is given by
\begin{equation}\label{extcurv:eq}
\begin{split}
&\begin{aligned}
\Theta^t_t &= -\left(1 + \frac{1}{r^2}\right)^{-\frac{1}{2}} \left(1 + \frac{m\gamma^6}{r^4}(3 + a^2 \sin^2 \theta + b^2\cos^2\theta)  \right) - \frac{4abq\gamma^6}{r^4}\\
 \Theta^\phi_\phi &= -\sqrt{1+1/r^2}  \left( 1 - \frac{m\gamma^6}{r^4}(3 a^2 \sin^2 \theta -b^2\cos^2\theta + 1)\right) +  \frac{4abq\gamma^6\sin^2\theta}{r^4}\\
\Theta^\psi_\psi &= -\sqrt{1+1/r^2}     \left( 1 - \frac{m\gamma^6}{r^4}          (3 b^2 \cos^2 \theta -a^2\sin^2\theta + 1) \right) +  \frac{4abq\gamma^6\cos^2\theta}{r^4}
\end{aligned}\\
&\begin{aligned}
\Theta^t_\phi &= \frac{2(2am +bq) \gamma^6 \sin^2 \theta}{r^4} &
\Theta^\phi_t &= -\frac{2(2am+bq) \gamma^6 }{r^4} \\
\Theta^t_\psi &= \frac{2(2bm +aq) \gamma^6 \cos^2 \theta}{r^4} &
\Theta^\psi_t &= -\frac{2(2bm+aq) \gamma^6 }{r^4}\\
 \Theta^\psi_\phi &= \frac{2(2abm +b^2q +a^2q) \gamma^6 \sin^2 \theta}{r^4} &
 \Theta^\phi_\psi &= \frac{2(2bam +b^2q +a^2q) \gamma^6 \cos^2 \theta}{r^4} \\
\Theta^\theta_\theta &= \frac{m\gamma^4}{r^4}\,,
\end{aligned}
\end{split}
\end{equation}
where $\gamma^2=\frac{1}{1- a^2 \sin^2 \theta -b^2\cos^2\theta}$.
Therefore
\begin{equation*}
\Theta = \Theta^{\alpha}_{\alpha} = 4 + \frac{1}{r^2}\,.
\end{equation*}
It is easily verified that $\Theta_\alpha^\beta$ is traceless when the $r$ dependent divergent terms are cancelled by the counter terms at the limit $r$ going to infinity. After cancelling the divergent terms and then normalising it according to \eqref{sangnya:eq} the stress tensor is given by
\begin{equation}\label{sten:eq}
\begin{gathered}
\begin{aligned}
 \Pi^{tt} &= \frac{m}{8\pi G_5}\left(\gamma^6(3+ a^2\sin^2\theta +b^2\cos^2\theta) - \frac{4abq}{m}\gamma^6\right)\\
    &= \frac{m}{8\pi G_5}\left(\gamma^4(4\gamma^2-1) - \frac{4abq}{m}\gamma^6\right)\\
 \Pi^{\phi\phi} &= \frac{m}{8\pi G_5}\left(\gamma^6
            \left(\frac{3a^2\sin^2\theta -b^2\cos^2\theta + 1}{\sin^2\theta}\right) -\frac{4abq}{m}\gamma^6\right)\\
   &= \frac{m}{8\pi G_5}\left(\gamma^4\left(4\gamma^2a^2 +
                               \frac{1}{\sin^2\theta}\right) -\frac{4abq}{m}\gamma^6\right)\\
\Pi^{\psi\psi} &= \frac{m}{8\pi G_5}\left(\gamma^6
            \left(\frac{3b^2\cos^2\theta -a^2\sin^2\theta + 1}{\cos^2\theta}\right) -\frac{4abq}{m}\gamma^6\right)\\
  &= \frac{m}{8\pi G_5}\left(\gamma^4\left(4\gamma^2a^2 +
                               \frac{1}{\cos^2\theta}\right) -\frac{4abq}{m}\gamma^6\right)
\end{aligned}\\
\begin{aligned}
 \Pi^{t\phi} &= \Pi^{\phi t} = \left(\frac{4m}{8\pi G_5}\right)\left(a -\frac{2bq}{m}\right)\gamma^6&
\Pi^{t\psi} &= \Pi^{\psi t} = \left(\frac{4m}{8\pi G_5}\right)\left(b -\frac{2aq}{m}\right)\gamma^6\\
\Pi^{\phi\psi} &= \Pi^{\psi \phi} = \left(\frac{4m}{8\pi G_5}\right)\left(ab -\frac{2(a^2 + b^2)q}{m}\right)\gamma^6&
 \Pi^{\theta\theta} &= \frac{m}{8\pi G_5}\gamma^4.
\end{aligned}
\end{gathered}
\end{equation}
As we have explained in \S\S \ref{sec:bheqch}, $q/m \sim 1/r_+$, and so all terms proportional to $q$ in the equation above are subdominant compared to terms proportional to $m$ in the fluid mechanical limit $r_+ \to \infty$. Dropping all $q$ dependent terms, we find \eqref{sten:eq} matches perfectly with the stress tensor as derived in \eqref{blkstrour:eq} and \eqref{chstr:eq} upon identifying $(\phi_1,\phi_2)=(\phi,\psi)$, $(\omega_1,\omega_2)=(a,b)$ and using \eqref{h3ch:eq}.

\section{Notation}\label{app:notation}

We work in the $(-+++)$ signature. $\mu,\nu$ denote space-time indices, $i,j=1 \ldots c$ label the $c$ different R-charges and $a,b=1 \ldots n$ label the $n$ different angular momenta. The dimensions of the AdS space is denoted by $D$ whereas the spacetime dimensions of its boundary is denoted by $d=D-1$. In this paper we consider fluids on $S^{D-2}\times \R$ or equivalently $S^{d-1}\times \R$. Here we present some relations which are useful in converting between $n$, $D$ and $d$:
\begin{equation*}
\begin{split}
 D&= d+1 = 2n+2-(D \mod 2) \\
 d&=D-1 =2n + (d \mod 2) \\
 n&=\left[\frac{D-1}{2}\right]=\left[\frac{d}{2}\right]
\end{split}
\end{equation*}
where $[x]$ represents the integer part of a real number $x$.

A summary of the variables used in this paper appears in table \ref{tab:notation}.

\begin{table}
\begin{center}
  \begin{tabular}{||r|l||r|l||}
    \hline
    Symbol & Definition & Symbol & Definition \\
    \hline
    $D$ & Dimension of bulk & $d$ & $D-1$, Dimension of boundary \\
    $G_D$ & Newton Constant in AdS$_D$& $n$ & $[d/2]$,  no.\ of commuting \\
    $c$ & no.\ of commuting R-charges &  & ~~angular momenta \\
    $R_\mathrm{AdS}$ & AdS radius (taken to be unity) & $V_{d}$ & Volume of
    $S^{d-1}$, $\frac{2\pi^{d/2}}{\Gamma(d/2)}$ \\
    $R_H,r_+$ & Horizon radius & $l_\mathrm{mfp}$ & Mean free path, $\eta /\rho$ \\
    \hline
%    $\floc$ & Plasma free energy & $f$ & Free energy density \\
    $\eloc$ & Fluid energy & $\rho$ & Proper density \\
    $\sloc$ & Fluid entropy & $s$ & Proper entropy density \\
    $\tloc$ & Fluid temperature & $\ploc$ & Pressure \\
    $\rloc_i$ & Fluid R-charge & $r_i$ & Proper R-charge density \\
    $\mu_i$ & Fluid chemical potential & $\Gamma$ & Thermodynamic potential \eqref{Gammadef:eq} \\
    $\Phi$ & Thermodynamic potential \eqref{phidef:eq} & $h$ & $\ploc/\tloc^d$, see \eqref{hdef:eq} \\
    $h_i$ & $\p h / \p\nu_i$ & $\nu_i$ & $\mu_i/\tloc$ \\
    \hline
%%break start
%  \end{tabular}
%
%  \begin{tabular}{||r|l||r|l||}
%    \hline
%    Symbol & Definition & Symbol & Definition \\
%    \hline
%%break end
    $T^{\mu\nu}$ & Stress tensor & $J^\mu_S$ & Entropy current \\
    $J^\mu_i$ & R-charge current & $u^\mu$ & $\dr x^\mu/\dr\tau =\gamma(1,\vec{v})$, fluid velocity \\
    $\omega_a$ & Angular velocities & $\gamma$ & $\prn{1-v^2}^{-1/2}$ \\
    $v^2$ & $\sum_a g_{\phi_a\phi_a}\omega_a^2$ & $P^{\mu\nu}$ & Projection tensor, $g^{\mu\nu}+u^\mu u^\nu$ \\
    $a^\mu,\vartheta,\sigma^{\mu\nu}$ & see \eqref{fluidtensors:eq} & $\zeta,\eta$ & Bulk, shear viscosity \\
    $q^\mu$ & Heat flux, see \eqref{heatcond:eq} & $\kappa$ & Thermal conductivity \\
    $q^\mu_{i}$ & Diffusion current, see \eqref{fluidtensors:eq} & $D_{ij}$ & Diffusion coefficients \\
    \hline
%%break start
%  \end{tabular}
%
%  \begin{tabular}{||r|l||r|l||}
%    \hline
%    Symbol & Definition & Symbol & Definition \\
%    \hline
%%break end
    $E,H$ & Total energy \eqref{charge:eq} & $S$ & Total entropy \eqref{charge2:eq} \\
    $L_a$ & Angular momenta \eqref{charge:eq} & $R_i$ & Total R-charges \eqref{charge2:eq} \\
    $\Omega_a$ & Angular velocities \eqref{chpotdef:eq} & $T$ & Overall temperature \eqref{chpotdef:eq} \\
    $\zeta_i$ & Overall chemical potentials \eqref{chpotdef:eq} & $\gpf$ & Partition function \eqref{gcp:eq} \\
    \hline
%%break start
%  \end{tabular}
%
%  \begin{tabular}{||r|l||r|l||}
%    \hline
%    Symbol & Definition & Symbol & Definition \\
%    \hline
% %%break end
    $\Theta^{\mu\nu}$ & Extrinsic curvature & $\Pi^{\mu\nu}$ & Boundary stress tensor \\
    $n^\mu$ & Unit normal to boundary & $h^{\mu\nu}$ & Induced metric of boundary \\
    $\tc$ & Integration constant \eqref{solution:eq} & $\kappa_i$ & Thermodynamic parameters \eqref{bbeosparam:eq} \\
    $m,q,s_i,H_i$ & Black hole parameters & $X,Y,Z$ & $1/(1+\kappa_i)$ \eqref{xyzkappa:eq} \\
    \hline
   %break start
  \end{tabular}
\end{center}
\caption{Summary of variables used. Numbers in parentheses refer to the equation where it is defined \label{tab:notation}}
\end{table}

%%%%%%%%%%%%%%%%%%%%%%%%%%%%%%%%%%%%%%%%%%%%%%%%%%%%%%%%%%%%%%%%%%%%%%%%%%
%\nocite{Lahiri:2007ae} \nocite{Son:2007vk} \nocite{Chong:2005da} \nocite{Chong:2005hr} \nocite{Chong:2006zx}
\bibliographystyle{utcaps}
%\bibliography{qft,string,adscft,gr,maths}
\bibliography{rotbhl-minimal}

\end{document}